\documentclass{aa}  

\usepackage{graphicx}	
\usepackage{amsmath}	
\usepackage{amssymb}	
\usepackage{txfonts}
\usepackage{lscape} 
 
\newcommand{\hii}{\,H\,{\small II}}

\begin{document}

   \title{Subaru Hyper-Supreme Cam observations of IC 1396:}

   \subtitle{Source catalogue, member population, and sub-clusters of the complex }

   \author{Swagat R Das\inst{1,2}\fnmsep \thanks{}
           , Saumya Gupta\inst{3,2}, Jessy Jose\inst{2}\thanks{}, Manash Samal\inst{4}, Gregory J. Herczeg\inst{5}
           , Zhen Guo\inst{6,7,8}, Surhud More\inst{9}, and Prem Prakash\inst{10}
          }

   \institute{Departamento de Astronomia, Universidad de Chile, Las Condes, 7591245 Santiago, Chile\\
              \email{swagat@das.uchile.cl / dasswagat77@gmail.com}
         \and
             Department of Physics, Indian Institute of Science Education and Research Tirupati, Yerpedu, Tirupati - 517619, Andhra Pradesh, India \\
             \email {jessyvjose1@gmail.com}
         \and
             Astronomy Unit, School of Physical and Chemical Sciences, Queen Mary University of London, Mile end Road, E14NS, London, UK
         \and
             Physical Research Laboratory (PRL), Navrangpura, Ahmedabad 380 009, Gujarat, India
         \and 
             Kavli Institute for Astronomy and Astrophysics, Peking University, Yi He Yuan Lu 5, Haidian Qu, Beĳing 100871, China
         \and 
             Instituto de Física y Astronomía, Universidad de Valparaíso, ave. Gran Bretaña, 1111, Casilla 5030, Valparaíso, Chile
         \and 
             Millennium Institute of Astrophysics,  Nuncio Monse{\~n}or Sotero Sanz 100, Of. 104, Providencia, Santiago,  Chile
         \and     
             Centre for Astrophysics Research, University of Hertfordshire, Hatfield AL10 9AB, UK
         \and 
             Inter University Centre for Astronomy and Astrophysics, Ganeshkhind, Pune 411007, India
         \and 
             Indian Institute of Technology Hyderabad, Kandi, Sangareddy, Telangana, India               
             }

   \date{}

 
  \abstract
   {Identifying members of star-forming regions is an initial step to analyse the properties of a molecular cloud complex. In such membership analysis, the sensitivity of a dataset plays a significant role in detecting stellar mass up to a specific limit, which is crucial for understanding various stellar properties such as disk evolution and planet formation across different environments.}
   {IC 1396 is a nearby classical \hii\ region dominated by feedback-driven star formation activity. In this work, we aim to identify the low-mass member populations of the complex using deep optical multi-band imaging with Subaru-Hyper Suprime Cam (HSC) over $\rm \sim 7.1~deg^2$ in $\rm r_2$, $\rm i_2$, and Y-bands. The optical dataset is complemented by UKIDSS near-infrared data in the J, H, and K bands. Through this work, we evaluate the strengths and limitations of machine learning techniques when applied to such astronomical datasets.}
   {To identify member populations of IC 1396, we employ the random forest (RF) classifier of machine learning technique. Random forest classifier is an ensemble of individual decision trees suitable for large, high-dimensional datasets. The training set used in this work is derived from previous Gaia-based studies, where the member stars are younger than $\sim$ 10~Myr. However, its sensitivity is limited to $\sim$ 20~mag in the $\rm r_2$ band, making it challenging to identify candidates at the fainter end. In this analysis, in addition to magnitudes and colours, we incorporate several derived parameters from the magnitude and colour of the sources to identify candidate members of the star-forming complex. By employing this method, we are able to identify promising candidate member populations of the star-forming complex. We discuss the associated limitations and caveats in the method and for improvment in future studies.}
   {In this analysis, we identify 2425 high-probability low-mass stars distributed within the entire star-forming complex, of which 1331 are new detections. Comparison of these identified member populations shows a high retrieval rate with Gaia-based literature sources, as well as sources detected through methods based on optical spectroscopy, {\it Spitzer}, $\rm H_{\alpha}/X-ray$ emissions, optical, and 2MASS photometry. The mean age of the member populations is $\rm \sim 2-4~Myr$, consistent with findings from previous studies. Considering the identified member populations, we present preliminary results by exploring the presence of sub-clusters within IC 1396, assessing the possible mass limit of the member populations, and providing a brief discussion on the star formation history of the complex. }
   {The primary aim of this work is to develop a method for identifying candidate member populations from a deep, sensitive dataset like Subaru-HSC by employing machine learning techniques. Although we overcome some limitations in this study, the method requires further improvements to address the caveats associated with such membership analysis.}

\titlerunning{Membership analysis of IC 1396 with Subaru-HSC}
\authorrunning{Das et al.}
   \maketitle
%

\section{Introduction} \label{intro}
Stars are the fundamental units of galaxies because they control the formation and evolution of their host galaxies. Most stars form in clustered environments \citep{2003ARA&A..41...57L} within giant molecular clouds (GMCs) through fragmentation and hierarchical collapse \citep{1981MNRAS.194..809L,2004ARA&A..42..211E,2004RvMP...76..125M,2007ARA&A..45..565M,2012MNRAS.426.3008K}. Since stars in a given cluster form from the same molecular cloud and roughly at the same time, the stars of a cluster have similar ages and chemical compositions. Thus, star clusters hold unique information regarding the kinematics and evolution of star-forming regions \citep{2019ApJ...871...46K,2019ApJ...870...32K,2020ApJ...900L...4P} as well as the origins and dynamics of host galaxies \citep{2008IAUS..246...13K,2016ApJ...828...75F}.

Star clusters are ideal laboratories for understanding the origin of the initial mass function (IMF) \citep{2006astro.ph.10687E,2010ARA&A..48..339B,2021MNRAS.504.2557D,2023arXiv230317424D}, disk evolution, and planet formation scenarios in clustered environments \citep{2018A&A...614A..22P,2019MNRAS.482..732C,2022EPJP..137.1071R,2023arXiv230518147D}, as well as mass segregation \citep{2000ASPC..211..215M,2002MNRAS.331..228D,2011A&A...532A.119O,2021A&A...645A..94N}. Many clusters are also associated with massive O and B-type stars, which disrupt their surrounding environment, controlling the star-formation ecosystem through the feedback effect of energetic stellar winds and \hii\ regions \citep{2012ApJ...755...20S,2014A&A...566A.122S,2016ApJ...822...49J,2017MNRAS.472.4750D,2021MNRAS.500.3123D,2022ApJ...926...25P}. The dynamics of clusters and their formation processes regulate the global outcome of star-formation efficiency and star-formation rate of a molecular cloud \citep{2012MNRAS.424..377D,2013MNRAS.430..234D,2015MNRAS.454..238W,2022AJ....164..129P}.

Thus, clusters serve as the black boxes of the parental molecular cloud, encapsulating all of its star-formation history. Analyzing young stellar clusters represents the initial step towards understanding how individual stars and protoplanetary disks evolve in a clustered environment, and how the outcomes of star-formation processes in a GMC depend on the stellar properties, kinematics, and feedback processes of associated clusters.

Therefore, the accurate detection of the member population of a star cluster becomes crucial to initiate such analyses. Membership analyses of clusters and star-forming complexes typically involve various techniques (\citealt{2023ApJ...948....7D}, and references therein), each with its own mass limit determined by the sensitivity of the respective dataset. For instance, the Gaia data \citep{2016A&A...595A...1G} have revolutionised the identification of clusters and their members in the solar neighbourhood \citep{2023ASPC..534...43Z}. Such studies utilise the successful application of various machine learning (ML) algorithms (e.g., \citealt{2018ApJ...869....9G,2018AJ....156..121G,2023ApJ...948....7D,2024MNRAS.tmp..456G}). With Gaia, identifying high-confidence members is feasible using astrometric parameters such as parallax and proper motions. However, the sensitivity of Gaia does not extend to the detection of very low-massive objects, including brown dwarfs ($\rm <0.08~M_{\odot}$) in relatively distant star-forming complexes ($\rm > 500~pc$, age $\rm \sim 2~Myr$, $\rm A_V \sim 2~mag$; \citealt{2015A&A...577A..42B}) (e.g., \citealt{2023arXiv230317424D}).

In this study, we focus on conducting a membership analysis of IC 1396, one of the most dominant Galactic bubbles powered by massive stars of the Trumpler 37 (Tr37) cluster \citep{1930LicOB..14..154T}, using deep optical data from the Subaru-Hyper Supreme Cam (HSC) and applying ML techniques. This represents the deepest optical photometric data available for this complex to date. Following our previous membership analysis of this complex using Gaia-DR3 \citep{2023ApJ...948....7D}, this work aims to identify more low-mass member stars within this star-forming complex.

This work is arranged as follows. We introduce the star-forming complex around Tr 37, the \hii\ region IC 1396, in Section \ref{det_ic1396}, along with the details of the membership analysis already carried out for this complex in the literature. The observational details and data reduction of Subaru-HSC data are given in Section \ref{obsrv}, including the details of complete catalogue preparation and its properties, along with details regarding archival datasets used in this work. Section \ref{hsc_imge} presents the properties of images obtained from the HSC observations.
We carry out the membership analysis in Section \ref{memb_analy} with the details of the ML techniques implemented in this study. Results regarding the member population, their properties, and sub-clusters detection are discussed in Section \ref{res}. Using the detected member population, we discuss the star-formation history of this complex, IC 1396, in Section \ref{diss}. We summarize our work in Section \ref{summ}.

\section{Details of the star-forming complex IC 1396} \label{det_ic1396}
\begin{figure*}
\centering
\includegraphics[scale=0.6]{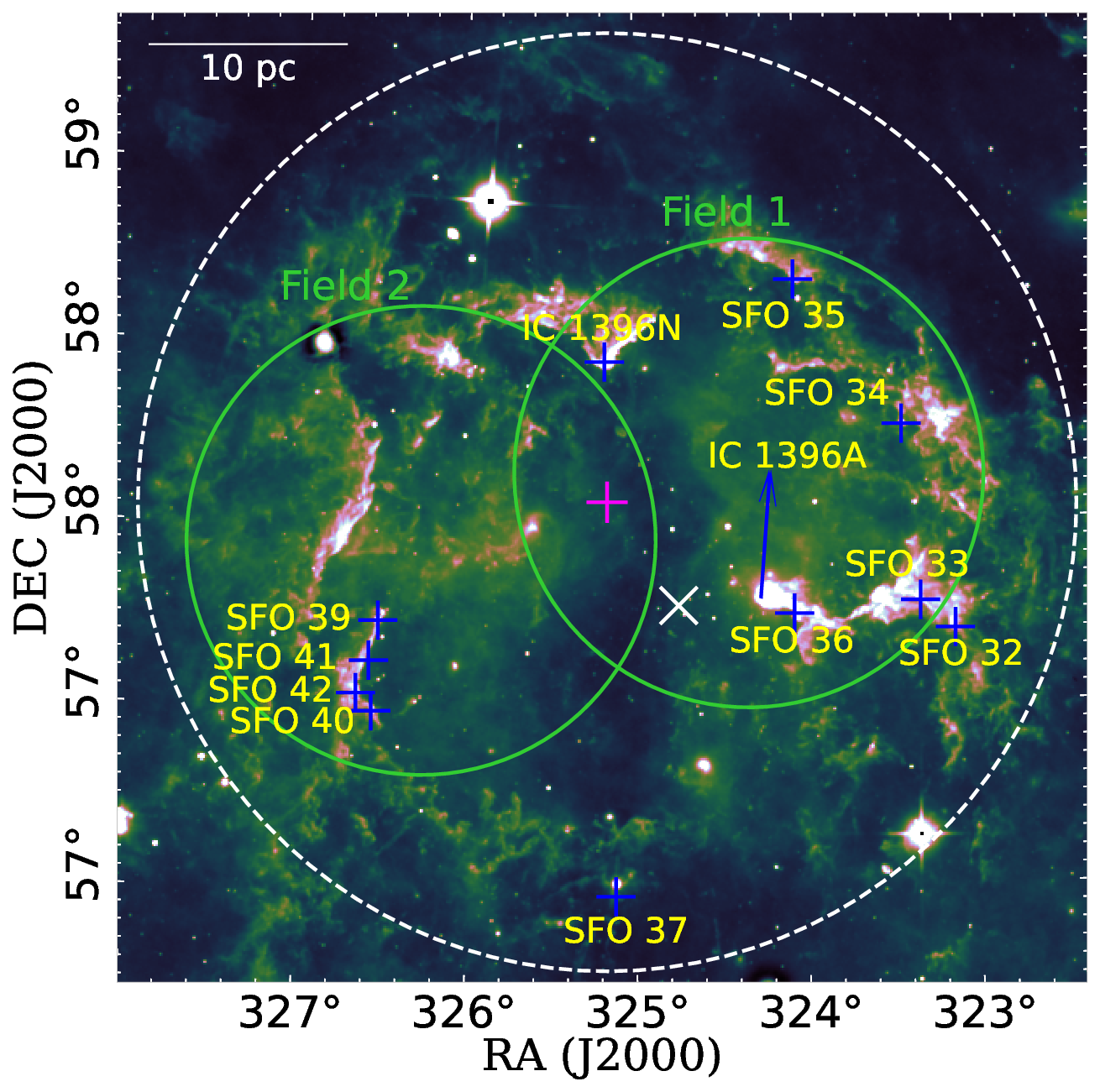}
\caption{The background colourscale represents the WISE $\rm 22~\mu m$ image of the IC 1396 complex. The position of the central exciting star(s) HD 206267 is denoted by a white cross mark ($\times$). The major globules identified by \citet{1991ApJS...77...59S} are indicated by blue `+' symbols along with their respective names. The dashed white circle with a radius of $1.5^\circ$, centred at $\rm \alpha = 21:40:39.28$ and $\rm \delta = +57:49:15.51$ (highlighted as a magenta `+' mark), fully encompasses the star-forming complex and serves as the region of interest for this study. The green circles represent the two fields observed with Subaru-HSC, each having a radius of $0.75^\circ$. A scale bar of 10~pc is provided in the top-left corner. }
\label{Tr37_wise}
\end{figure*}

IC 1396 (S131; \citealt{1959ApJS....4..257S}) is a textbook example of classical \hii\ regions (Fig. \ref{Tr37_wise}), with a simple, circular morphology. This is one of the most well-studied \hii\ regions of the Milky Way Galaxy. IC 1396 is located in the Cepheus OB2 region \citep{1999AJ....117..354D} and has relatively low ($\rm A_V<5$) foreground reddening \citep{2005AJ....130..188S,2012MNRAS.426.2917G,2012AJ....143...61N}, associated with the cluster Tr 37 \citep{1998AJ....116.2423P,1995ApJ...447..721P,1998ApJ...507..241P}. This star-forming complex is ionized by massive multiple stellar systems (e.g., HD 206267; \citealt{1995Obs...115..180S}) of spectral type $\rm O5 - O9 V$, located at its near centre \citep{2012A&A...538A..74P,2020A&A...636A..28M}.

Several distance estimations have been made for the complex from several previous works. Based on main-sequence (MS) fitting, \citet{2002AJ....124.1585C} obtain a distance of 870 pc, and later \citet{2019A&A...622A.118S} estimate a distance of $945^{+90}_{-73}$ pc based on Gaia-DR2. Two most recent distance estimations are $925\pm73$ pc \citep{2023AA...669A..22P} based on Gaia-EDR3 and $917\pm2.7$ pc \citep{2023ApJ...948....7D} based on Gaia-DR3. Within errors, all the distances fairly agree with each other. In this work, we use the distance of 917 pc obtained by \citet{2023ApJ...948....7D}.

This star-forming complex is relatively young, with an estimated age of $2-4$ Myr \citep{2005AJ....130..188S} based on spectroscopically identified members and some pre-main sequence (PMS) tracks. The recent analysis based on Gaia-EDR3 and Gaia-DR3 also yields a similar average age for this complex \citep{2023AA...669A..22P,2023ApJ...948....7D}.

IC 1396 is one of the dominant feedback-driven star formation complexes in the Milky Way. The feedback-driven activity of the central massive star(s) in the complex leads to the formation of several exciting structures such as bright-rimmed clouds (BRCs; \citealt{1991ApJS...77...59S}), fingertip structures, and elephant trunk structures. These structures are distributed in and around the complex \citep{1991ApJ...370..263S, 2005A&A...432..575F, 2012MNRAS.421.3206S}. The prominent BRCs of the complex IC 1396 A and IC 1396 N are often considered as the best examples of feedback-driven star formation \citep{2004AJ....128..805S,2006ApJ...638..897S,2007ApJ...654..316G,2010ApJ...717.1067C,2013A&A...559A...3S,2014MNRAS.443.1614P,2014A&A...562A.131S,2019A&A...622A.118S}.

In Fig. \ref{Tr37_wise}, we present the entire star-forming complex with a WISE $\rm 22~\mu m$ image. The energetic ultraviolet (UV) radiation and stellar wind of central massive star(s) shape the complex, which is seen as a cavity of radius $\sim 1.5^{\circ}$ clearing up most of the gas and dust and leading to the formation of BRCs, fingertip structures, and elephant trunk structures around the boundary. By correctly identifying the stellar members, we can better understand the star formation history of the complex. Many studies have already been carried out to find the associated young stellar object (YSO) populations of the complex based on various datasets. In this work, we identify the low-mass member stars up to the brown dwarf using the much deeper Subaru-HSC optical catalogue.

\subsection{Members from previous studies} \label{memb_lit}
This section briefly describes details of the member populations of those past studies. The population of IC 1396 consists of members identified using several datasets and methodologies, which is summarized in Table \ref{work_mem}. All the different studies have covered different regions of the star-forming complex. Along with the reference, we provide the method of the work, the area covered by the individual studies, and the number of stars retrieved by the individual works. A total of 2178 member stars were detected towards IC 1396 from all the non-Gaia based studies in the literature.

Gaia-based member populations have also been carried out for this complex. \citet{2018AA...618A..93C} conducted a membership analysis of a large sample of 1229 clusters of the Milky Way using Gaia-DR2 data and detected 460 candidate member stars towards IC 1396. \citet{2023AA...669A..22P} identified an additional 334 sources from Gaia-EDR3 catalogue. The work of \citet{2023ApJ...948....7D} identified 1243 stars as members of the star-forming complex from Gaia-DR3 data. All the Gaia-based studies have revealed 1468 member stars of IC 1396.

The most recent study by \citet{2024MNRAS.tmp..456G} identifies brown dwarfs within the central $\rm 22\arcmin$ radius of the complex. 
In this work, we obtain the deepest membership catalogue of the complex within the circular region of radius $1.5^\circ$, as shown in Fig. \ref{Tr37_wise}. The region of interest in this work is similar to the area covered in several previous studies \citep{2011MNRAS.415..103B,2012AJ....143...61N,2023ApJ...948....7D}. In Section \ref{comp_lit}, we compare the members identified in this work with those in the literature.

\begin{landscape}
\begin{table}
\centering
\caption{Area covered and candidate members obtained in previous works of literature.}
\label{work_mem}
\begin{tabular}{cccccc}
\hline 
Work & Method & No. of identified members & Radius (degree) & RA (J2000) & DEC (J2000) \\
\hline

\citet{2002AJ....124.1585C}   & Optical Spectroscopy & 66  & 0.5  & 21:39:09.89 & +57:30:56.07 \\
\citet{2006AJ....132.2135S}   & Optical Spectroscopy & 172 & 0.6  & 21:37:54.41 & +57:33:15.32 \\
\citet{2013AA...559A...3S}    & Optical Spectroscopy & 67  & 0.25 & 21:37:03.17 & +57:29:05.43 \\

\citet{2004ApJS..154..385R}   & {\it Spitzer} MIR & 17  & 0.12  & 21:36:33.09 & +57:29:13.83 \\
\citet{2006ApJ...638..897S}   & {\it Spitzer} MIR & 57  & 0.15 & 21:36:39.73 & +57:29:28.45 \\
\citet{2009ApJ...702.1507M}   & {\it Spitzer} MIR & 69  & 0.15 & 21:36:36.32 & +57:29:54.78 \\
\citet{2013AJ....145...15R}   & {\it Spitzer} MIR & 14 & 0.50 & 21:33:28.23 & +58:04:59.51 \\

\citet{2011MNRAS.415..103B}   & $\rm H_{\alpha}$ emission & 158 & 1.5 & 21:40:00.43 & +57:26:42.60 \\
\citet{2012AJ....143...61N}   & $\rm H_{\alpha}$ emission & 639 & 1.4 & 21:39:48.76 & +57:30:31.56 \\

\citet{2007ApJ...654..316G}   & X-ray emission & 24  & 0.1  & 21:40:36.73 & +58:15:37.51 \\
\citet{2009AJ....138....7M}   & X-ray emission & 39  & 0.15 & 21:38:54.67 & +57:29:17.61 \\
\citet{2012MNRAS.426.2917G}   & X-ray emission & 457 & 0.25 & 21:37:05.85 & +57:32:30.06  \\

\citet{2021AJ....162..279S}   & NIR, MIR and X-ray & 421 & 0.37 & 21:33:59.30 & +57:29:30.76\\

\citet{2004AJ....128..805S} & Optical photometry, spectroscopy and variability  & 53  & 0.35 & 21:39:06.36 & +57:32:56.86 \\
\citet{2005AJ....130..188S} & Optical/2MASS photometry, spectroscopy, and variability & 249 & 0.65 & 21:39:28.19 & +57:31:51.63 \\
\citet{2010ApJ...710..597S} & Optical photometry and spectroscopy & 102 & 0.50 & 21:38:08.05 & +57:31:02.52 \\
\citet{2019ApJ...878....7M} & NIR variability & 359 & 0.60 & 21:35:35.94 & +57:40:02.68 \\

\citet{2018AA...618A..93C}   & Gaia-DR2  & 460  & 0.7 & 21:38:58.80 & +57:30:50.40 \\
\citet{2023AA...669A..22P}   & Gaia-EDR3 & 334  & 2.0 & 21:38:56.86 & +57:44:35.81 \\
\citet{2023ApJ...948....7D}  & Gaia-DR3  & 1243 & 1.5 & 21:40:39.28 & +57:49:15.51 \\

\citet{2024MNRAS.tmp..456G}  & Subaru-HSC & 458 & 0.37 & 21:38:57.62 & +57:29:20.54 \\
\hline
\end{tabular}
\\ The methods given in column 2 present the dominant mechanisms of the respective works. For \citet{2023AA...669A..22P}, we used sources detected only from the Gaia-EDR3 data set. Complete details of the methods used in the studies can be found in the respective papers.  
\end{table}
\end{landscape}

\section{Observation, data reduction, and catalogue preparation} \label{obsrv}
\subsection{Details of observations} \label{obs_det}
The Subaru Telescope \citep{2004PASJ...56..381I} is an 8.2 m class telescope operated by the National Astronomical Observatory of Japan (NAOJ). The Hyper Supreme Cam\footnote{More details regarding the HSC instrument can be found at https://www.subarutelescope.org/Observing/Instruments/HSC/index.html.} (HSC; \citealt{{2018PASJ...70S...1M},{2018PASJ...70S...2K},{2018PASJ...70...66K},{2018PASJ...70S...3F},{2018PASJ...70S...6H},{2018PASJ...70S...7C}}) is a 1.77 $\rm deg^2$ mosaic CCD camera mounted at the prime focus of the telescope. The HSC consists of a total of 116 2k$\times$4k CCDs, out of which 104 CCDs cover a $1.5^{\circ}$ field-of-view in diameter with a pixel scale of 0.17$\arcsec$. A single exposure of the HSC is termed a {\it visit}, which is a pair of single exposures of the same pointing processed at the earliest stages. The datasets are identified by the combination of the {\it visit} ID and CCD ID.

The wide-field imaging of the region towards IC 1396 was conducted using Subaru-HSC on 17 September 2017 (PI: Jessy Jose, S17B0108N) in excellent imaging conditions (seeing $\sim0.5-0.7\arcsec$ and $1.0\leqslant$airmass$\leqslant2.0$). This observation comprised two pointings, each with a radius of $0.75^{\circ}$ in three filters ($\rm r_2,~ i_2,~ \text{and} ~Y$)\footnote{Wavelength of the three bands are 622.6~nm, 776.7~nm, and 1005.1~nm and their wavelength coverage are 550--695~nm, 695--845~nm, and 930--1070~nm in $\rm r_2$, $\rm i_2$, and Y-bands respectively \citep{2018PASJ...70...66K}.}. The two pointings of these observations are depicted in Fig. \ref{Tr37_wise}. Details regarding the number of frames and integration time in each filter are provided in Table \ref{hsc_obs}.

\begin{table*}
\centering
\caption{Details of number of exposures and integration time.}
\label{hsc_obs}
\begin{tabular}{ccccc}
\hline
&  \multicolumn{2}{c}{\underline{~~~~~~~~~~~~~~~~~~~~~~~~~~~~~~~$\rm Field 1$~~~~~~~~~~~~~~~~~~~~~~~~~~~~~~~}} & \multicolumn{2}{c}{\underline{~~~~~~~~~~~~~~~~~~~~~~~~~~~~~~~$\rm Field 2$~~~~~~~~~~~~~~~~~~~~~~~~~~~~~~~}} \\
Filter & No. of Exposures & Total integration time (s) & No. of Exposures & Total integration time (s) \\
\hline
$\rm r_2$ & 14+3 & 4290 & 14+3 & 4290 \\
$\rm i_2$ & 8+3  & 2475 & 8+3  & 2475 \\
$\rm Y$   & 4+3  & 621  & 5+3  & 761  \\
\hline                
\end{tabular}
\\ Numbers in second and fourth columns coressponds to long + short exposure frames. \\
Each long exposure frames of $\rm r_2~and~i_2$ bands has exposure time of 300~s and $\rm Y$ band has 150~s exposure frames. \\
Each short exposure frame has exposure time of 30~s, 25~s, and 7~s in $\rm r_2$, $\rm i_2$, and Y-bands, respectively. 
\end{table*}

\subsection{Data reduction} \label{data_reduc}
The observed raw data was downloaded from STARS\footnote{https://stars2.naoj.hawaii.edu/} (Subaru Telescope archive system). We processed the observed raw data using the HSC Pipeline (\textsc{hscPipe}\footnote{\rm https://hsc.mtk.nao.ac.jp/pipedoc/pipedoc\_6\_e/}) version 6.7. Further details on \textsc{hscPipe} V6.7 can be found in the description of data reduction in \citet{2021MNRAS.508.3388G} for Cygnus OB2 complex. This data reduction followed the same procedure to obtain the final catalogue for the IC 1396 complex. In the following, we briefly explain the reduction process of \textsc{hscPipe}.

The complete data reduction process with the pipeline consists of multiple steps: (1) single-visit processing, (2) joint calibration, (3) coaddition, and (4) coadd processing/multiband analysis.
During the initial step of single-visit processing, the pipeline corrects each science frame by performing overscan subtraction, bias correction, dark subtraction, flat-fielding, and fringe subtraction. The pipeline also identifies and interpolates defects such as bad pixels, cross-talk, and saturated pixels through Instrument Signature Removal (ISR).
Following this, both the astrometry and photometry calibration processes commence. For the calibration process, \textsc{hscPipe} first selects point sources above a $50\sigma$ threshold and then performs calibration using the Pan-STARRS DR1 PV3 reference catalogue. By default, \textsc{hscPipe} employs the `Optimistic B' matching algorithm for this process, which is suitable for low-density fields such as extragalactic regions. However, it performs poorly for high-density regions \citep{2021MNRAS.508.3388G}. Given that our observed region is a highly dense Galactic star-forming region, we opt for the `Pessimistic B' algorithm.

Then the process of coaddition is performed to create a deeper image from multiple images or {\it visits}. In this step, the pipeline warps the {\it visits} to the SkyMap coordinate system (warp) and then coadds all {\it visits} together using the WCS and flux scale determined by mosaicking. Additionally, this step applies sky correction, writing a new background model.
Following coaddition, \textsc{hscPipe} conducts multiband analysis to generate the final photometric catalogue for each band. During this stage, the pipeline extracts sources from the coadded images using a $5\sigma$ threshold value and retrieves photometry catalogues from the coadded images in each filter. To facilitate this process, footprints, which are regions above the threshold level and consist of many discrete sources, are generated. Following the methodology outlined in \citet{2021MNRAS.508.3388G}, we maintain the footprint size at $10^{10}$ pixels for the $\rm i_2$ and Y filters, and $\rm 10^{11}$ pixels for the $\rm r_2$ filter to ensure maximum detection in the presence of nebulosities.
Subsequently, we follow the same point source selection procedure by applying certain flags, as described by \citet{2021MNRAS.508.3388G}. Details regarding the catalogue flags are provided in \citet{2018PASJ...70S...5B}, and \citet{2021MNRAS.508.3388G} explains how applying certain flags removes spurious sources from the final catalogue. To prepare the final catalogue, we retain sources with magnitude errors $<0.2$, ensuring that more than 90\% of sources lie within that error value.

To compile the final HSC source catalogue of IC 1396, we merged the individual photometric catalogues generated in the three bands. Our merged catalogue contains a total of 859867 sources with photometry in all three bands. The final catalogue tends to saturate at approximately 14~mag in the $\rm r_2$ and $\rm i_2$ bands and around 12~mag in the Y-band, similarly for the Cygnus OB2 complex \citep{2021MNRAS.508.3388G}. As depicted in Fig. \ref{Tr37_wise}, the HSC observations do not cover the entire $1.5^{\circ}$ circular region (as indicated by the white dashed circle) of the star-forming complex IC 1396. To account for stars in the region not covered by HSC observations and stars above the saturation limit of HSC, we incorporate stars from the Pan-STARRS catalogue, the details of which are provided in the following section.
It is important to note that the inclusion of stars from the Pan-STARRS catalogue enhances the detection of member populations beyond the HSC survey area. We like to mention that preliminary results on the central part of the complex using this deep HSC photometry has been published in \citet{2024MNRAS.tmp..456G}, which reports the presence of low-massive objects like brown dwarfs. Further details regarding membership analysis of the entire complex are explained in subsequent sections.

\subsection{Data quality and completeness} \label{quality}

\begin{figure*}
\centering
\includegraphics[scale=0.32]{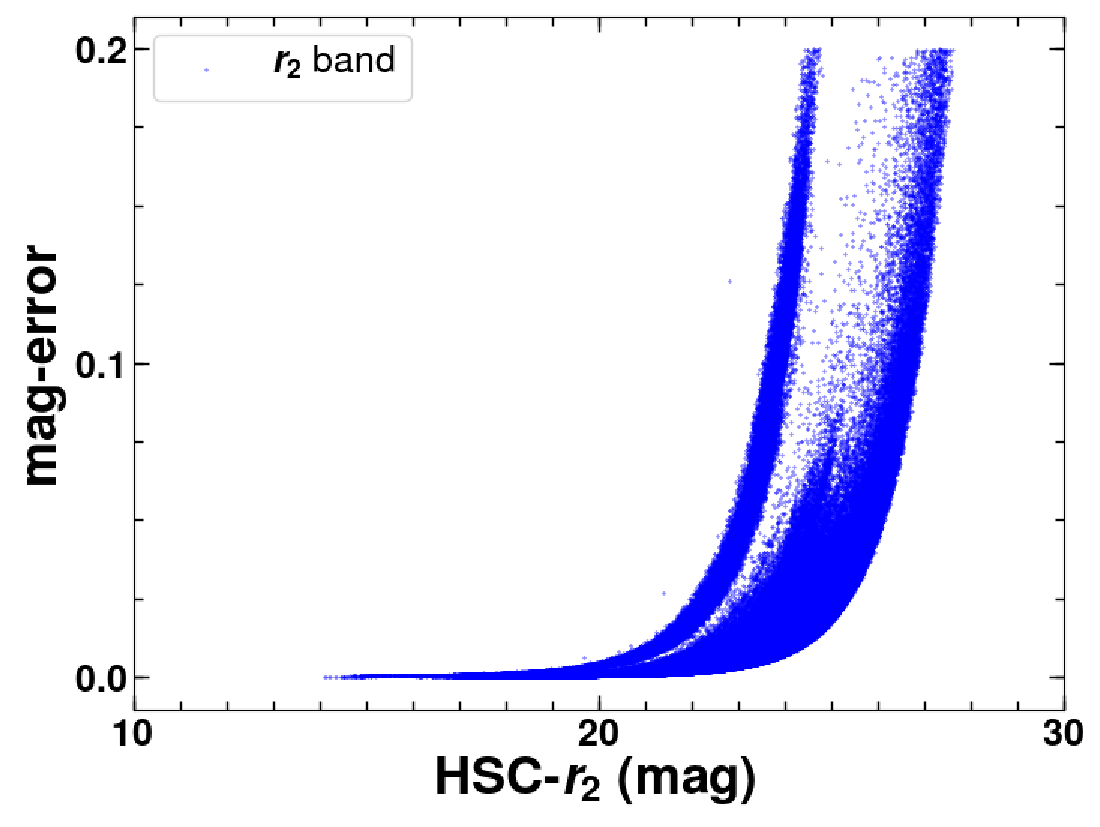}
\includegraphics[scale=0.32]{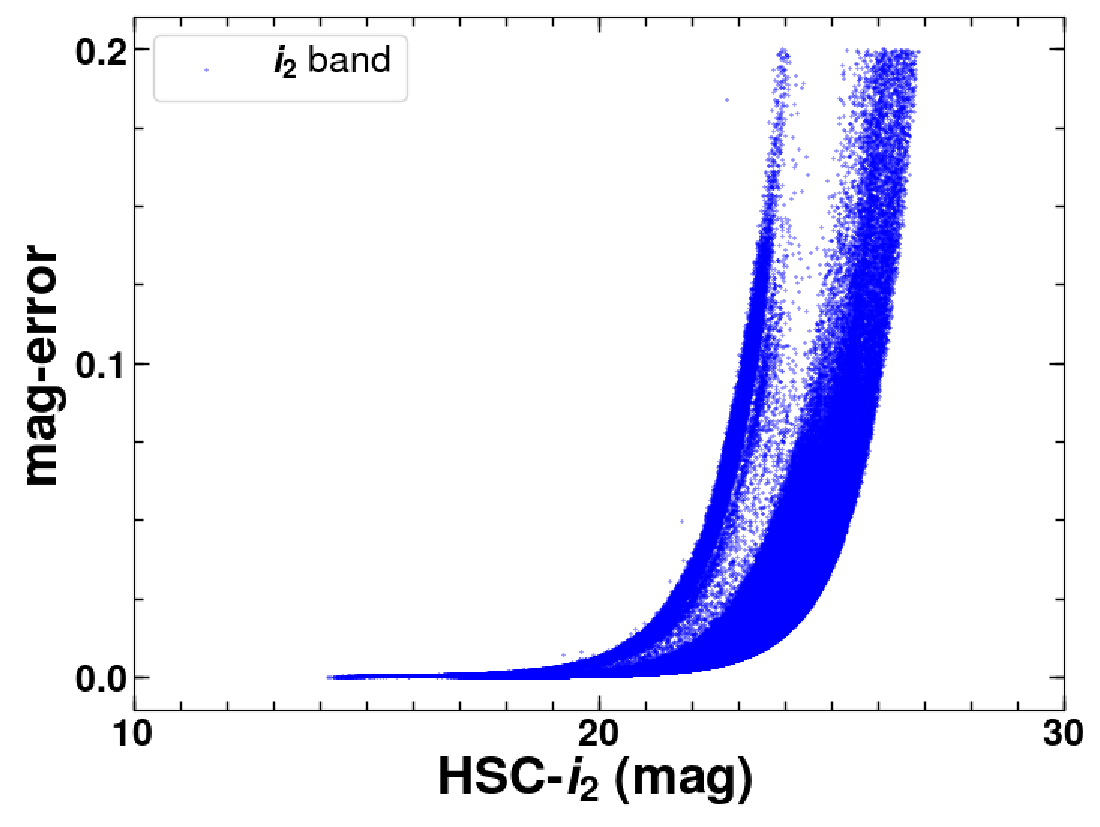}
\includegraphics[scale=0.32]{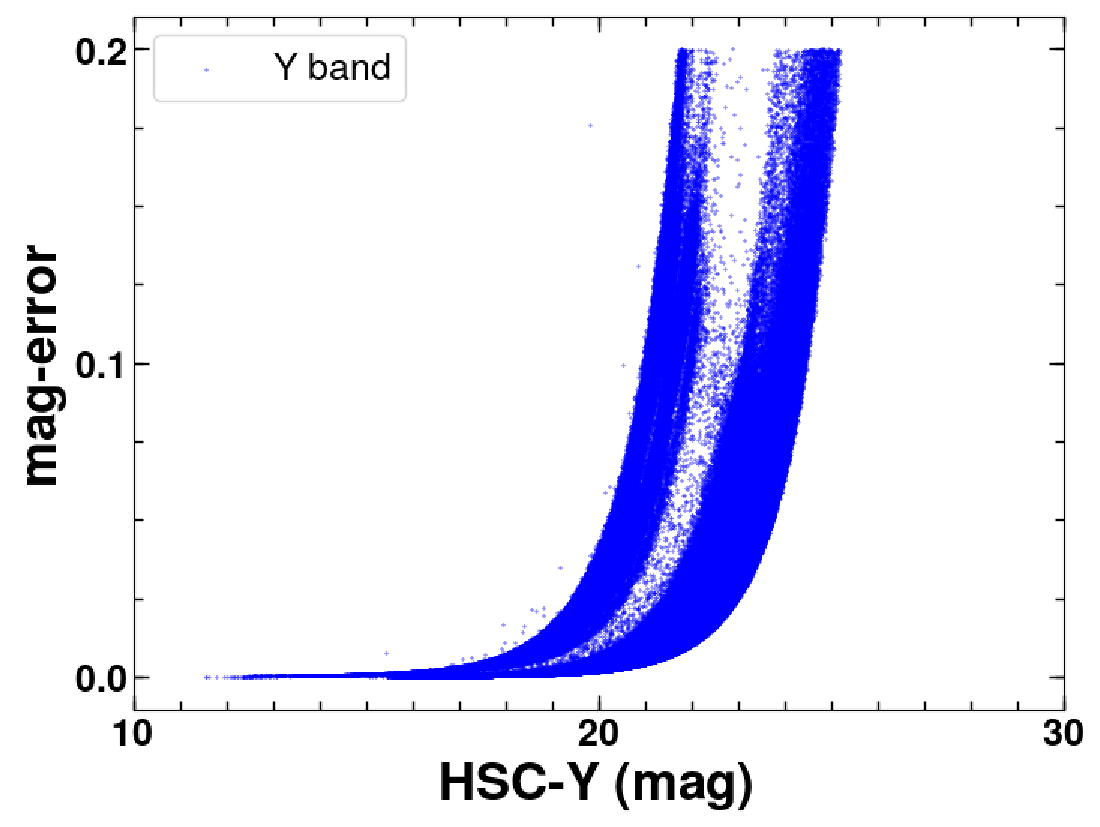}
\caption{Plots showing the distribution of photometric magnitude with respect to its error in individual HSC bands. The two separate populations are due to the photometry obtained from the long and short exposure frames.}
\label{mag_err}
\end{figure*}

\begin{figure}
\centering
\includegraphics[scale=0.30]{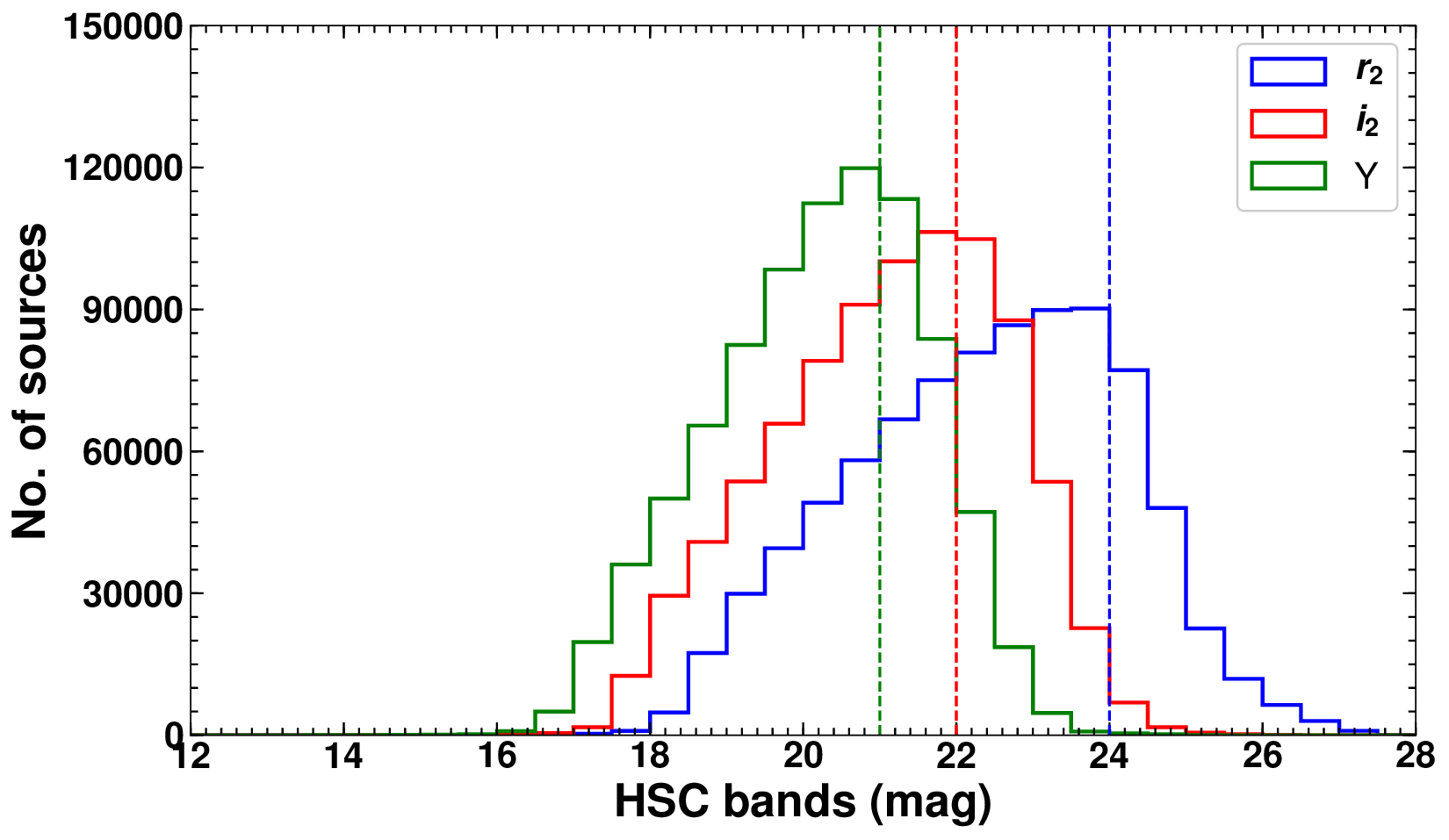}
\caption{Histogram distribution of photometric magnitudes of the 859867 stars with three HSC bands. The dashed line marks the turnover point corresponding to its 90\% completeness in HSC bands. The bin size of the plot is 0.5~mag. }
\label{mag_hist}
\end{figure}

In Fig. \ref{mag_err}, we present the magnitude distribution relative to error for the three bands. This plot illustrates two distributions primarily due to the photometry obtained from the long and short exposure frames. We also assess the quality of the retrieved photometric catalogue.

In Fig. \ref{mag_hist}, we depict the histogram distributions of the photometric magnitudes in individual bands. The limiting magnitudes are 27.6, 26.6, and 25.2 in the $\rm r_2$, $\rm i_2$, and Y bands, respectively. The HSC observations of the Cygnus OB2 complex also exhibit similar magnitude limits \citep{2021MNRAS.508.3388G}. We determine the 90\% completeness limits to be 24, 22, and 21 mag in the $\rm r_2$, $\rm i_2$, and Y bands, respectively.

\subsection{Archival data sets} \label{archive}
In this study, we have additionally used datasets from other archives within the circular region of radius $1.5^\circ$ centred at $\alpha = 21:40:39.28$ and $\delta = +57:49:15.51$ (as indicated by the white dashed circle in Fig. \ref{Tr37_wise}). Below, we present brief details of these datasets.

\subsubsection{Optical data from Pan-STARRS} \label{pan-star}
We use optical data in the $\rm r_2$, $\rm i_2$, z, and Y-bands\footnote{Wavelength of the four bands are 617~nm, 752~nm, 866~nm, and 962~nm and their wavelength coverage are 550--689~nm, 690--819~nm, 818--922~nm, and 918--1001~nm in $\rm r_2$, $\rm i_2$, z, and Y-bands, respectively \citep{2012ApJ...750...99T}.} from the archive of the Pan-STARRS1 survey (PS1). For detailed information regarding the survey and data release, please refer to \citet{2016arXiv161205560C,2020ApJS..251....7F}. We downloaded the PS1 catalogue from the Vizier online data archive\footnote{https://vizier.cds.unistra.fr/viz-bin/VizieR?-source=II/349} \citep{2017yCat.2349....0C} within the circular region of radius $1.5^\circ$ (refer to Fig. \ref{Tr37_wise}).

For our analysis, we retained only the data of good quality based on specific criteria. Firstly, all stars must have photometric values in all four bands $\rm r_2$, $\rm i_2$, z, and Y-bands. Additionally, the quality flag\footnote{https://outerspace.stsci.edu/display/PANSTARRS/PS1+Object+Flags} of all stars should fall within the range of 16 to 64, indicating good quality stars. To ensure stars with good photometric quality, we selected stars with an error in all bands less than 0.2 and photometric magnitude difference\footnote{https://outerspace.stsci.edu/display/PANSTARRS/PS1+Comparison+ \\of+different+photometric+measures} in PSFmag and Kronmag less than 0.05. Applying these constraints enabled us to obtain 800125 stars of good photometric quality for further analysis.

To merge the PS1 catalogue with the HSC catalogue, we initially convert the PS1 stars into the HSC photometric system using the conversion equations provided in \citet{2021MNRAS.508.3388G}. During this merging process, we prioritise the photometry from HSC for stars that are common in both the HSC and PS1 catalogues. Additionally, we include additional stars from PS1 in regions not covered by the HSC observations. It is worth noting that the completeness of that region tends towards the brighter end due to Pan-STARRS photometry. As a result of this merging process, we incorporate 462252 additional stars from PS1 into the HSC catalogue. Consequently, the combined catalogue encompasses 1322119 stars within the IC 1396 complex. For simplicity, henceforth, we refer to the complete HSC plus PS1 catalogue as the HSC catalogue.

\subsubsection{NIR data from UKIDSS and 2MASS} \label{nir_data}
In this study, we also use near-infrared (NIR) datasets in the JHK\footnote{In UKIDSS, wavelength of the three bands are 1.2483~$\rm \mu m$, 1.6316~$\rm \mu m$, 2.2010~$\rm \mu m$ and wavelength coverage are 1.169--1.328~$\rm \mu m$, 1.492--1.748~$\rm \mu m$, and 2.029--2.380~$\rm \mu m$, in J, H, and K bands, respectively \citep{2007MNRAS.379.1599L}. In 2MASS, wavelength of the bands are 1.25~$\rm \mu m$, 1.65~$\rm \mu m$, and 2.16~$\rm \mu m$ in J, H, and K bands, respectively \citep{2006AJ....131.1163S}.} bands from the archives of the UKIRT Infrared Deep Sky Survey (UKIDSS; \citealt{2007MNRAS.379.1599L}) and the Two Micron All Sky Survey (2MASS; \citealt{2006AJ....131.1163S}). We obtained stars from both catalogues within the same circular region of radius $1.5^{\circ}$.

We downloaded the UKIDSS\footnote{http://wsa.roe.ac.uk/dbaccess.html}11PLUS Galactic Plane Survey (GPS) data. This data has a resolution of approximately $1^{\prime\prime}$ and a typical limiting magnitude of K = 18.4~mag \citep{2006MNRAS.367..454H}, with a saturation limit of K $\sim$ 13~mag \citep{2008MNRAS.391..136L}. To ensure that only stars of good quality are retained, we followed the criteria discussed in \citet{2008MNRAS.391..136L} and \citet{2021MNRAS.504.2557D}. We filtered out sources classified as `noise' in the catalogue and retained those with {\tt mergedClass} values of -1 or -2, indicating a star or probable star, respectively. Additionally, to account for duplicate catalogue entries, we used the attribute {\tt PriOrSec}. We retained stars with a magnitude in the K-band greater than 13~mag.

To account primarily for the brighter end of the UKIDSS catalogue, we use NIR data in the JHK bands from the archive of 2MASS and merged it with the UKIDSS catalogue. Prior to merging, we converted the 2MASS photometry into the UKIDSS photometry system using the conversion equations provided by \citet{2001AJ....121.2851C}. For simplicity, we refer to this combined NIR catalogue as the UKIDSS catalogue.

There are 802100 common stars between the HSC and UKIDSS catalogues\footnote{The catalogues are cross-matched within a matching radius of $1\arcsec$} with photometry magnitudes in six bands. For simplicity in further analysis, we refer to this as the HSC+UKIDSS catalogue, which we utilize for membership analysis. Including UKIDSS data enhances the effectiveness and reliability of detecting member stars within the star-forming complex.

\begin{figure*}
\centering
\includegraphics[scale=0.41]{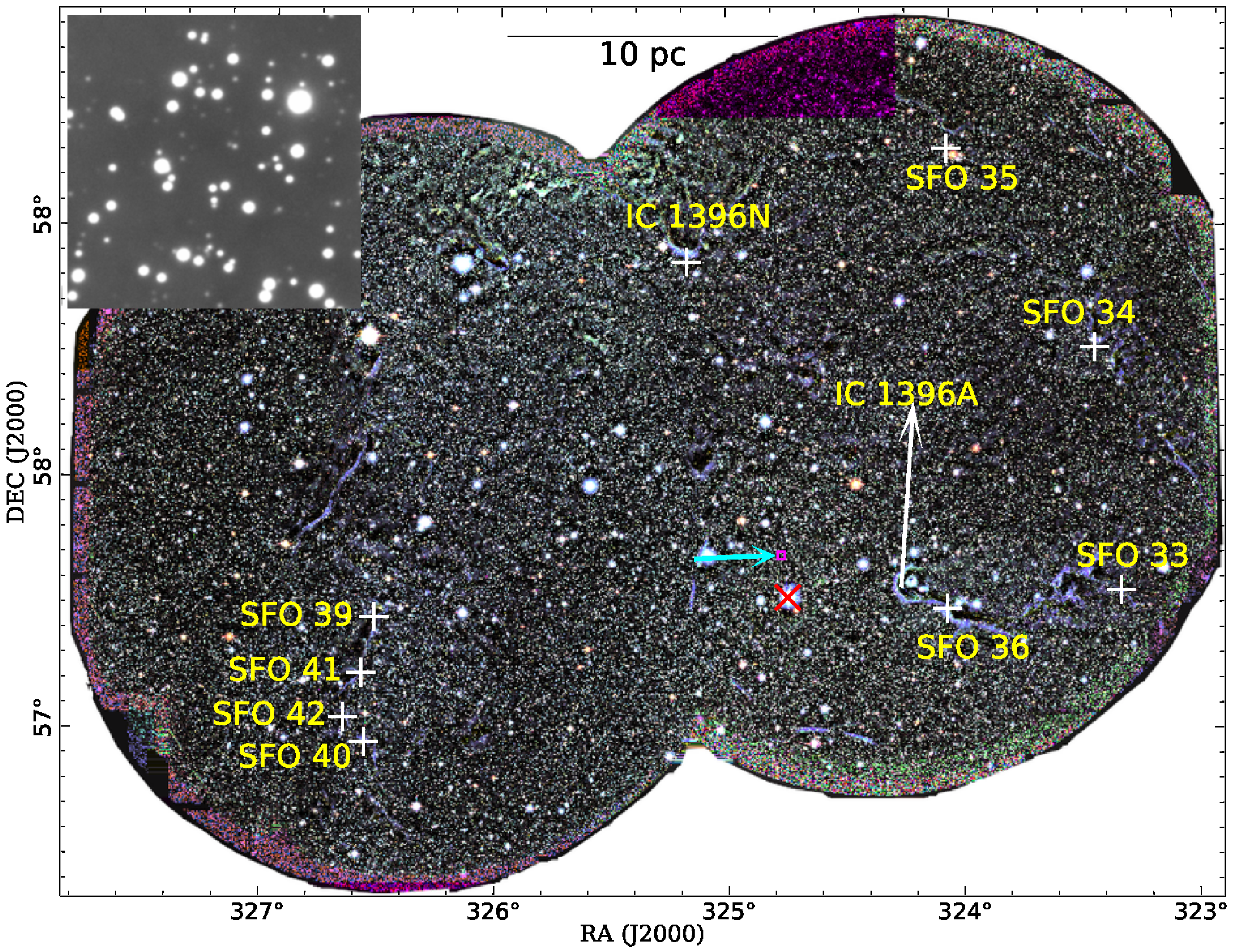}
\caption{Colour-composite (Y-red, $\rm i_2$-green, and $\rm r_2$-blue) image showing the entire area covered by the Subaru-HSC observations of IC 1396. The position of the central massive star(s) HD 206267 is denoted by a red cross mark ($\times$). Prominent BRCs of the complex are marked similarly to those in Fig. \ref{Tr37_wise}. The image appears pink, primarily around the border and towards the northwest region. This is mainly due to missing data in one of the $\rm r_2$ or $\rm i_2$ bands. In the northwest region, the pink colour is due to the missing $\rm i_2$ band, likely caused by bad pixels in that band. However, the final catalogue does not include sources affected by bad pixels or regions, and only stars with photometry in all three bands are included and spurious sources have been removed by applying certain flags (see Section \ref{data_reduc}).  {\it Inset:} A zoomed-in region in the $\rm r_2$ band, covering a $1\arcmin\times1\arcmin$ area centred at $\alpha = 21:39:04.34$ and $\delta = +57:35:10.95$, demonstrates the impressive resolution of the HSC observations. Location of this region is shown as a magenta square and highlighted through the cyan arrow. A scale bar of 10~pc is shown at the top of the image. }
\label{rgb_full}
\end{figure*}

\section{HSC image of IC 1396} \label{hsc_imge}
In Fig. \ref{rgb_full}, we present the colour-composite image of the entire region observed by Subaru-HSC. The renowned BRCs of the complex, including IC 1396 A and IC 1396 N, are visible in the image. The image unveils the diffuse nebulosity and distinct, bright rims of the BRCs, predominantly in the $\rm r_2$ band. Additionally, sharp nebular features are observable in the peripheral regions of the \hii\ region.

\begin{figure}
\includegraphics[scale=0.36]{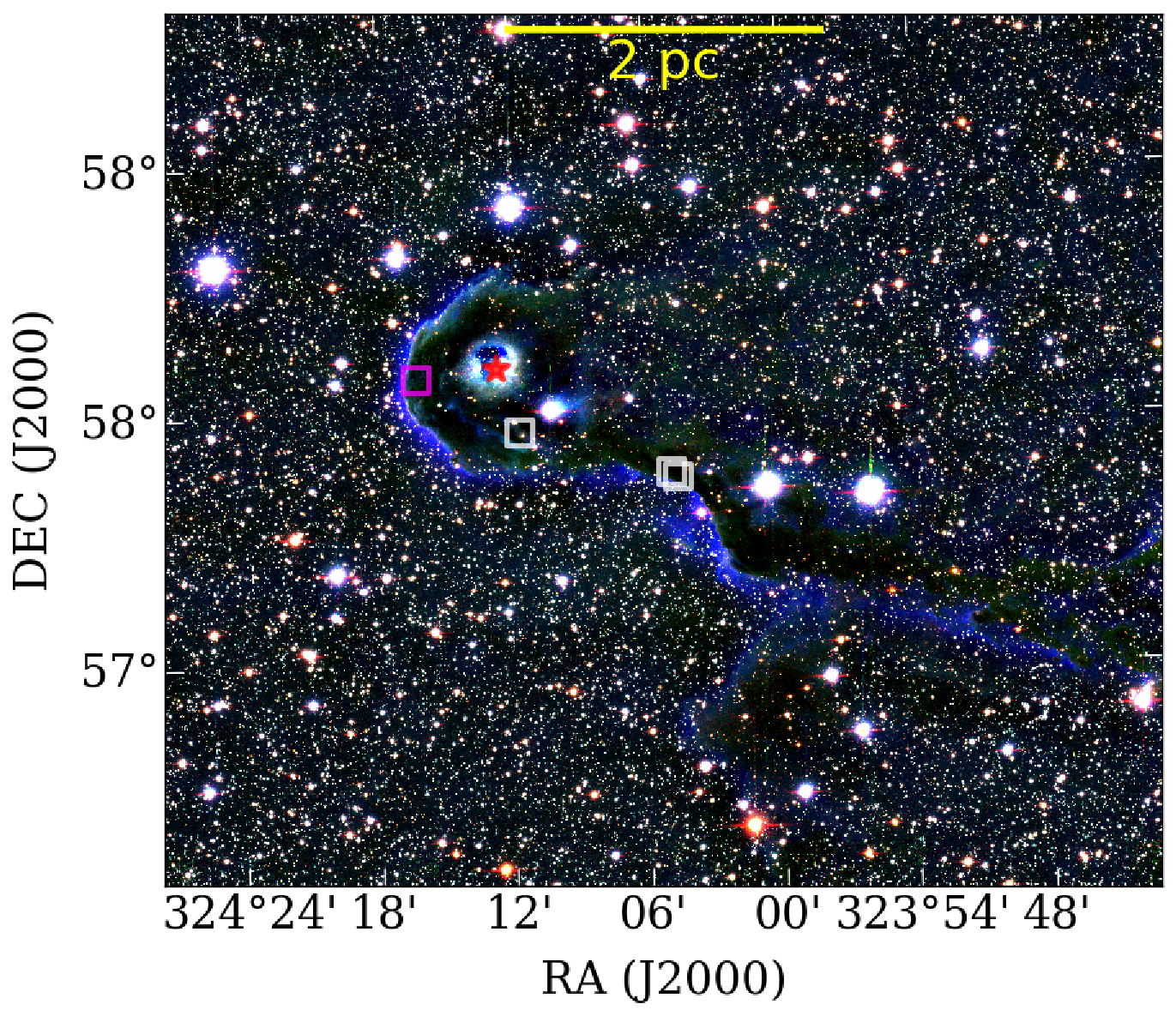}
\includegraphics[scale=0.4]{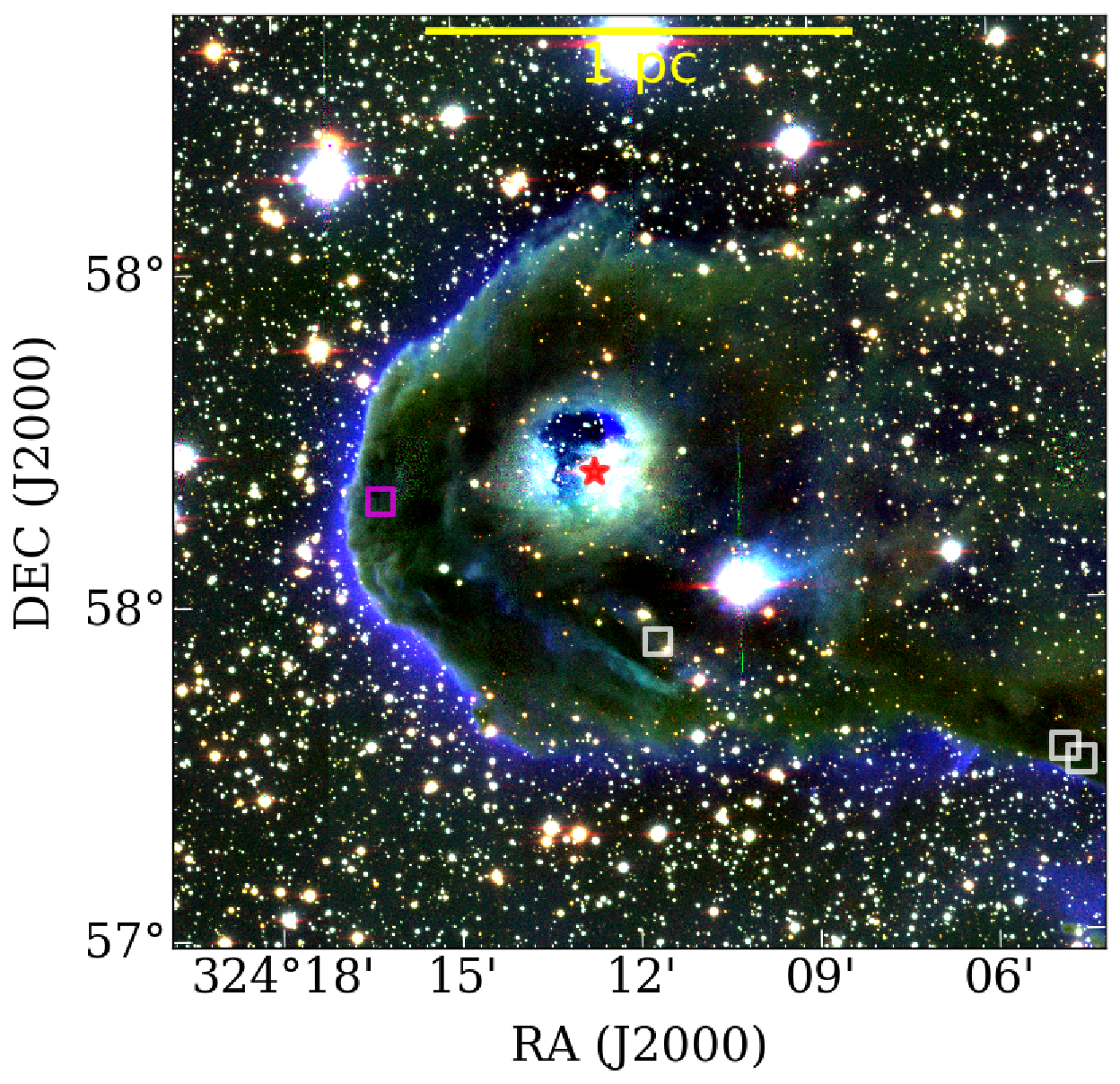}
\caption{Same as Fig. \ref{rgb_full}, but for the region towards the BRC IC 1396 A. {\it Top}: A larger view of BRC IC 1396 A. {\it Bottom}: A zoom-in view of the region around the head of the BRC complex. In these figures, we have also overplotted a few young sources \citep{2014A&A...562A.131S}, shown as squares. The Magenta square is Class 0, and the white squares are Class I stars. The red star marks the location of Herbig Ae star V 390 Cep. Coordinates of these stars are taken from \citet{2014A&A...562A.131S}. Scale bars of 2~pc and 1~pc are shown on the top and bottom images, respectively. }
\label{rgb_ic1396a}
\end{figure}

Fig. \ref{rgb_ic1396a} illustrates the colour-composite image depicting the area surrounding the BRC IC 1396 A. The signatures of feedback-driven activity within the star-forming complex are discernible across all the HSC bands, with the $\rm r_2$ band primarily highlighting the head of IC 1396 A alongside the extended tail of the BRC. The morphology of the BRC in $\rm r_2$ corresponds to the WISE $\rm 22~\mu m$ infrared image, revealing both the head and tails of the structure (refer to Fig. \ref{Tr37_wise}). Specifically, the $\rm r_2$ band is ideal for examining the signatures of feedback and shocks-driven activity, as this band covers the $\rm H_{\alpha}$ line (656.3~nm), which is a tracer of ionised gas \citep{2022ApJ...926...25P}.

In the close-up view of the BRC head (Fig. \ref{rgb_ic1396a}), sharp diffuse emissions are prominently visible around the head, accompanied by numerous small-scale filamentary structures behind the sharp, bright rim of the BRC head. The formation of the head in IC 1396 A arises from combined feedback-driven activity, primarily influenced by the massive star(s) HD 206267 and the winds from the Herbig Ae star V 390 Cep. A distinctive cavity surrounding the Herbig Ae star is observed in HSC images, attributed to local feedback-driven activity, wherein the energetic radiations from the Herbig Ae star disperse the gas, resulting in a void visible in the HSC images. Such a property of the Herbig Ae star has also been observed previously, where it was reported that, due to feedback-driven activity, this intermediate-mass star is creating an ionisation hole around its vicinity \citep{2019A&A...622A.118S}.

A Class 0 star was identified within the BRC head through {\it Herschel} PACS data, supporting the notion of multiepisodic star formation activity within the BRC. \citet{2014A&A...562A.131S} speculated that the Class 0 star might have been triggered via radiative-driven implosion (RDI) induced by the massive star(s) HD 206267. However, the Class 0 star remains undetected in HSC images because it is deeply embedded, as the HSC images primarily capture diffuse emissions from the warm dust surrounding the BRC head.

In Fig. \ref{rgb_imags}, we display the colour-composite images of several intriguing regions within the star-forming complex. We showcase the central part of the complex, focusing on the area surrounding the massive star(s) HD 206267. As anticipated, this region is densely populated, with the central massive star(s) exerting significant influence over the surrounding star-formation activity. Notably, this central area hosts the prominent central cluster of IC 1396. Additionally, this figure highlights the BRC complexes IC 1396 N, SFO 35, and IC 1396 G.

In addition to point sources, the HSC images capture the bright diffuse emissions emanating from the BRCs. The HSC $\rm r_2$ image also unveils intriguing morphologies, including diffuse nebulae, filaments, and proplyds. The filter coverage of $\rm r_2$ band includes the $\rm H_\alpha$ emission line at 656.3~nm, making it well-suited for tracing proplyds.

\section{Membership analysis} \label{memb_analy}
This study utilises the ML algorithm to identify the low-mass membership population within the IC 1396 complex, using the HSC+UKIDSS catalogue. \citet{2023ApJ...948....7D} provided a concise review of membership analysis methodologies found in the literature and determined the member population of this complex from Gaia-DR3 using RF classifier of the ML algorithm.

Random-forest (RF; \citealt{breiman2001random, DBLP:journals/corr/abs-1201-0490}) classifier is part of the supervised ML algorithms, and due to its efficiency, RF classifier has been widely used in astronomy \citep{2011MNRAS.414.2602D, 2013MNRAS.435.1047B, 2017ApJ...843..104L,2018PASJ...70S..39L,2018MNRAS.476.3974P,2018AJ....156..121G,2018ApJ...869....9G,2021arXiv210305826M,2024MNRAS.tmp..456G}. A proper training set is essential for this ML algorithm to work efficiently. Additionally, the parameters used to filter the member population from the field population play a crucial role in such analyses.

In this study, we use the RF classifier to obtain the membership population of the whole IC 1396 complex. We describe the whole procedure in the following.

\subsection{Data set preparation for ML}
\subsubsection{Preparing the test data set} \label{rem_conta_gaia}
Before proceeding with the entire ML procedure, it is sensible to filter out as many contaminated field stars as possible from the HSC+UKIDSS catalogue. Utilizing Gaia-DR3 data can aid in the removal of these contaminated stars. Previous Gaia-based studies of this region have demonstrated that the member population of the IC 1396 complex typically falls within the parallax range of $\sim$0.8-1.6~mas \citep{2019A&A...622A.118S,2023AA...669A..22P,2023ApJ...948....7D}. Therefore, stars lying outside of this parallax range are likely contaminants and should be excluded.

Initially, we obtain the Gaia-DR3 catalogue within the area of interest. To ensure the inclusion of only stars with good photometric quality, we set a criterion of $\rm G\leqslant19~mag$. Among these stars, 186485 exhibit parallax values outside the range of $\sim$0.8-1.6~mas. These stars are presumed to be contaminants, and their counterparts in the HSC+UKIDSS catalogue can be identified and removed. Subsequently, we identified 164505 matches within the HSC+UKIDSS catalogue and eliminated them. After the removal of these contaminated stars, we are left with 637595 stars in the HSC+UKIDSS catalogue, which constitutes the dataset used to derive the member population of IC 1396.

\subsubsection{Preparing the training data set} \label{train_set}
A suitable training set is crucial for effectively using the RF classifier algorithm. \citet{2023ApJ...948....7D} derived the membership catalogue of 1243 stars for the entire complex from Gaia-DR3 data. In this study, we use the HSC+UKIDSS counterparts of the 1243 Gaia-based members. These counterparts constitute the member population of our training set. As the training set also requires a non-member population, we utilise the HSC+UKIDSS counterparts of the Gaia-based non-member population identified by \citet{2023ApJ...948....7D}. However, caution is necessary in selecting the non-member population.

\begin{figure*}
\centering
\includegraphics[scale=0.45]{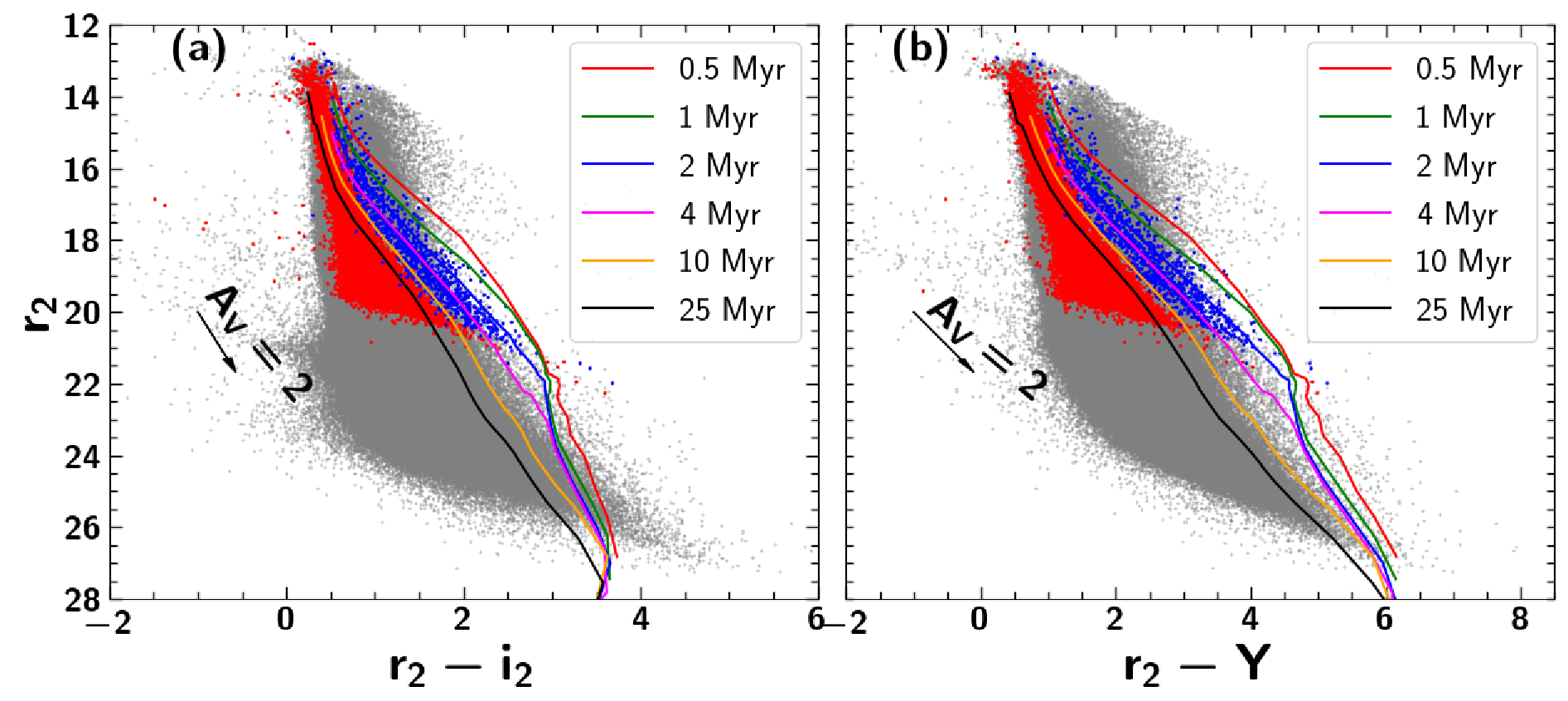}
\caption{$\rm r_2$ vs $\rm r_2-i_2$ and $\rm r_2$ vs $\rm r_2-Y$ CMD plots are shown in this figure. The grey dots are the complete HSC+UKIDSS 802100 sources, the red dots are the non-member population, and the blue dots are the candidate members of the training set. Isochrones of age 0.5, 1, 2, 4, 10, and 25~Myr from \citet{2015A&A...577A..42B} are plotted after correcting for visual extinction of $\rm A_V=1~mag$ and distance of 917~pc. Reddening laws of \citet{2019ApJ...877..116W} are used to correct the extinction of isochrones. The length and direction of the reddening vector of $\rm A_V=2~mag$ is shown as the black arrow. Details regarding the use of the extinction and distance values are explained in Section \ref{age_compl}. }
\label{cmd_train}
\end{figure*}

\citet{2023ApJ...948....7D} illustrated the distribution of the non-member population on the colour-magnitude diagram (CMD). The distribution on the CMD (see their Fig. 15), indicates that stars with $\rm P_{RF}\leqslant10$ fall primarily to the left of the 10~Myr isochrone. In this analysis, we consider these stars as the more reliable non-member population for our training set. Consequently, our training set comprises 16430 stars, including 1080 candidate members and 15350 non-member stars. Fig. \ref{cmd_train} displays the distribution of the training set stars on the $\rm r_2$ vs $\rm r_2-i_2$ and $\rm r_2$ vs $\rm r_2-Y$ CMDs, illustrating the distinction between the two populations. As visible from the plot, the magnitude of training set sources extends up to $\sim$20~mag in the $\rm r_2$ band. This is primarily due to limitations in the photometry of the Gaia catalogue.

Many candidate member populations have been reported in the literature, identified by various methods, including Gaia data (see Section \ref{memb_lit}). However, for the training set, we use only the Gaia DR3-based member and non-member sources from \citet{2023ApJ...948....7D}. The primary reason for this choice is that, in membership analysis using an ML algorithm, a proper training set requires both member and non-member populations. Although member populations are available from other Gaia-based studies, there is no corresponding non-member population. Additionally, there is significant overlap between all the Gaia-based candidate member catalogues, indicating that the additional Gaia sources are not crucial.

The non-Gaia-based studies report a large number of candidate member populations. Sources identified by different methods have varying properties. The literature-based sources include objects such as variable stars and evolved stars. However, in this analysis, we do not use them for our training set. This is because there is no adequate non-member population available, and several sources show a wide spread on the CMDs, which could indicate evolved older diskless stars. In such cases, preparing a proper corresponding non-member catalogue is highly desirable. Due to these potential issues, we use the results of \citet{2023ApJ...948....7D} to prepare the training set for this analysis.

\subsection{RF classifier training procedure} \label{rf_class} 
The next phase of this analysis involves training the machine using the training set and then applying it to the complete HSC+UKIDSS catalogue to identify the candidate members of IC 1396. In this ML technique, we require parameters that possess distinctive features to successfully extract members from the extensive pool of field stars. For instance, in the Gaia-based membership analysis conducted by \citet{2023ApJ...948....7D}, certain colour parameters and proper motions played a significant role in membership identification. In our scenario, we need to identify parameters that can facilitate the more efficient retrieval of the HSC+UKIDSS-based member population.

In our case, we have spatial parameters (RA and DEC) and photometric magnitudes in the HSC and UKIDSS bands ($\rm r_2$, $\rm i_2$, and Y; J, H, and K). In the CMD plot (Fig. \ref{cmd_train}), we illustrate the distribution of the training set stars. However, due to the sensitivity limitations of Gaia, the training set is restricted to magnitudes up to $\rm \sim 20~mag$ in the $\rm r_2$ band. Consequently, in this analysis, we cannot heavily rely on photometric magnitudes alone since the final result might be biased towards the magnitude range of the training set.

The next set of parameters we consider are the colour terms. However, using only colour parameters for extracting the member population is insufficient. Therefore, in addition to the colour terms, we need to incorporate a few other parameters that can assist us in effectively identifying member stars. Hence, we aim to derive several additional parameters for all stars based on their photometric magnitudes and colours.

Below, we provide details of the new parameters. We opt not to utilise the spatial coordinate parameters (RA and DEC) as they may introduce biases towards certain parts of the complex.

\subsubsection{New parameters used in RF classifier} \label{new_param}
After several attempts, we have derived the following parameters. The utilization of these parameters yields a satisfactory outcome in our analysis. We briefly describe these parameters below.

\noindent{\it First set of new parameters:}\\
In this work, we derive a set of parameters using the relation AAB = A/(A-B), where A and B represent magnitudes in bands A and B, respectively. Therefore, in this relation, we derive the parameter by dividing a magnitude by the colour term of a band. For instance, RRY can be defined as $\rm r_2 / (r_2-Y)$, and similarly for other terms. By employing this approach, we obtain twelve parameters from the HSC+UKIDSS dataset.

\noindent{\it Second set of new parameters:}\\ 
Similarly, we derive another set of parameters using a relation defined as $\rm AAB1 = A-(A-B)^2$, where A and B represent magnitudes in bands A and B, respectively. In this case, the parameter is defined as the subtraction of the square of a colour from the magnitude of a band. For example, $\rm RRY1 = r_2 - (r_2-Y)^2$. By employing this approach, we obtain another twelve parameters through various combinations of magnitudes and colours from the HSC+UKIDSS catalogue.

\begin{figure}
\centering
\includegraphics[scale=0.6]{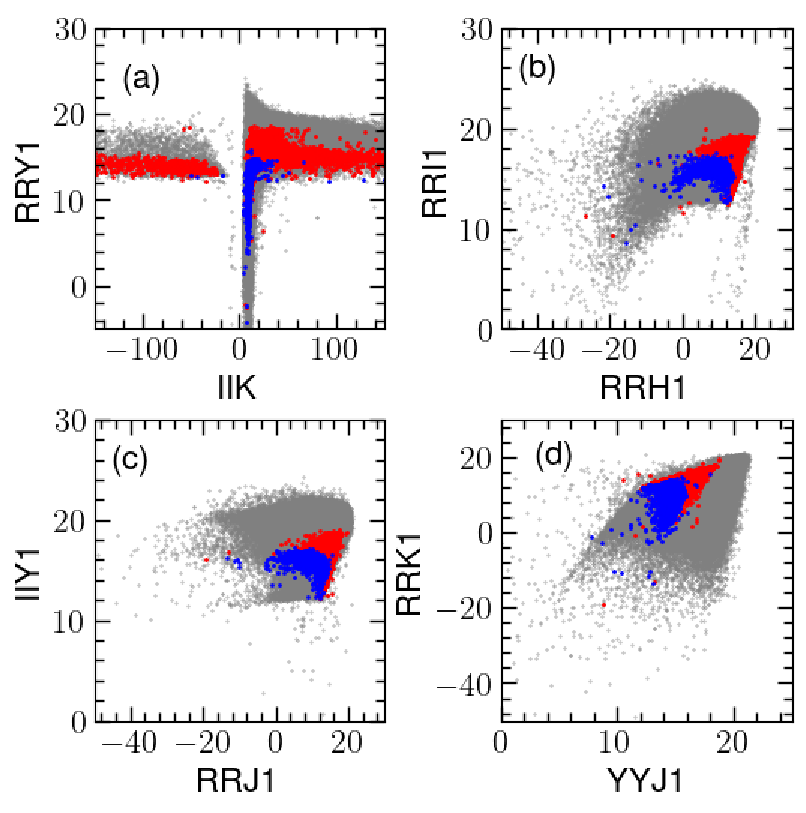}
\caption{Pair plots between some new parameters derived in this work are shown in this figure. The plots show the zoom-in regions covering the maximum number of sources. The dots and meaning of the colours are the same as in Fig. \ref{cmd_train}. }
\label{pair_const}
\end{figure}

Due to the combination of both colour and photometric magnitudes, these new parameters provide distinct information for each star. The values of these parameters vary significantly for a pair of stars with very different magnitudes and colours, and this unique property further aids in their effective segregation. The utility of these parameters in distinguishing between member and non-member populations can be observed in Fig. \ref{pair_const}, where we illustrate pair plots of some of these parameters. Additionally, pair plots of several more combinations of these parameters are presented in Fig. \ref{pair_const_ten}.

The effectiveness of ML techniques arises from the characteristics embedded within parameters of the training set, and the inclusion of the derived parameters proved instrumental in identifying member populations within the complex. The resulting outcomes are statistically influenced by the composition of the training dataset; thus, the selection of an appropriate training set holds paramount importance in ML endeavours. Subsequent sections of the analysis explain how these parameters significantly contribute to achieving satisfactory results.

\subsubsection{Total list of parameters} \label{42parm_list}
In this analysis, we have several parameters. These include three HSC photometric magnitudes ($\rm r_2$, $\rm i_2$, and Y), three UKIDSS photometric magnitudes (J, H, and K), twelve colour parameters ($\rm r_2-Y$, $\rm r_2-I$, $\rm i_2-Y$, $\rm r_2-J$, $\rm r_2-H$, $\rm r_2-K$, $\rm i_2-J$, $\rm i_2-H$, $\rm i_2-K$, $\rm Y-J$, $\rm Y-H$, and $\rm Y-K$), the first set of twelve new parameters (RRY, RRI, IIY, RRJ, RRH, RRK, IIJ, IIH, IIK, YYJ, YYH, and YYK), and the second set of twelve new parameters (RRY1, RRI1, IIY1, RRJ1, RRH1, RRK1, IIJ1, IIH1, IIK1, YYJ1, YYH1, and YYK1). In total, we have forty-two parameters for use in this analysis.

\subsection{Applying the RF algorithm} \label{app_rf}
We have already discussed that our current training set obtained from previous Gaia analysis lacks coverage towards the fainter end of HSC bands. Therefore, our initial objective is to strengthen our training set, ensuring it encompasses both member and non-member populations across the entire photometric range of HSC bands. The main rationale behind this effort is that although we have derived a set of new parameters and can obtain member sources using them, the results are still distorted towards the fainter end of HSC bands. This bias occurs due to the prevalence of field stars compared to member populations and due to the absence of training set towards the fainter end.

To address this issue, we initially implement the RF algorithm on the HSC+UKIDSS catalogue corresponding to a small circular region with a radius of $30\arcmin$, centred at the location of the massive star(s) HD 206267. As the central cluster is situated around the massive star(s), conducting the analysis in this limited region aids in more effectively distinguishing the member population from the field star population. The selection of a $30\arcmin$ radius is based on multiple trials, where we have determined it to be the optimal choice. Within this radius, the dominance of field stars is relatively reduced, facilitating a better separation of the cluster population from the field population.

\subsubsection{Result from the $30\arcmin$ circle} 
We begin by training the machine using a training set corresponding to this $30\arcmin$ circular region. Initially, there are 14685 HSC+UKIDSS sources within this circular region, of which 3089 sources are eliminated following the same process outlined in Section \ref{rem_conta_gaia}. The remaining 11596 sources are utilised for the analysis.

\begin{figure*}
\centering
\includegraphics[scale=0.45]{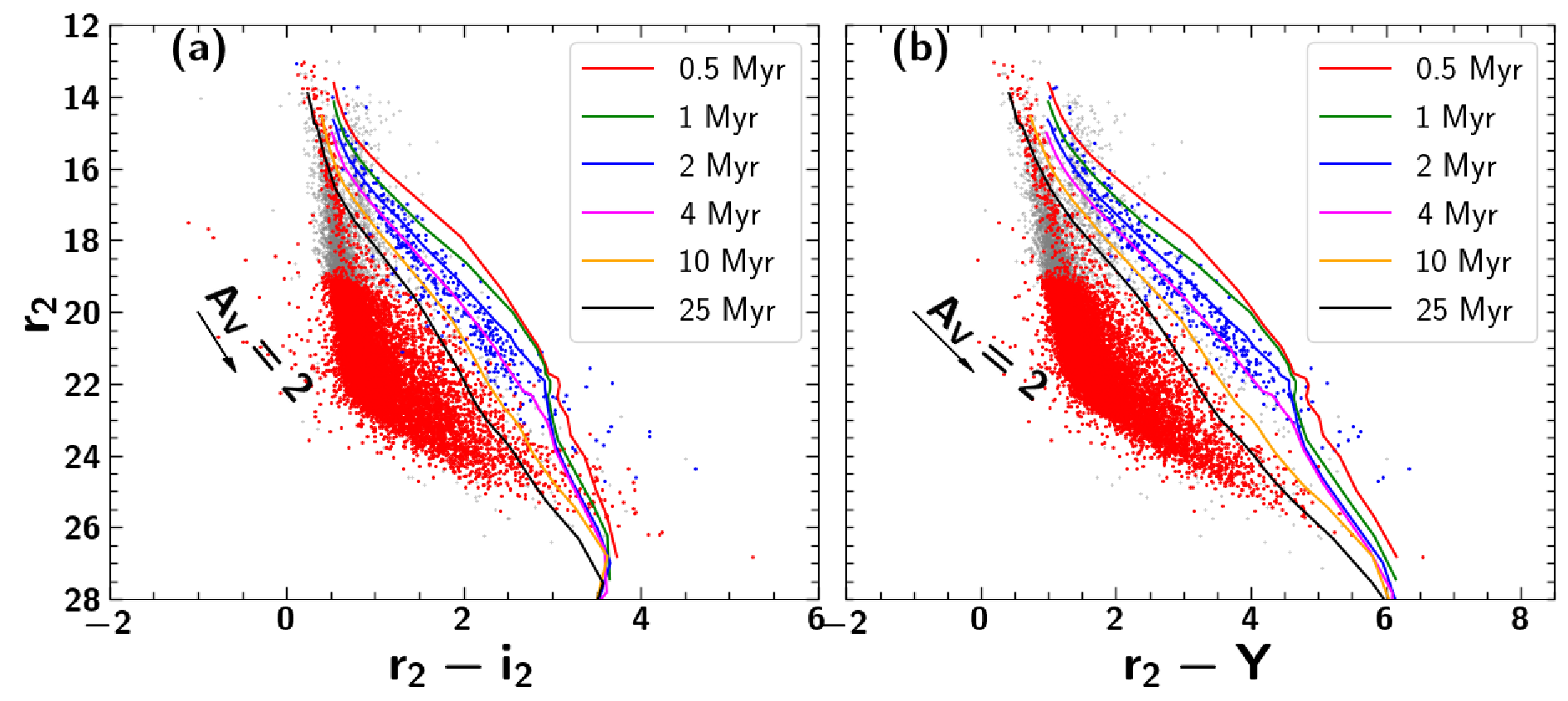}
\caption{CMD plots of $\rm r_2$ vs $\rm r_2-i_2$ (a) and $\rm r_2$ vs $\rm r_2-Y$ (b) for sources within the $30\arcmin$ circle. Grey dots are the total 14685 HSC+UKIDSS stars within the circle. The blue dots are the 399 stars detected as cluster population by the RF method, and the red dots are the 11057 stars with $\rm P_{RF} \leqslant0.5$. The isochrone curves and the black arrow have the same meaning as in Fig. \ref{cmd_train}. }
\label{cm_30arcmin}
\end{figure*}

Given that the colour range of training set stars is limited compared to the entire dataset, at this stage, we utilise only the twenty-four newly derived parameters. Further details regarding the efficiency test of the machine are explained in Appendix \ref{rf_eff}. The primary three leading parameters during the training procedure are RRK1, RRH1, and RRJ1. Subsequent to training the machine, we apply the RF classifier to the 11596 HSC+UKIDSS stars and identify 399 stars with $\rm P_{RF}\geqslant0.8$, constituting the member populations within the $30\arcmin$ circular region. The reliability of the retrieved member population is evident from the CMD plots illustrated in Fig. \ref{cm_30arcmin}. The distribution of these 399 sources in the CMDs conforms to a pre-main sequence locus. Out of the 11596 stars, there are 11057 stars with $\rm P_{RF} \leqslant0.5$. These sources are most likely non-members or field stars within the circular region.

At this stage, this is not a definitive detection of cluster members due to the limitations of the training set and the exclusion of some parameters. However, applying the RF classifier to this small central region helps us obtain a member and non-member catalogue whose magnitudes and colours align with the complete HSC+UKIDSS catalogue. Now, we merge these derived member and non-member sources within the $30\arcmin$ circular region with the Gaia-based training set. Through this process, our final training set comprises 27423 stars, including 1250 member stars and 26173 non-member stars. We use this training set to retrain the machine and subsequently apply it across the entire HSC+UKIDSS catalogue of 637595 sources (see Section \ref{rem_conta_gaia}), as described in the next section.

\subsection{Final result from RF classifier} \label{rf_res}

\begin{figure}
\includegraphics[scale=0.4]{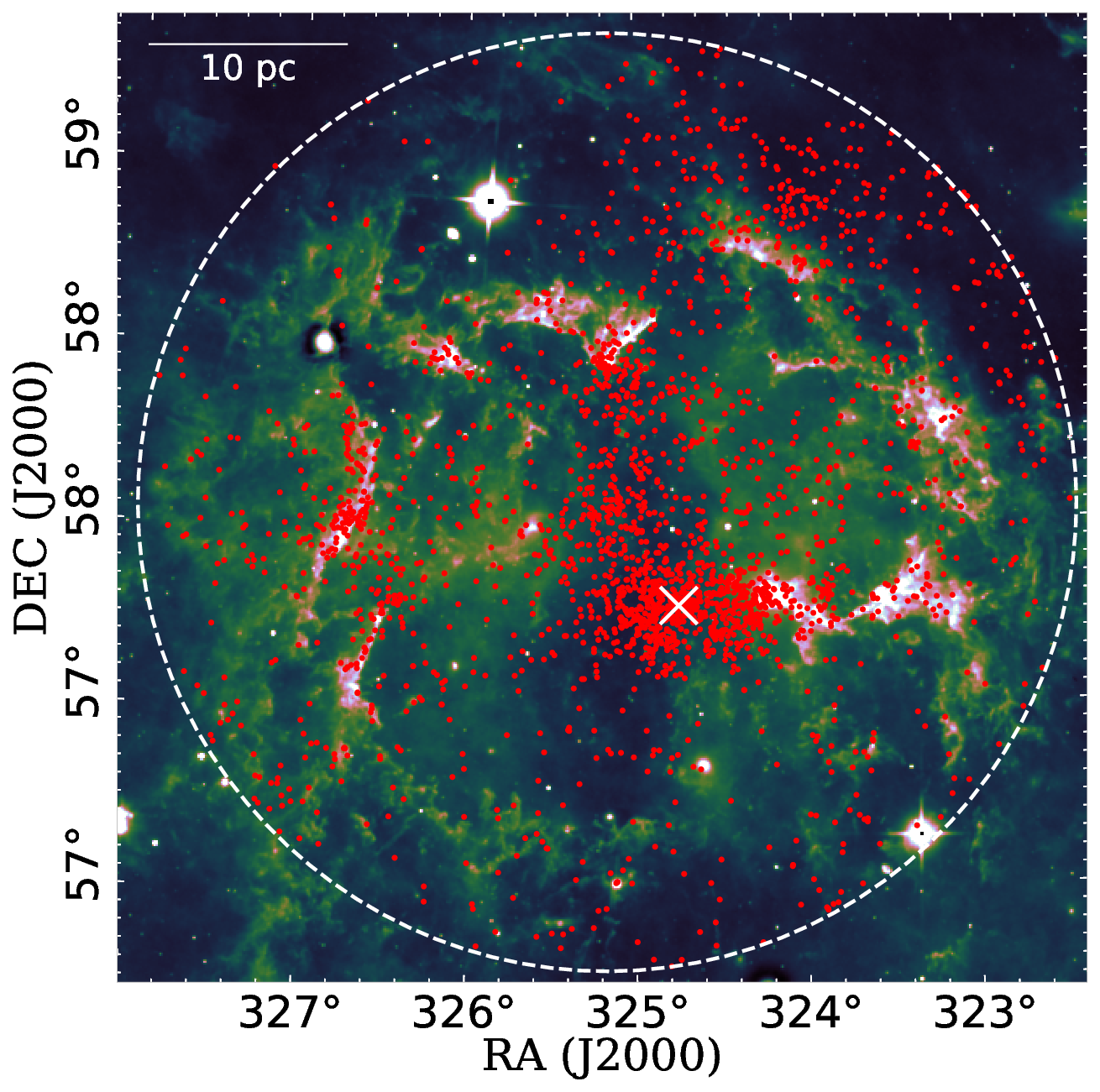}
\caption{The spatial distribution of the 2425 candidate members identiﬁed from the RF method in this work on the WISE $\rm 22~\mu m$ band. The white ``$\times$'' symbol shows the position of the massive central star(s) HD 206267. }
\label{hsc_memb}
\end{figure}

After establishing a suitable training set, we initiate the ML procedure to identify member stars throughout the entire IC 1396 complex. The magnitude and colour range of the final training dataset aligns closely with the test dataset. Therefore, we train the machine using all parameters, including the six photometric magnitude bands, twelve colour parameters, and twenty-four new parameters. This comprehensive training ensures the robustness of the machine with forty-two input parameters.

The efficiency test of the machine with the RF classifier is elaborated in Appendix \ref{rf_eff}. The first five parameters with the highest importance are IIH1, RRY1, IIK1, IIY, and RRH. This highlights the critical role of these new parameters in segregating the member and non-member populations.

The machine demonstrates an efficiency\footnote{The efficiency of a machine learning training procedure indicates how effectively the model is learning from the data. Typically, an accuracy of more than 90\% is considered indicative of a well-performing training process, suggesting that the model is generalizing well to unseen data while maintaining a low error rate.} of $99.6\%$ during its training. Following the successful training of the machine, we apply the RF classifier to the entire HSC+UKIDSS catalogue and identify 2425 stars with $\rm P_{RF}\geqslant0.8$. Details of the coordinates, photometric magnitudes in HSC+UKIDSS bands, and the $\rm P_{RF}$ value of these 2425 candidate members are provided in Table \ref{star_list}.

As our membership analysis relies on the photometry of sources, we cannot eliminate the possibility of a small fraction of contamination in the membership catalogue. Spectroscopic and proper-motion observations of the sources would be beneficial in further confirming their membership. Additionally, due to limitations on the input parameters, and the training set, the possibility of a small fraction of undetected of candidate members, particularly towards the fainter end, cannot be ruled out.

Furthermore, the HSC observations do not cover the entire $1.5^{\circ}$ circular region, as illustrated in Fig. \ref {Tr37_wise}. Consequently, the member population outside of the HSC observed area will have sensitivity corresponding to the Pan-STARRS catalogue.

Fig. \ref{hsc_memb} displays the spatial distribution of 2425 stars on the WISE $\rm 22~\mu m$ image. This distribution is consistent with earlier studies \citep{2012AJ....143...61N,2023ApJ...948....7D,2023AA...669A..22P}.  Most stars form a diagonal pattern from north to south, connecting the BRC IC 1396 N with the central cluster. The western part of the star-forming complex exhibits slightly higher stellar density. Additionally, a small fraction of stars appear randomly distributed throughout the complex with minor clumping. This spatial distribution suggests the presence of several sub-clusters associated with the complex. Previous studies have also identified sub-clusters in IC 1396 using member stars detected from various surveys \citep{2012AJ....143...61N,2023AA...669A..22P,2023ApJ...948....7D}. In the next section, we explore into the analysis of derived properties of the member stars and identify the sub-clusters within the IC 1396 complex.

\begin{table*}
\tiny
\caption{Details of the candidate members identified in this work using the RF method. The table provides the positions, magnitude values, and their errors in HSC and UKIDSS bands along with $\rm P_{RF}$ values of 2425 stars identified with the RF classifier in this work. The `*' mark in column 1 refers to new detections in this work.}
\label{star_list}
\begin{tabular}{cccccccccccccccc}
\\ \hline

No. & RA (2000) & DEC (2000) & $\rm r_2$ & $\rm r_2$-err & $\rm i_2$ & $\rm i_2$-err & Y     & Y-err & J & J-err & H & H-err & K     & K-err & $\rm P_{RF}$\\
         & (degree)  & (degree)   & (mag)     & (mag)         & (mag)     & (mag)         & (mag) & (mag) & (mag)     & (mag)         & (mag)     & (mag)         & (mag) & (mag) &\\
\hline
1 & 324.5679 & 58.2440 & 19.555 & 0.0002 & 18.055 & 0.0002 & 16.964 & 0.0002 & 15.330 & 0.0051 & 13.872 & 0.0023 & 12.795 & 0.0024 & 0.880 \\ 
2* & 324.5860 & 58.3625 & 19.916 & 0.0003 & 18.011 & 0.0002 & 16.983 & 0.0003 & 15.481 & 0.0056 & 14.756 & 0.0044 & 14.437 & 0.0077 & 0.930 \\ 
3* & 324.2975 & 58.4395 & 21.283 & 0.0007 & 19.092 & 0.0003 & 17.488 & 0.0003 & 15.864 & 0.0071 & 15.238 & 0.0064 & 14.704 & 0.0102 & 1.000 \\ 
4* & 324.4215 & 58.4946 & 19.621 & 0.0003 & 18.008 & 0.0002 & 16.681 & 0.0002 & 15.169 & 0.0044 & 14.481 & 0.0035 & 13.970 & 0.0057 & 0.990 \\ 
5* & 324.5790 & 58.5278 & 20.373 & 0.0005 & 18.443 & 0.0003 & 17.265 & 0.0004 & 15.552 & 0.0800 & 14.772 & 0.0990 & 14.552 & 0.1100 & 0.870 \\ 
6 & 324.4456 & 58.5364 & 19.921 & 0.0004 & 17.863 & 0.0003 & 16.768 & 0.0003 & 15.335 & 0.0049 & 14.696 & 0.0041 & 14.330 & 0.0075 & 1.000 \\ 
7* & 324.3399 & 58.6861 & 24.388 & 0.0156 & 21.320 & 0.0022 & 19.187 & 0.0018 & 17.264 & 0.0259 & 16.490 & 0.0221 & 15.895 & 0.0303 & 0.800 \\ 
8* & 324.4855 & 58.6527 & 22.100 & 0.0016 & 19.492 & 0.0005 & 17.747 & 0.0005 & 15.997 & 0.0089 & 15.253 & 0.0074 & 14.767 & 0.0112 & 1.000 \\ 
9 & 324.4721 & 58.6248 & 19.757 & 0.0003 & 17.734 & 0.0002 & 16.555 & 0.0002 & 14.965 & 0.0043 & 14.212 & 0.0032 & 13.806 & 0.0052 & 1.000 \\ 
10* & 324.4924 & 58.6258 & 20.803 & 0.0006 & 18.537 & 0.0003 & 17.323 & 0.0004 & 15.737 & 0.0073 & 15.061 & 0.0063 & 14.672 & 0.0103 & 0.910 \\ 
\hline
\end{tabular}
\\(This table is available in its entirety as online material. A portion is shown here for guidance regarding its form and content.)
\end{table*}

\subsection{Impact of extinction in the analysis} \label{impact_AV}
Extinction has not yet been considered as an input parameter in this analysis, although variable extinction could potentially be present along the line of sight towards the star-forming complex. Therefore, it is important to test the impact of extinction in this membership analysis by incorporating visual extinction ($\rm A_V$) alongside the other parameters. For this purpose, we obtained the $\rm A_V$ values (both mean and median) towards the location of each star in our HSC+UKIDSS catalogue from the Bayestar19\footnote{More details about the dust maps and their availability can be found at the following online links: http://argonaut.skymaps.info/usage and https://dustmaps.readthedocs.io/en/v1.0.7/maps.html} 3-D dust reddening map \citep{2019ApJ...887...93G}. This dust reddening map, derived with the inclusion of Gaia parallaxes and stellar photometry from Pan-STARRS1 and 2MASS, created a 3-D dust map of the sky, out to a distance of a few kiloparsecs.

\begin{figure}
\centering
\includegraphics[scale=0.36]{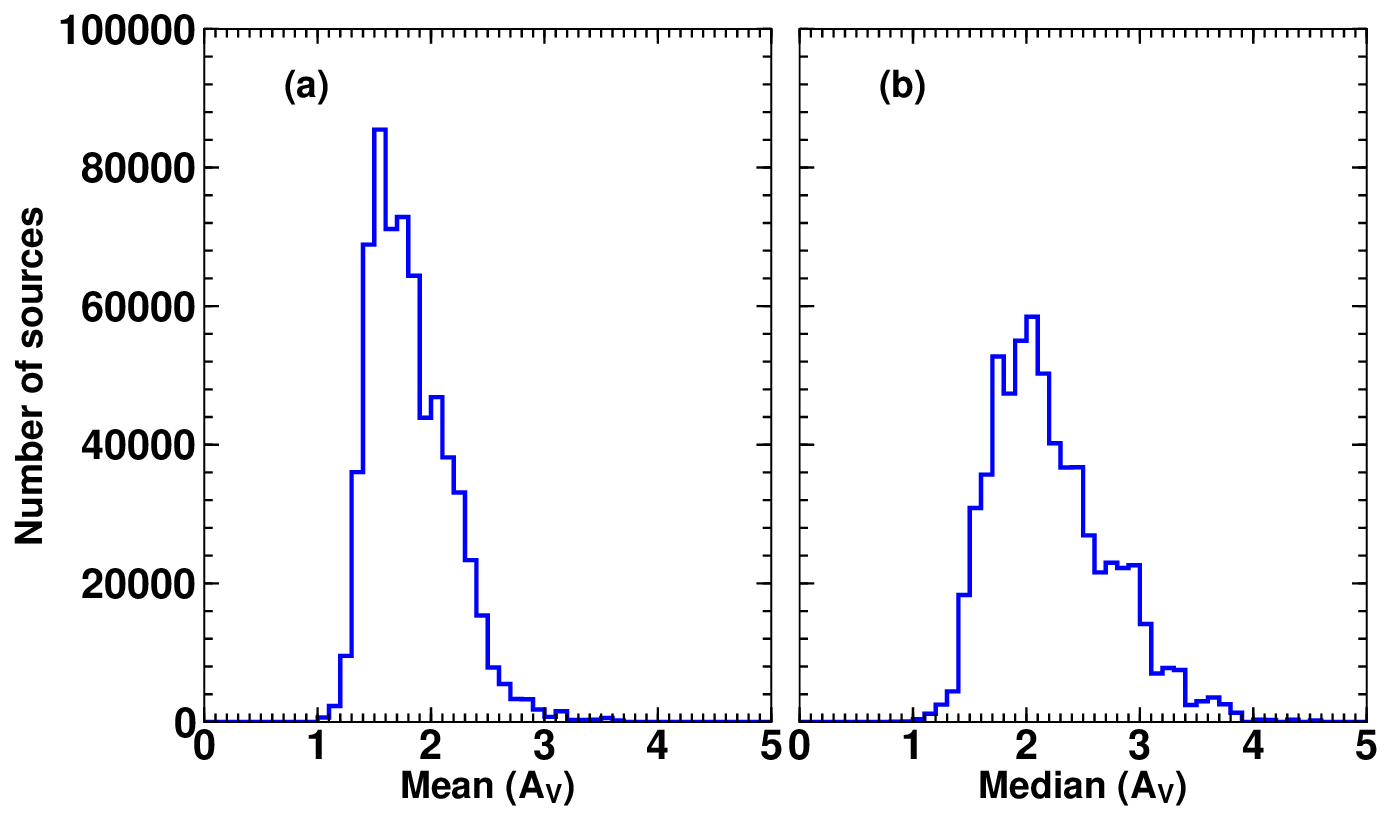}
\caption{Histogram distributions of the mean and median values of $\rm A_V$ for the HSC+UKIDSS stars. The bin size of the plots is 0.1. }
\label{hist_AV}
\end{figure}

In Fig. \ref{hist_AV}, we show the histogram distribution of both mean and median values of $\rm A_V$ obtained for all stars. The mean and median values of $\rm A_V$ lie in the range of 0.96–3.69 and 0.86–4.53, respectively, consistent with the extinction values reported in the literature for the star-forming complex \citep{2002AJ....124.1585C,2005AJ....130..188S,2012AJ....143...61N,2023AA...669A..22P}.

To assess the impact of extinction on the analysis, we incorporated a mean $\rm A_V$ for each star and repeated the entire analysis. This resulted in 2310 sources with $\rm P_{RF} \geqslant 0.8$, approximately 92\% of which match the existing catalogue of 2425 member stars. The final results differ by only 5\% compared to our previous findings. Fig. \ref{cm_hsc_AV} illustrates the distribution of the 2310 stars alongside the 2425 stars on CMDs, further demonstrating the minimal impact of including $\rm A_V$ in the analysis. We also tested the results using the median $\rm A_V$ values. However, this adjustment also did not lead to any significant changes, with the final result comprising 2241 stars, approximately 93\% of which match the results obtained without considering $\rm A_V$.

In this exercise, we observe that $\rm A_V$ does not have major impact on the training of the machine. The primary reason for this could be the use of the newly derived parameter sets. As several combinations of magnitude and colours are involved in their derivation, these parameters help in reducing the temperature-extinction degeneracy. Consequently, these parameters hold significantly greater importance in the analysis compared to extinction. Additionally, it is important to note that the Bayestar19 dust map has a resolution of $\rm 3.4\arcmin-13.7\arcmin$ \citep{2015ApJ...810...25G,2018MNRAS.478..651G}, which is substantially larger than the resolutions of the datasets used in this analysis. As a result, stars in our catalogue that fall within the same pixel will share the same $\rm A_V$ values, potentially introducing bias for some sources. However, this analysis demonstrates that the inclusion of $\rm A_V$ does not significantly impact the final outcome. For further analysis in the paper, we used the 2425 candidate member stars initially obtained without considering $\rm A_V$.

\subsection{CMDs of member stars} \label{prop}
\begin{figure}
\includegraphics[scale=0.45]{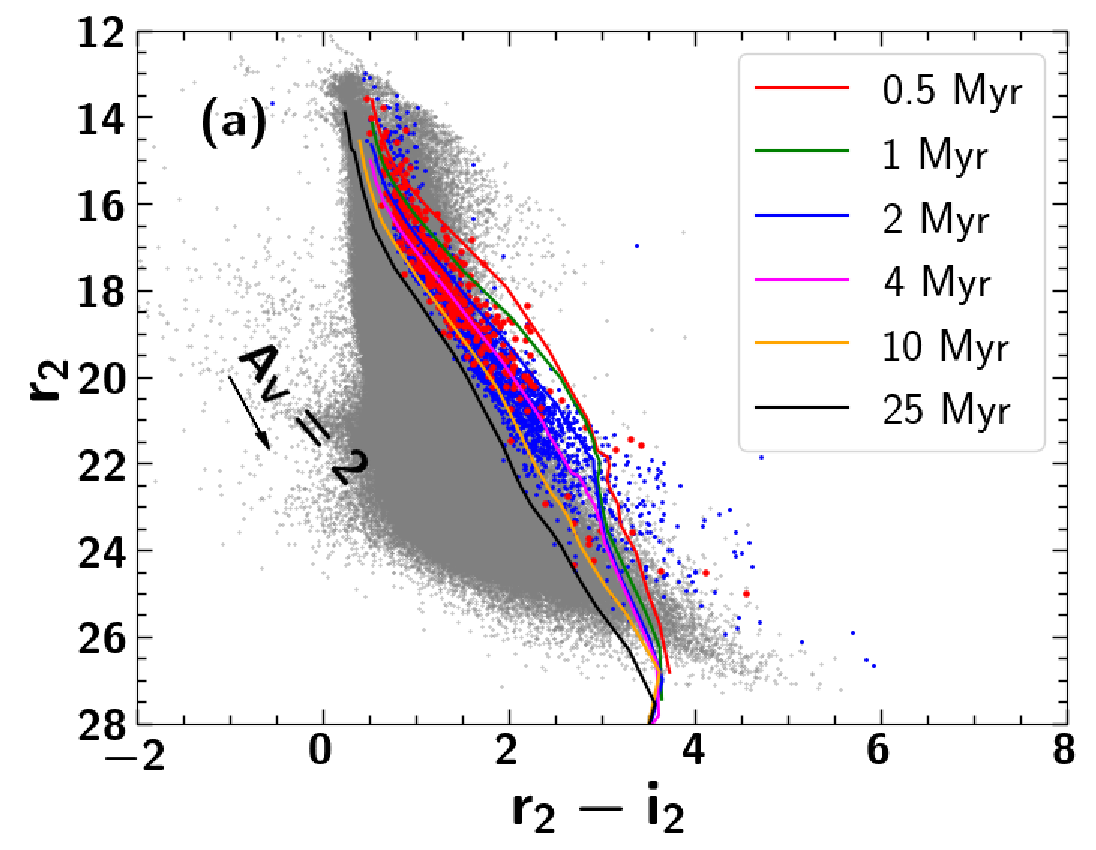}
\includegraphics[scale=0.45]{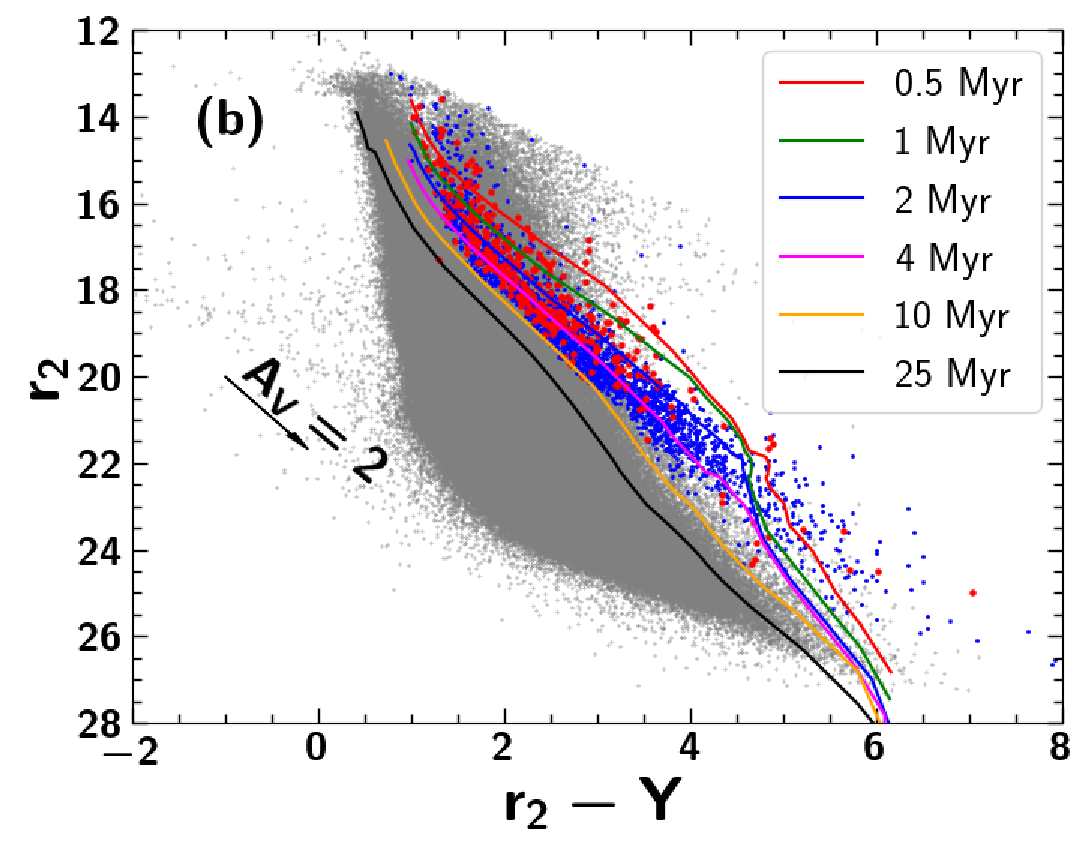}
\caption{CMD plots of $\rm r_2$ vs $\rm r_2-i_2$ (a) and $\rm r_2$ vs $\rm r_2-Y$ (b). Grey dots are the full HSC+UKIDSS 802100 stars, and the blue dots are the 2425 stars detected in this work. The red dots represent the 403 non-Gaia-based literature member stars retrieved in this work. The isochrone curves and the black arrow have the same meaning as in Fig. \ref{cmd_train}.}
\label{cm_hsc}
\end{figure}

In Fig. \ref{cm_hsc}, we present the $\rm r_2$ vs $\rm r_2-i_2$ and $\rm r_2$ vs $\rm r_2-Y$ CMD plots. These plots show the distribution of the 2425 candidate members of IC 1396 identified in this study. The plots demonstrate that the detected 2425 stars closely follow to the anticipated pre-main-sequence locus, with the majority of candidate members situated near $\rm 2-4~Myr$ isochrones. Comparison with the CMDs presented in Fig. \ref{cmd_train} suggests that the distribution of this member population matches with the Gaia-DR3-based stars. This further confirms the authenticity of the member populations identified in this study. In the subsequent sections, we compared our identified membership catalogue with the literature to further assess its reliability and the effectiveness of the method adopted in this work.

\subsection{Comparison with literature} \label{comp_lit}
Section \ref{memb_lit} provides a brief overview of the literature-based membership analyses conducted for the star-forming complex IC 1396. There are 2178 stars (see section \ref{memb_lit}) reported from various studies, excluding Gaia data. Analyses utilising Gaia data have reported a total of 1468 member stars in this complex. We compare these two literature catalogues with the member stars identified in our study.

\subsubsection{Comparison with non-Gaia based literature sources} \label{comp_non-gaia}
First, we compare with the non-Gaia-based catalogue of 2178 stars. Out of which 1224 stars have HSC+UKIDSS counter parts; however of these, 514 stars are not considered for comparison because their Gaia parallaxes do not fall within the valid range of the star-forming complex, as explained in Section \ref{rem_conta_gaia}. Eventually, out of these 2178 stars, 710 have corresponding entries in the final HSC+UKIDSS catalogue, upon which we have applied the RF method. We find that out of these 710 stars, we retrieve 403 stars ($\sim 57\%$). Upon closer inspection (see Fig. \ref{cm_hsc}), it becomes evident that the majority of the stars retrieved in this work belong to the brighter end of the HSC catalogue (around 20 mag in the $\rm r_2$ band). Due to the higher sensitivity of Subaru-HSC observations, many fainter sources were detected in this work compared to previous studies. A few stars from the 403 stars are found towards the fainter end, with magnitudes reaching up to 26 mag in the $\rm r_2$ band. Additionally, we can clearly see that the locus of the literature-based member population overlaps with the locus of the candidate member stars retrieved in this work, enhancing the reliability of this membership analysis.

In Table \ref{detail_compa_list}, we provide a detailed comparison of our identified sources with member populations identified through various methods in the literature. The table shows that, except for the literature-based variability sources \citep{2019ApJ...878....7M} and the YSOs detected through NIR, MIR and X-ray \citep{2021AJ....162..279S}, sources detected through all other methods show a high retrieval percentage. Although the YSOs from \citet{2021AJ....162..279S} also include X-ray detections, they represent a small fraction of the NIR population. Of their 139 sources (Table \ref{detail_compa_list}) used in our RF analysis, only 18 sources ($\sim$13\%) have X-ray counterparts. Therefore, most of the YSOs not retrieved in our work from \citet{2021AJ....162..279S} and \citet{2019ApJ...878....7M} are of NIR origin. A large fraction of sources from spectroscopy-based studies \citep{2002AJ....124.1585C,2006AJ....132.2135S,2013AA...559A...3S} and other surveys ({\it Spitzer}, $\rm H_{\alpha}$, and X-ray) are retrieved in this work. This higher retrieval suggest the effectiveness of the method adopted in this work.

Out of the 710 stars, 353 lie within the 90\% completeness limit, which corresponds to 18.5 mag in the $\rm r_2$ band. Of these, we successfully retrieve 273 stars ($\sim$77\%), indicating the robustness of the method adopted in this work. Among the remaining 80 stars within the 90\% completeness range, many are located to the left of the 10 Myr isochrone, while those that fall to the right of the 10 Myr isochrone still have $\rm P_{RF} \geqslant 0.5$ (see Fig. \ref{cmd_307}). Beyond the 90\% completeness magnitude, only a few stars from the literature are retrieved by our method, and these lie to the right of the 10 Myr isochrone.

In the $\rm r_2$ band, 22 mag corresponds to a mass of approximately $\rm 0.1~M_{\odot}$ (see Section \ref{age_compl}), roughly marking the boundary between stars and brown dwarfs. Beyond 22 mag in the $\rm r_2$ band, we identify 253 stars, indicating a significant improvement in the detection of very low-mass objects within the star-forming complex. However, from non-gaia based literature sources, only 12 such very low mass stars are retrieved by us.

The remaining 307 stars, which are not recovered have a membership probability of less than $80\%$. The random distribution of these 307 stars on CMDs (refer to Fig. \ref{cmd_307}) indicates their poor membership probability based on this membership analysis. Most of them are observed to fall on the left side of the 10~Myr isochrone on the CMDs. These 307 stars do not satisfy the properties of membership. This is because, in the current method, the machine is trained with an input training set that focuses on the population of the complex, with a potential age of less than 10 Myr. The majority of these 307 stars belong to the older population of the star-forming complex. This can be considered as a potential limitation of the method, as it is not able to identify the older populations of the complex, which we discuss briefly in Section \ref{point_rememb}. Most of the 307 stars have been detected through observations targeting variable stars \citep{2019ApJ...878....7M} and YSOs from \citet{2021AJ....162..279S}. These 307 stars only include a small fraction of stars identified from spectroscopic studies and observations involving $\rm H_\alpha$ and X-ray. Thus, most of the 307 stars are of NIR origin, and hence spectroscopic studies will be helpful in revealing their true nature.

\begin{landscape}
\begin{table}
\caption{Detailed comparison of the membership catalogue with literature-based catalogues obtained from individual methods. Column 1 specifies the method used to obtain candidate members. Columns 2 and 3 list the total number of stars and their HSC+UKIDSS counterparts, respectively. Column 4 provides the number of stars rejected based on the Gaia parallax condition, as explained in Section \ref{rem_conta_gaia}. Column 5 indicates the number of stars actually used in the RF analysis. Columns 6 and 7 show the number of stars recovered and not recovered by our current method, respectively. Column 8 lists the references for the individual methods. }
\label{detail_compa_list}
\begin{tabular}{cccccccc}
\\ \hline
Method & Total no.  & HSC+UKIDSS  & No. of stars rejected  & No. of stars  & No. of stars  & No. of stars  & References \\
       & of  stars  & counterpart & due to parallax        & used in RF    & recovered     & not recovered & \\
\hline
\multicolumn{8}{c}{Non-Gaia based literature sources} \\
\hline
Optical Spectroscopy          & 301 & 206  & 54   & 152  & 132 (87\%)  & 20  &  C02; S06b; S13b \\
{\it Spitzer} MIR             & 87  & 44   & 17   & 27   & 20 (74\%)   & 7   &  R04; S06a; M09; R13 \\
$\rm H_\alpha$ emission       & 621 & 421  & 201  & 220  & 196 (89\%)  & 24  &  B11; N12 \\
X-ray emission                & 518 & 247  & 75   & 172  & 132 (77\%)  & 40  &  G07; ME09; G12 \\
NIR, MIR, and X-ray     & 421 & 144  & 5    & 139  & 21 (15\%)   & 118 &  S21 \\
Optical/2MASS photometry, spectroscopy and variability  & 263 & 184  & 44   & 140  & 113 (81\%)  & 27  &  S05; S05; S10 \\
NIR variability               & 359 & 323  & 173  & 150  & 53 (35\%)   & 97  & M19 \\  \\

All the above                 & 2178 & 1224 & 514 & 710  & 403 (57\%)  & 307 &  -- \\    
\hline
\multicolumn{8}{c}{Gaia-based literature sources} \\
\hline
Gaia-Dr2/EDR3/DR3             & 1468 & 1217 & --  & 1217 & 1020 (84\%)  & 197 & C18; P23; D23 \\
\hline
\end{tabular}
\\
C02-\citet{2002AJ....124.1585C}; S06b-\citet{2006AJ....132.2135S}; S13b-\citet{2013AA...559A...3S}; R04-\citet{2004ApJS..154..385R}; S06a-\citet{2006ApJ...638..897S}; M09-\citet{2009ApJ...702.1507M}; R13-\citet{2013AJ....145...15R}; B11-\citet{2011MNRAS.415..103B}; N12-\citet{2012AJ....143...61N}; G07-\citet{2007ApJ...654..316G}; ME09-\citet{2009AJ....138....7M}; G12-\citet{2012MNRAS.426.2917G}; 
S21-\citet{2021AJ....162..279S}; S04-\citet{2004AJ....128..805S}; S05-\citet{2005AJ....130..188S}; S10-\citet{2010ApJ...710..597S}; M19-\citet{2019ApJ...878....7M}; C18-\citet{2018AA...618A..93C}; P23-\citet{2023AA...669A..22P}; D23-\citet{2023ApJ...948....7D} 
\end{table} 
\end{landscape}

\subsubsection{Comparison with Gaia-based literature sources} \label{compa_gaia}
Next, we compare our membership catalogue with all the Gaia-based member stars. There are 1468 Gaia-based member stars detected in IC 1396. Out of these, we have 1217 stars with HSC+UKIDSS counterparts over which the RF method has been applied. Our comparison reveals that out of these 1217 stars, 1020 ($\sim 84\%$) stars are retrieved in this work. This high retrieval of the Gaia-based member stars further confirms the reliability of the candidate member stars obtained in this work. This is because the Gaia-based member population offers a higher reliability of candidate membership due to their astrometric parameters, such as parallax and proper motions. The mean, median, and standard deviation in parallax of the Gaia-retrieved sources are [1.089$\pm$0.004, 1.078, 0.116]~mas, respectively. Similar statistics are [-1.179$\pm$0.002, -1.193, 0.284]~$\rm mas~yr^{-1}$ and [-4.277$\pm$0.004, -4.473, 0.684]~$\rm mas~yr^{-1}$ for $\rm \mu_{\alpha} \cos \delta$ and $\rm \mu_{\delta}$, respectively. These values agree with the results obtained by \cite{2023ApJ...948....7D} from their Gaia-DR3 member stars. The statistics of parallax for the retrieved 1020 Gaia members also match other works \citep{2019A&A...622A.118S,2023AA...669A..22P,2023ApJ...948....7D}.

The remaining 197 Gaia stars that were not recovered by us are primarily ($\sim$ 52\%) from the survey of \citet{2018AA...618A..93C}, in which they included all stars with a membership probability above 50\%. This current study could not recover their low-probability stars, which possibly have a higher chance of contamination. Of the 1080 Gaia-based sources used in the training set, 994 ($\sim 92\%$) stars are retrieved with $\rm P_{RF}\geqslant80\%$. For the remaining 86 stars, a large fraction of stars still have $\rm P_{RF}\geqslant50\%$.

In classification tasks, the RF classifier computes the probability of a data point belonging to each class by averaging the probabilities assigned by individual decision trees. In the context of the membership analysis, the RF classifier assigns a probability to each star being a member of the IC 1396 complex based on its photometric properties and other derived parameters. The probability represents the confidence of the model in the classification of the star as a member or non-member, with higher probabilities indicating greater confidence in the classification. However, due to the statistical nature involved in assigning the probabilities, a small degree of mismatch between the final outcome and the training set is still expected, as observed in this study.

\begin{figure}
\centering
\includegraphics[scale=0.7]{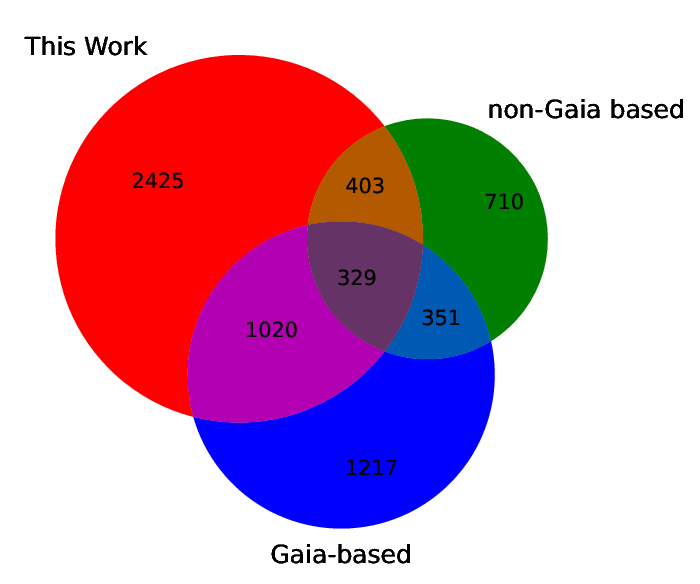}
\caption{Venn diagram summarizing the comparison between the member population of this work with the stars from several other surveys and with the Gaia-based member stars \citep{2018AA...618A..93C,2023AA...669A..22P,2023ApJ...948....7D}. }
\label{venn}
\end{figure}

For easier understanding, we present the entire comparison of catalogues in the form of a Venn diagram in Fig. \ref{venn}. Out of the 2425 candidate member stars, 1331 are newly identified in this study. In Table \ref{star_list}, the new member stars are marked with an asterisk symbol. Additionally, in one of the most recent analyses of the complex, \citet{2024MNRAS.tmp..456G} conducted a search for low-mass objects within a $22\arcmin$ region centred around the central massive star HD 206267. They employed ML techniques on the HSC datasets and identified a population of 458 stars within that specific region. In our work, we are able to retrieve approximately 86\% of their sources, and the small discrepancy is primarily due to the different approaches employed by both analyses.

\subsection{Caveats and points to ponder}\label{point_rememb}
In this section, we discuss the limitations or caveats associated with our membership analysis. As previously mentioned, conducting a membership analysis of a star-forming complex based solely on photometric information presents challenges. The primary difficulty arises due to the unavailability of a training set extending to the fainter end of the HSC+UKIDSS catalogue. To address this issue, we have derived several new parameters using the magnitudes and colours of the sources. These new parameters, when combined with magnitudes and colours, enable us to identify a reliable member population from a large pool of sources.

It is likely that some candidate members may remain undetected, especially at the fainter end. This is primarily due to the lack of sources in the training set at the fainter end and in addition to the consideration of a high probability of $\rm P_{RF}\geqslant0.8$. This lack of sources in the training set towards the fainter end thereby introduces a bias. Hence, statistically, the chance of non-detection of candidate members is higher at the fainter end compared to the brighter end. This limitation affects the detection of very low-mass objects such as brown dwarfs in the star-forming complex.

Furthermore, this analysis only identifies the young sources of the complex lying rightward of the 10~Myr isochrone. The older populations of the complex of $>$ 10~Myr (if any), or the YSOs that are scattered on the CMD and lying leftward of the 10~Myr isochrone would have potentially been removed in our analysis (See Appendix \ref{cmd_ext_307}). There are several reasons for the broad distribution of young sources on the CMD, such as differential reddening, variability, age spread, etc (e.g., \citealt{2017ApJ...836...98J,2017MNRAS.468.2684P}). Such members are difficult to detect in this analysis due to a lack of proper knowledge about these sources and the absence of a reliable training set.

The non-detection of the older population can also be attributed to the training set used in this work (see Section \ref{train_set}), which includes Gaia-based stars with ages less than 10~Myr, thus introducing a potential bias. The outcome of ML algorithms is highly dependent on the nature of the training set. Therefore, the detection of member populations younger than 10 Myr and possibly undetected stars towards the fainter end can be attributed to the bias present in the training set. However, we assume that the bias is less significant, as only a relatively small percentage of spectroscopic (13\%) and X-ray (23\%) stars lie to the left of the 10 Myr isochrone. We are keen on improving such membership analysis by addressing the known limitations in future studies.

It is worth noting that integrating UKIDSS with the HSC catalogue enhances the reliability of our membership detection. Therefore, we advocate for incorporating more bands with photometry to yield more dependable results from such analyses. The rationale for using additional wavelength bands is similar to spectral energy distribution (SED) analysis, where more data points contribute to improved accuracy. In a similar study, \citet{2023AJ....166...87R} identified YSO populations in IC 417 by analysing the SED obtained from datasets across several optical and infrared bands. Although our approach differs, we have retrieved the reliable member populations of IC 1396.

Additionally, it is true that even though the final membership detection is more reliable, it is less complete because the final HSC+UKIDSS catalogue is restricted to the common sources from both catalogues. Hence, if a source is missing in either one of the catalogues, for example, the UKIDSS catalogue, it will also be missing from the final HSC+UKIDSS catalogue and eventually from the final candidate member detection. This can also be considered as a limitation in such membership analysis.

It is important to note that the stars in the training set are available up to $\sim$ 20~mag in the $\rm r_2$ band. Additionally, other non-Gaia-based studies in the literature primarily correspond to the brighter end (see Fig. \ref{cm_hsc}). Of the 2425 candidate stars identified in this work, 1461 are brighter than 20~mag in the $\rm r_2$ band. Moreover, there is a high retrieval rate with Gaia-based literature sources, as well as sources detected through methods based on optical spectroscopy, {\it Spitzer}, $\rm H{\alpha}$/X-ray emissions, optical, and 2MASS photometry (see Section \ref{comp_lit}). Therefore, these 1461 stars have a higher confidence level for being members of the star-forming complex. The distribution of the 2425 candidate populations in the CMD (see Fig. \ref{cm_rry}) suggests the presence of very low-mass stars (e.g. brown dwarfs).

It is also important to consider that the identified candidate population, particularly at the low-mass end, may include contamination from different origins. For example, very faint foreground stars and distant background objects may be contaminating the retrieved sample. The nebular emission, particularly the bright rims near the BRCs, is associated with line emissions such as $\rm H{\alpha}$ and [S II] \citep{2009A&A...504...97B,2023AA...669A..22P}. These line emissions may blend with the stars and alter their apparent properties. The nebular emissions may also appear as compact, star-like sources. Additionally, shock emissions near the BRCs \citep{2001A&A...376..553N,2002ApJ...573..246B,2006A&A...449.1077C,2023AA...669A..22P} might further affect the observed characteristics of the stars. Such contaminants are difficult to filter out using a photometry-based membership analysis.

Therefore, additional studies, such as spectroscopic and proper motion analyses, will be essential to further confirm the membership of the identified candidate population in this work, including brown dwarfs. A detailed analysis of these brown dwarfs will provide crucial insights into the formation of low-mass objects in clustered environments within the star-forming complex.

As explained in previous sections, we initially applied the RF technique to a $30\arcmin$ circular area in our methodology, which facilitated the enhancement of our training set. While deriving the member stars, we utilised twenty-four and forty-two parameters in the $30\arcmin$ circle and the entire complex, respectively. However, we experimented with reducing the parameters by leveraging the correlation matrix and retaining only the highly uncorrelated parameters that would provide maximal information. It is worth noting that, through this approach, our results varied by approximately ten per cent. We opted to use more parameters to ensure the retrieval of the most reliable outcome.

The applicability of this method employed in this work may vary from one star-forming region to another and for different datasets. Nevertheless, this method serves as a helpful or guiding tool in such membership analyses. Our intention is to apply this methodology to other star-forming regions using diverse datasets.

\section{Results} \label{res}

\subsection{Age and mass completeness} \label{age_compl}

\begin{figure}
\centering
\includegraphics[scale=0.5]{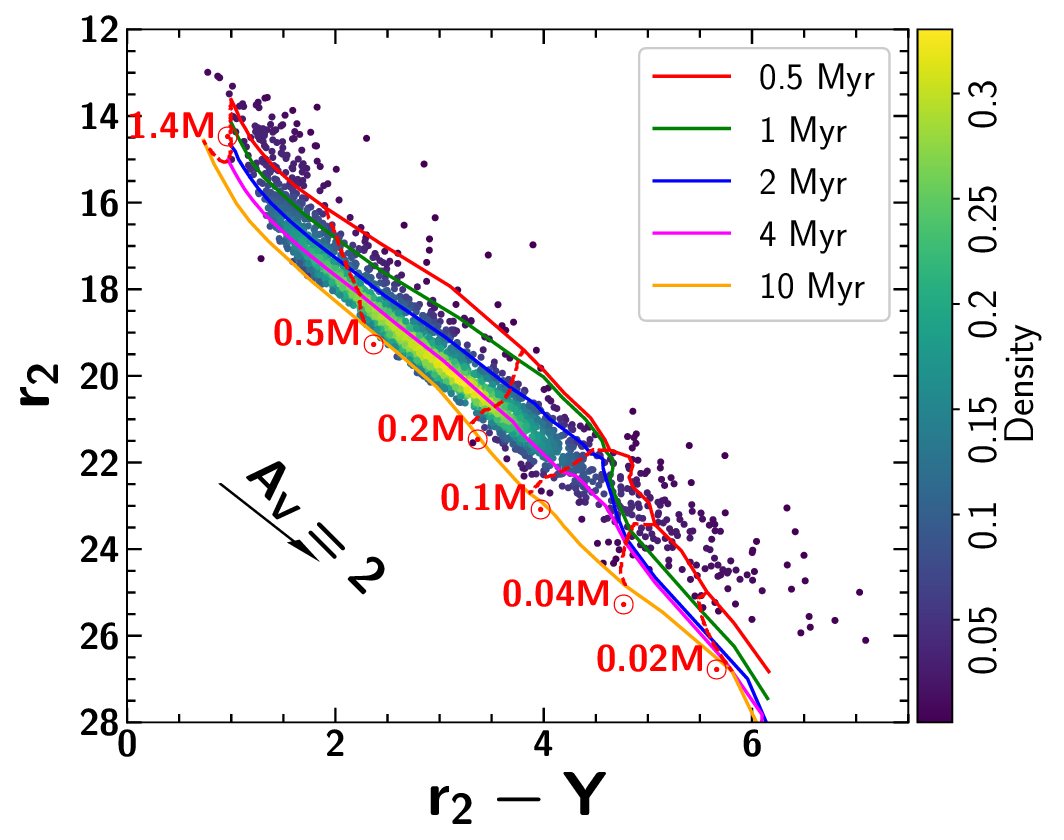}
\caption{$\rm r_2$ vs $\rm r_2-Y$ CMD of the 2425 candidate members of IC 1396 detected in this work. Isochrones and the black arrow have the same meaning as in Fig. \ref{cmd_train}. On the CMD, we show evolutionary tracks of mass 0.02, 0.04, 0.1, 0.2, 0.5, and $\rm 1.4~M_{\odot}$ \citep{2015A&A...577A..42B}. }
\label{cm_rry}
\end{figure}

Previous studies have provided an age estimate of $\rm \sim 2-4~Myr$ for IC 1396 \citep{2005AJ....130..188S,2023AA...669A..22P,2023ApJ...948....7D}. From the identified candidate members, we also attempt to derive an age estimate of the complex. Additionally, to gain insight into the mass limit of the candidate population, we use all the candidate members identified in Section \ref{rf_res}, and the $\rm r_2$ vs $\rm r_2-Y$ CMD is shown in Fig. \ref{cm_rry}. After correcting for extinction and distance using reddening laws of \citet{2019ApJ...877..116W}, we plot isochrones along with the mass evolutionary tracks from \citet{2015A&A...577A..42B}. Earlier works have reported sparse visual extinction towards IC 1396. Using NIR and optical data, \citet{2005AJ....130..188S} estimated an average visual extinction of the entire complex to be $\rm A_V = 1.5\pm0.5$ mag, which also matches estimations from a few other studies \citep{2002AJ....124.1585C,2012AJ....143...61N}. A more recent analysis of the complex derived an average extinction of $\rm A_V = 1.4\pm0.52$ mag from Gaia-EDR3 colours of the member stars \citep{2023AA...669A..22P}.

In the star-forming complex, the feedback effect of central massive stars clears the dust and gas mass and creates a cavity, which can be seen in the WISE $\rm 22~\mu m$ image (see Fig. \ref{Tr37_wise}). Thus, the inner part of the complex is expected to be associated with lesser visual extinction compared to the surrounding regions. Due to these reasons, \citet{2023ApJ...948....7D} have used the minimum extinction value of $\rm A_V = 1$ mag obtained from \citet{2012AJ....143...61N}. In this analysis, we have also used the same extinction value and the distance of 917 pc derived from Gaia-DR3-based member stars to correct the isochrones.

The Fig. \ref{cm_rry} presents the density distribution of the 2425 candidate members on the CMD. The plot shows that the maximum number of sources lie within an age range of $\rm \sim 2-4~Myr$, and the mean age would be $\rm \sim 3~Myr$. This matches with all the previous studies \citep{2005AJ....130..188S,2023AA...669A..22P,2023ApJ...948....7D}. However, a detailed SED analysis will help deriving a more accurate age estimation for individual stars and the whole complex.

In Fig. \ref{cm_rry}, we have overplotted evolutionary tracks of various masses starting from $\rm 0.02~M_{\odot}$. In this section, we derive a quantitative estimate of the limiting mass and the 90\% mass completeness limit of the member populations. We use the HSC-$\rm r_2$ to estimate them since its depth is maximum compared to the other two bands. HSC-$\rm r_2$ band magnitude limit of the membership catalogue is $\rm \sim 27~mag$. Using an isochrone of age $\rm 2~Myr$, this magnitude limit in the $\rm r_2$ band corresponds to a mass $\rm \sim 0.02~M_{\odot}$. Detection of low-mass objects of IC 1396 becomes possible due to the higher sensitivity of the Subaru-HSC observations. Next, we derive the 90\% mass completeness of the member population. In Section \ref{quality}, we have derived the 90\% completeness of the HSC catalogue to be $\rm \sim 24~mag$ in $\rm r_2$ band. Considering the isochrone of 2~Myr, the 24~mag in $\rm r_2$ corresponds to a mass of $\rm \sim 0.04-0.05~M_{\odot}$. This analysis shows possible detection of very low-mass objects (e.g., brown dwarfs) in IC 1396. It is to be noted that this membership analysis includes the UKIDSS catalogue with the HSC catalogue, so the final HSC+UKIDSS has 90\% completeness of $\sim$22~mag in $\rm r_2$ band, and so, the 90\% mass completeness is $\rm \sim 0.1~M_{\odot}$.

Note that the isochrones plotted in Fig. \ref{cm_rry} do not correctly fit the low-mass stars and deviate for mass below 0.1~$\rm M_\odot$. The origin of this discrepancy is still unclear and probably an inherent limitation of isochrone models for low-mass stars \citep{2013MNRAS.434.3236K,2020ApJ...901...49L}. This discrepancy between observations and multiple isochrone models has also been reported in several observational analysis  such as in Taurus, Ophiuchus, Chamaeleon, Serpens-south, NGC 2244 etc. (e.g., \citealt{2013MmSAI..84..931M,2017ApJ...849...63R,2018ApJ...858...41Z,2020ApJ...892..122J,2023A&A...677A..26A}). Such caveats in the isochrone models can affect determining the individual properties of stars. However, the overall cluster properties, such as age and mass, will not be affected because those properties are mainly constrained by the location of the majority of sources in the CMD.

\subsection{Sub-clustering in IC 1396} \label{sub_clust}
\begin{figure}
\centering
\includegraphics[scale=0.4]{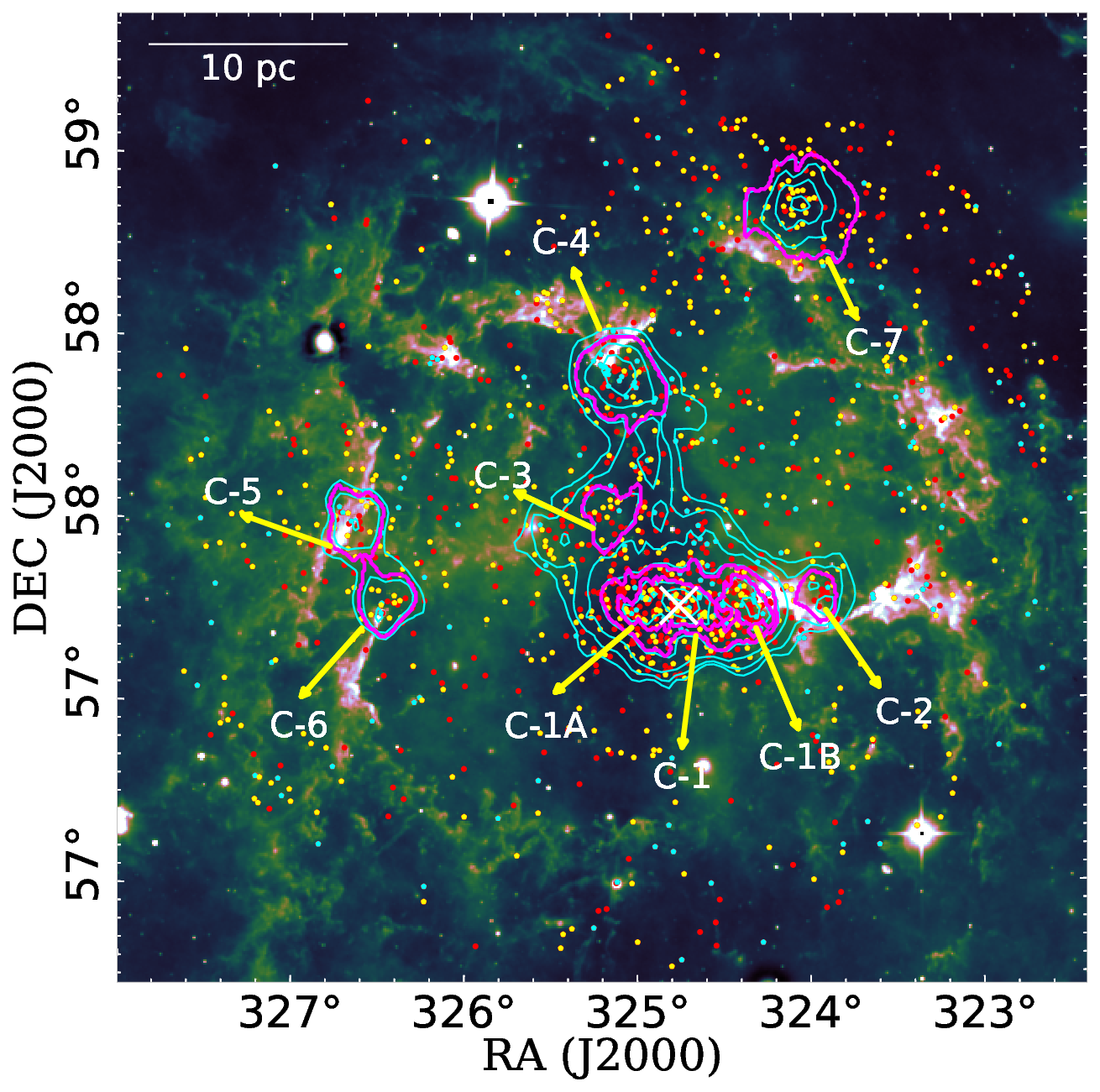}
\caption{The colourscale is the WISE $\rm 22~\mu m$ band image. The spatial distribution of the 1461 members stars as coloured dots.  All the stars are divided into three age groups based on their location on the $\rm r_2$ vs $\rm r_2-Y$ CMD. Out of 1461 stars, 218 have age $<$1~Myr (cyan dots), 568 have age $>$4~Myr (yellow dots), and the remaining 675 stars have ages between 1 and 4~Myr (red dots).  
The cyan contours are from the stellar density map generated from the 1461 candidate members. The contour levels are 1.5, 2, 3, 5, 10, 15, and 30 $\rm stars~ pc^{-2}$. The location of the central massive star(s) HD 206267 is marked as a white `$\times$' symbol. The clusters detected in this work are shown as  magenta curves, along with their nomenclature highlighted through yellow arrows. }
\label{sd_clust}
\end{figure}

Previous studies have identified several sub-clusters within the IC 1396 complex \citep{2012AJ....143...61N,2023AA...669A..22P,2023ApJ...948....7D}. In this section, we aim to identify the sub-clusters associated with IC 1396, using the more reliable set of 1461 member stars (see Section \ref{point_rememb}), whose spatial distribution (see Fig. \ref{sd_clust}) reveals the presence of multiple sub-clusters within the star-forming complex. To accomplish this, we first generate a surface density map using the member stars and applying the nearest neighbour (NN) method \citep{1985ApJ...298...80C,2011AN....332..172S}. This method enables us to derive the j-th nearest neighbour density of a star using the following expression.

\begin{equation}
\rm \rho_j~ = ~ \frac{j-1}{S(r_j)}
\end{equation}

\noindent where $\rm r_j$ is the distance to the j-th nearest neighbour, and $\rm S(r_j)$ is the surface area with radius $\rm r_j$. Many studies have indicated that $\rm j = 20$ is an optimal value for cluster identification \citep{2008MNRAS.389.1209S,2017MNRAS.465.4753R,2021MNRAS.504.2557D}. In our analysis, we also use $\rm j = 20$ and obtain the nearest neighbour density distribution of member stars. From the derived density values of all member stars, we generate a stellar density map with a pixel size of 0.1 pc ($20.5^{\prime\prime}$). In Fig. \ref{sd_clust}, we present the distribution of stellar density in the form of contours overlaid on the WISE 22 $\rm \mu m$ map. The lowest contour is set at 1.5 stars $\rm pc^{-2}$, representing the threshold above which the maximum number of sources is observed. These stellar density contours also reveal the presence of sub-clusters within the IC 1396 complex.

Using the {\it astrodendro} algorithm \citep{2019ascl.soft07016R} of Python, we identify sub-clusters within IC 1396. This algorithm constructs tree structures starting from the brightest pixels of the dataset and progressively includes fainter ones. The input parameters for the algorithm are the threshold flux value (minimum value), a contour separation value (minimum delta), and the minimum number of pixels. In this analysis, we set the threshold and minimum delta to 1.5 and 0.5 stars $\rm pc^{-2}$, respectively. For potential cluster detection, we specify the minimum number of pixels as 200. These parameter settings, determined through multiple trials, enable optimal cluster detection. Applying the {\it astrodendro} algorithm with these inputs, we identify 9 leaf structures in the dataset, which correspond to the sub-clusters of the IC 1396 complex.

In Fig. \ref{sd_clust}, we illustrate the detected sub-clusters along with their respective names. All identified sub-clusters are found to be associated with the BRCs and PDRs of IC 1396. Notably, we detect cluster C-1 around the central massive star(s) HD 206267, which exhibits two distinct sub-structures labelled as C-1A and C-1B. Several of these sub-clusters, including C-1A, C-1B, C-2, C-4, and C-6 have been reported in previous studies \citep{2012AJ....143...61N,2023ApJ...948....7D}. For instance, sub-cluster C-4 is associated with BRC IC 1396 N, while C-6 is linked with BRCs SFO 39 and SFO 41. Moreover, \citet{2023ApJ...948....7D} identified a sub-cluster near BRC SFO 35 based on Gaia-DR3 member stars, which again detected by us.

In our study, we detect a few new sub-clusters within the IC 1396 complex. We identify sub-cluster C-3 along the diagonal line connecting the central cluster and IC 1396 N. Furthermore, a new sub-cluster, {C-5} is detected towards the eastern edge of IC 1396, linked with BRC IC 1396 G \citep{1956BAN....13...77P}.

Table \ref{stat_cluster} provides statistics including the radius, number of stars, and average stellar density of the sub-clusters. To calculate the physical radius ($\rm R_{cluster} = (A_{cluster}/\pi)^{0.5}$; \citealt{2017MNRAS.472.4750D}), we utilize the apertures obtained from {\it astrodendro}. Area of each sub-cluster ($\rm A_{cluster}$) is determined as the product of the number of pixels (N) and the area of each pixel ($\rm A_{pixel}$). The central cluster C-1 is the largest, with a radius of 3.06~pc, hosting the maximum number of stars (312), and with an average stellar density of 10.62~$\rm stars~pc^{-2}$. Sub-clusters C-1A and C-1B, nestled within the central cluster, exhibit the highest stellar densities among all sub-clusters, with values of 15.32 and 13.95~$\rm stars~pc^{-2}$, respectively. Considering the stellar density and distribution of sub-clusters, we explore the star-formation history of the complex in subsequent sections.

\begin{table*}
\centering
\caption{The radius, number of stars and average stellar density of the sub-clusters associated with IC 1396.}
\label{stat_cluster}
\begin{tabular}{ccccc}
\\ \hline \hline
Cluster & Radius & No. of stars & Average surface density  & Average volume density\\
        & (pc)   &              &  $\rm (stars~pc^{-2})$   & $\rm (stars~pc^{-3})$ \\
\hline
C-1  & 3.06 & 312 & 10.62 & 2.61 \\
C-1A & 1.70 & 139 & 15.32 & 6.76 \\
C-1B & 1.21 & 64  & 13.95 & 8.66 \\
C-2  & 1.07 & 17  & 4.73  & 3.32 \\
C-3  & 1.38 & 25  & 4.15  & 2.24 \\
C-4  & 2.17 & 56  & 3.77  & 1.30 \\
C-5  & 1.57 & 22  & 2.83  & 1.35 \\
C-6  & 1.52 & 21  & 2.91  & 1.44 \\
C-7  & 2.62 & 56  & 2.59  & 0.74 \\
\hline
\end{tabular}
\end{table*}

\section{Star formation history of IC 1396} \label{diss} 
IC 1396 is one of the nearby and most well-studied star-forming complexes, where the feedback-driven star-formation activity dominates the entire complex. The strong stellar wind from the central massive star(s) creates a giant cavity of radius $\sim1.5^{\circ}$ ($\rm \sim 25~pc$) by clearing most of the gas and dust from the inner region. At the same time, the energetic UV radiations shape the surrounding region by creating several structures such as BRCs, filaments, and fingertip structures within and around the complex \citep{1991ApJ...370..263S, 2005A&A...432..575F, 2012MNRAS.421.3206S}. Star-formation activity of the individual BRCs and also the entire complex have been carried out in many past studies \citep{2008AJ....135.2323I,2014A&A...562A.131S,2014MNRAS.443.1614P,2023AA...669A..22P,2023ApJ...948....7D}. In this section, we attempt to briefly discuss the over-all star formation history of the entire complex.

\subsection{IC 1396: an expanding bubble} \label{kinematic}
\begin{figure*}
\centering
\includegraphics[scale=0.52]{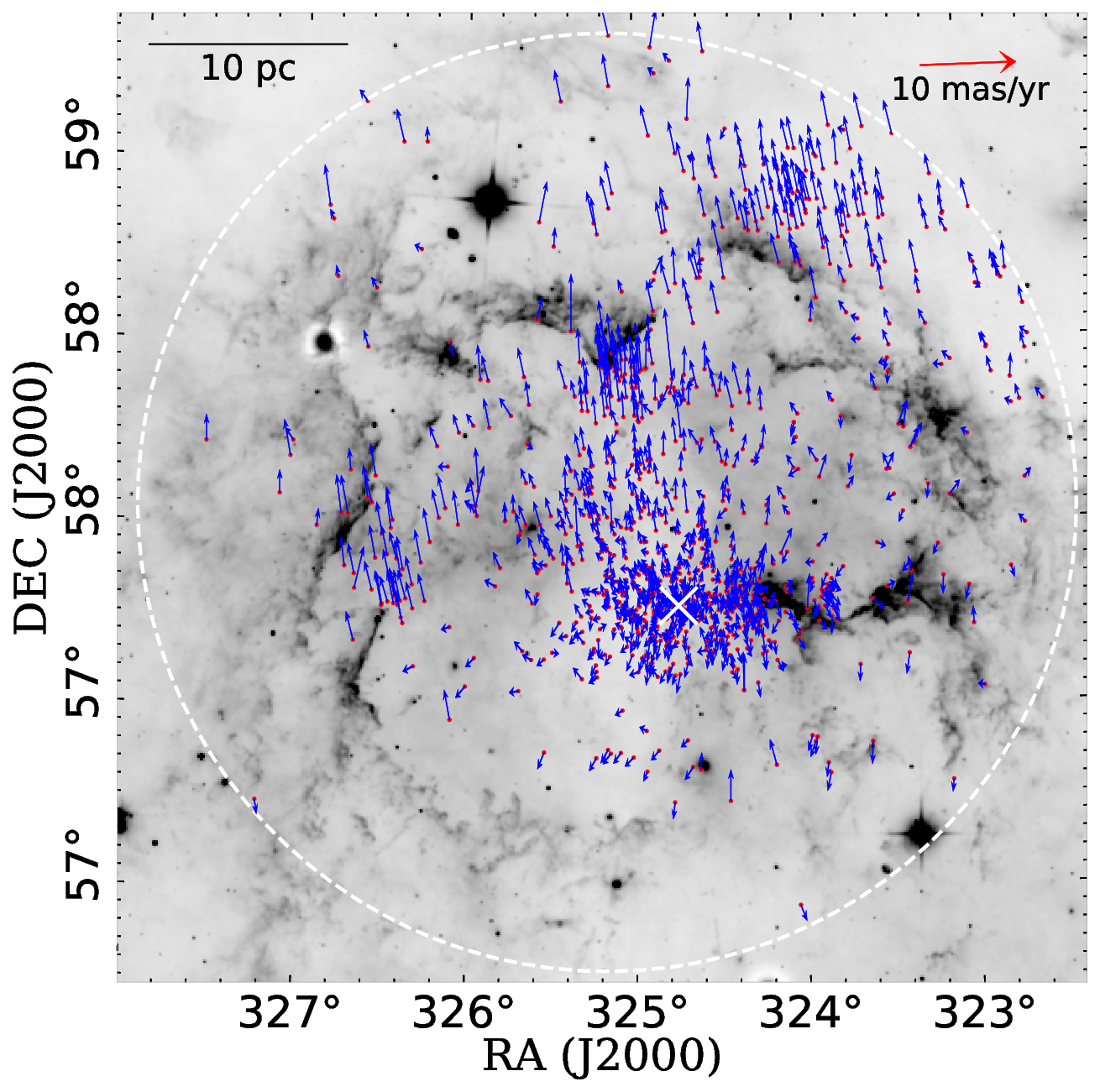}
\caption{The grayscale image represents the WISE $\rm 22~\mu m$ band data, overplotted with the velocity dispersion in proper motion of the 1020 Gaia counterparts of the candidate members. The red dots indicate the locations of these 1020 member stars. The blue arrows illustrate the direction and magnitude of the proper motion vectors relative to the average proper motions of sub-cluster C-1. The location of the central massive star(s), HD 206267, is marked with a white `$\times$' symbol. The scale bar and reference arrow for the proper motion vectors are shown in the panel.}
\label{plt_pm}
\end{figure*} 

\begin{figure}
\centering
\includegraphics[scale=0.232]{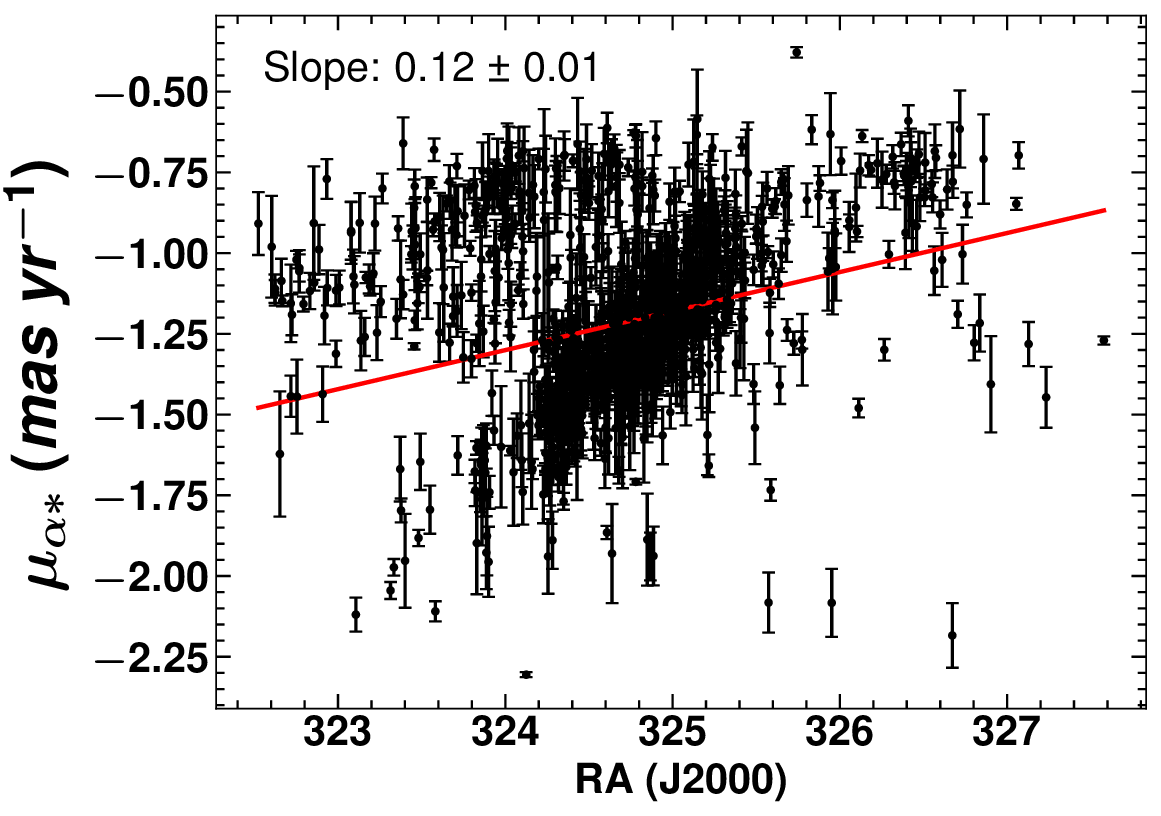}
\includegraphics[scale=0.232]{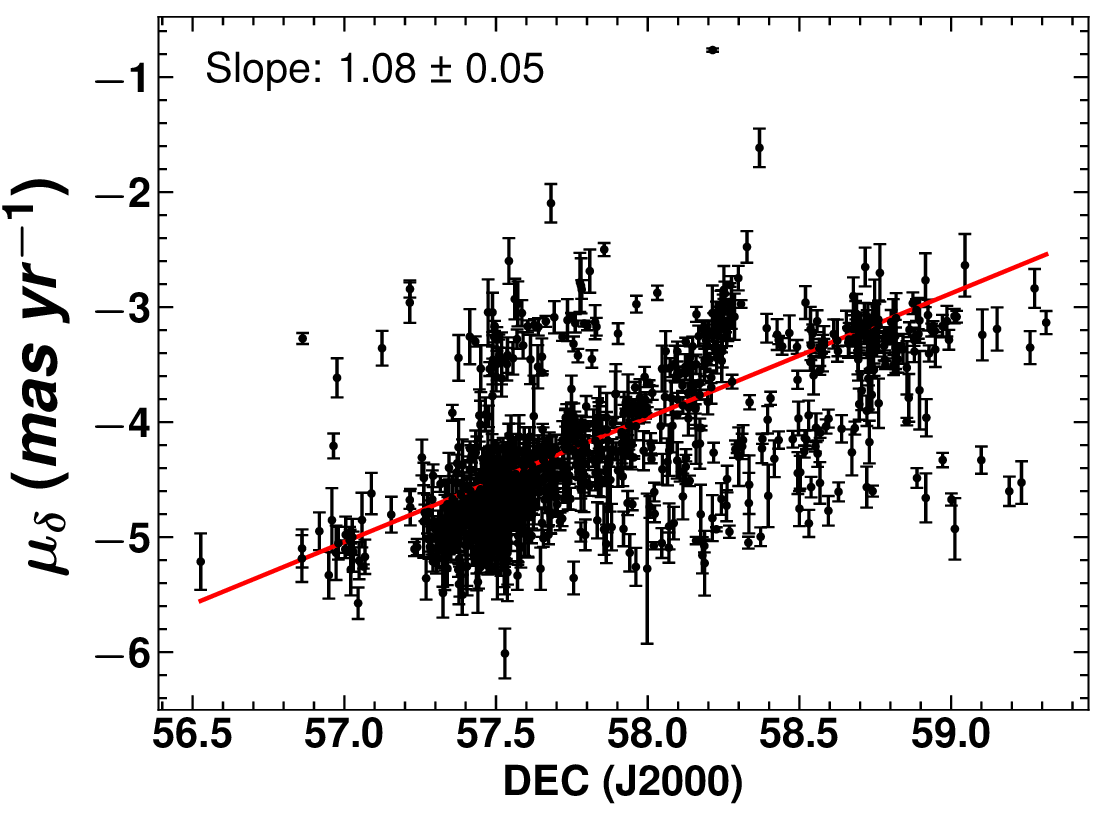}
\caption{Variation of RA with respect to $\mu_{\alpha*}$ (left) and DEC with respect to $\mu_{\delta}$ (right). A linear fit is applied, with the slopes labelled on each panel. The fit is shown as the red line.}
\label{plt_radec_pm}
\end{figure}

In Fig. \ref{plt_pm}, we show the distribution of 1020 candidate members that have Gaia counterparts. The direction of the proper motions relative to the central sub-cluster C-1 is illustrated, indicating a random motion of the stars. Previous studies have shown that the complex IC 1396 is gravitationally unbound and is expanding \citep{1995ApJ...447..721P,2023AA...669A..22P}. In this section, we explore the kinematics of IC 1396 using the astrometric details of these 1020 stars.

The variation of RA with respect to $\mu_{\alpha} \cos \delta$ ($\rm \mu_{\alpha*}$) and DEC with respect to $\rm \mu_{\delta}$ shows a positive correlation (Fig. \ref{plt_radec_pm}). This correlation suggests the presence of velocity gradient in IC 1396 and indicates the expanding nature of the complex, as observed in other regions (e.g., \citealt{2019MNRAS.486.2477W}). To further investigate the expanding feature of IC 1396, we conduct a simple 2-D analysis, as a 3-D analysis is not feasible due to the lack of radial velocities for all 1020 stars.

For this analysis, we first translate the celestial coordinates to the 2-D Cartesian coordinate system. Both positions and velocities in the Cartesian coordinates can be derived using the following equations \citep{2018A&A...616A..12G}.

\begin{align}
\begin{cases}
x = \cos \delta \sin(\alpha - \alpha_C) \\
y = \sin \delta \cos \delta_C - \cos \delta \sin \delta_C \cos(\alpha - \alpha_C) \\
\mu_x = \mu_{\alpha*} \cos(\alpha - \alpha_C) - \mu_\delta \sin \delta \sin(\alpha - \alpha_C) \\
\mu_y = \mu_{\alpha*} \sin \delta_C \sin(\alpha - \alpha_C) \\
\quad + \mu_\delta (\cos \delta \cos \delta_C + \sin \delta \sin \delta_C \cos(\alpha - \alpha_C))
\end{cases}
\end{align}

\noindent where, $\alpha_C$ and $\delta_C$ are the mean RA and DEC (324.668; 57.486) of stars within sub-cluster C-1, used as the reference position. The linear velocities ($\mu_x, \mu_y$) are derived as $\mu_x = (-k/\omega)\mu_{\alpha*}$ and $\mu_y = (k/\omega)\mu_{\delta}$, where $\omega$ is the parallax of each star and $k = 4.74$ is the conversion factor from $\rm mas~yr^{-1}$ to $\rm km~s^{-1}$ at a distance of 1 kpc. We derived the positions and velocities for all stars in the X and Y directions using these equations.

After deriving the individual velocities in the X and Y directions, the relative velocities $\mathbf{v}$ ($v_x, v_y$) of each star with respect to the mean values of C-1 are obtained. The directions of the relative velocities indicate whether a star is moving towards or away from C-1. The final expansion velocity can be obtained as $v_{\text{exp}} = \mathbf{v} \cdot \hat{\mathbf{r}}$, where $\hat{\mathbf{r}}$ is the radial unit vector. If a star moves outward from C-1, then $v_{\text{exp}}$ will be positive; otherwise, it will be negative.

\begin{figure}
\centering
\includegraphics[scale=0.45]{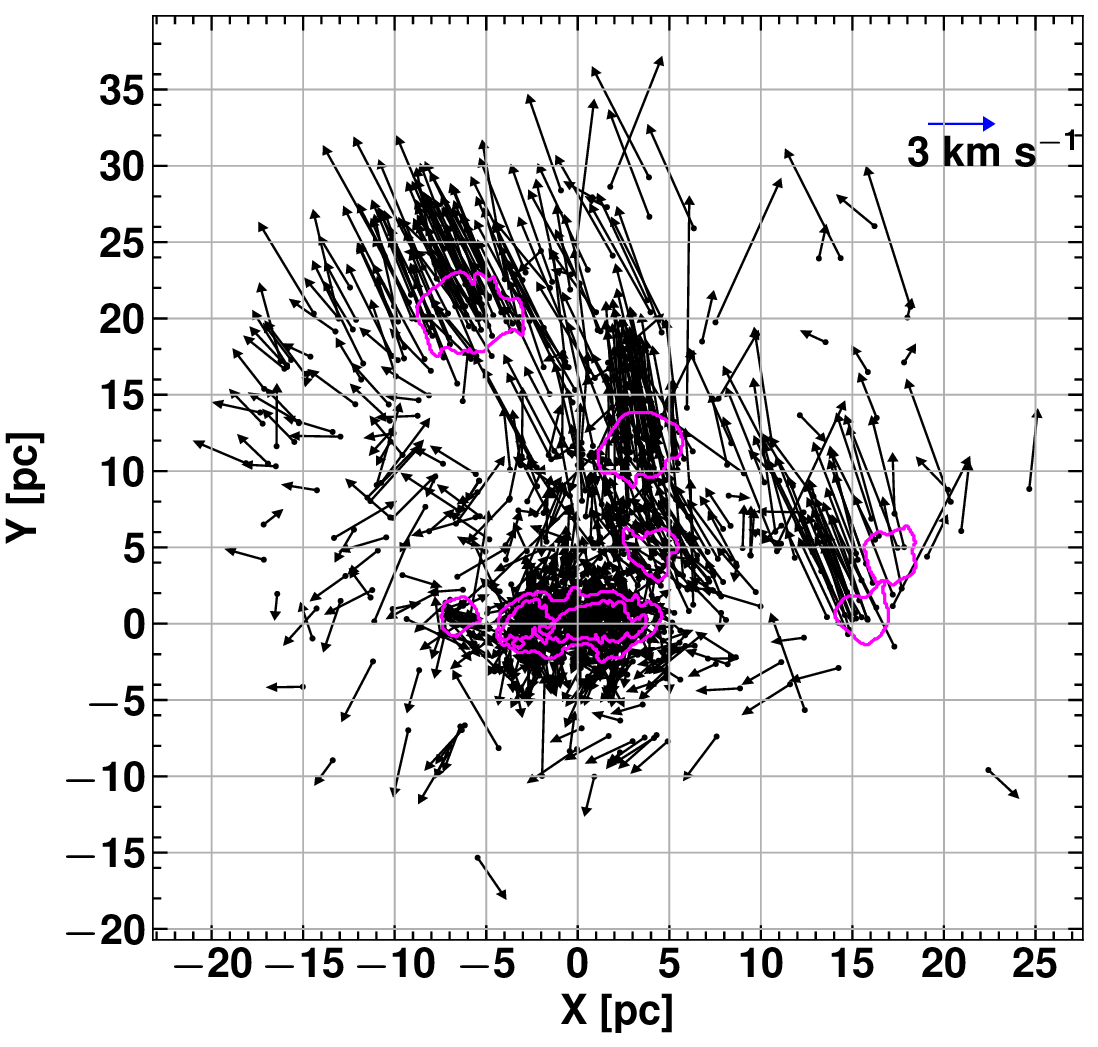}
\caption{Relative velocity vectors (magnitude and direction) for all stars with respect to the mean value of the central sub-cluster C-1. Magenta curves indicate the boundaries of the sub-clusters identified in this work. A reference velocity of $\rm 3~km~s^{-1}$ is shown as a blue arrow at the top right of the panel.}
\label{plt_rel_vel}
\end{figure}

\begin{figure}
\centering
\includegraphics[scale=0.45]{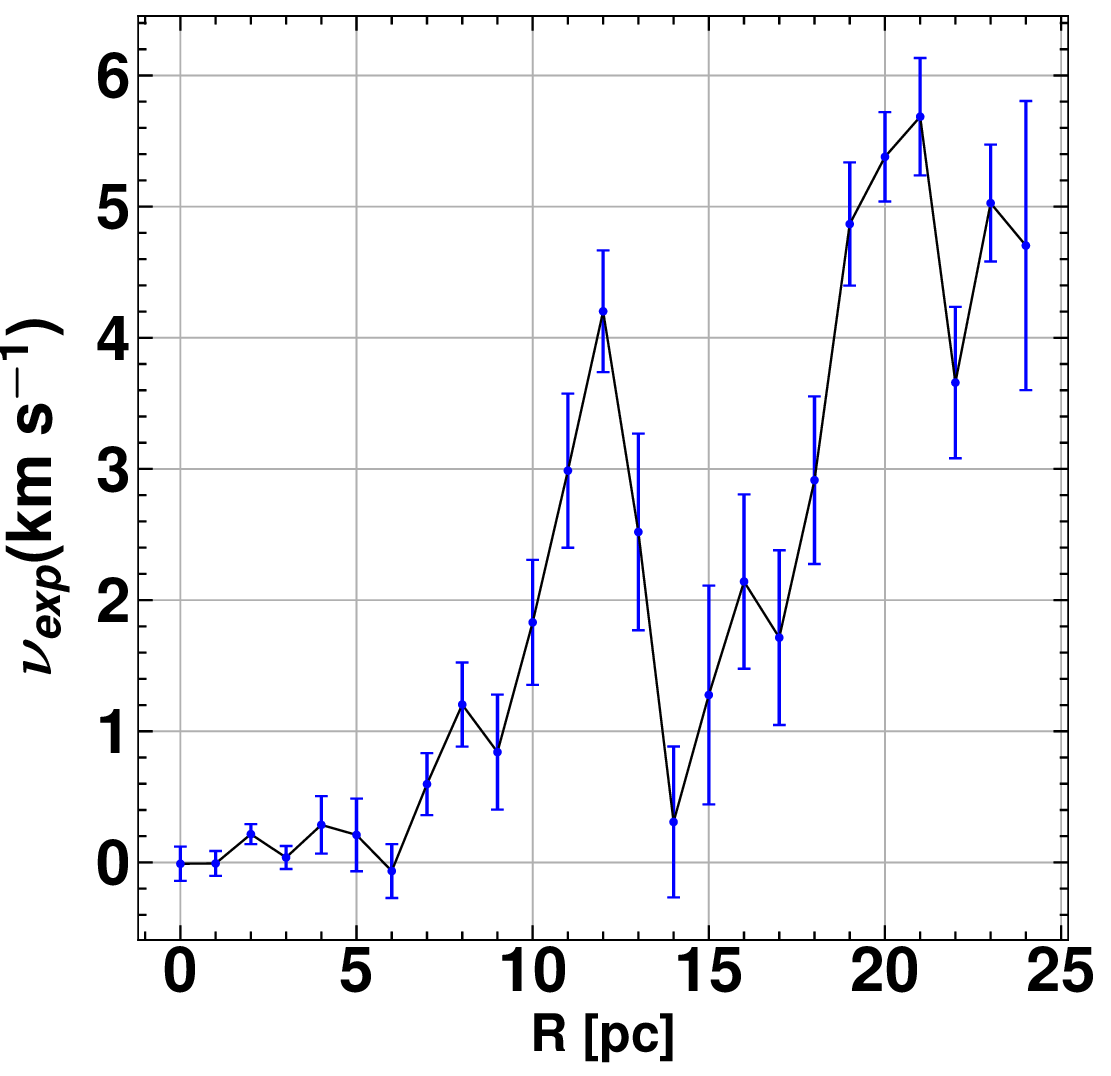}
\caption{Variation of $v_{\text{exp}}$ as a function of distance from the reference location (324.668; 57.486) within the central sub-cluster C-1. The distance is divided into bins of 1~pc. For each bin, the mean value of $v_{\text{exp}}$ is presented, with error bars estimated through the standard error of the mean of the $v_{\text{exp}}$ values within the distance bin.  }
\label{plt_vexp}
\end{figure}

In Fig. \ref{plt_rel_vel}, we show the distribution of the 1020 stars in the Cartesian coordinate system along with their relative velocities. As seen from the plot, the majority of stars are moving away from C-1, thereby providing evidence of the expansion of the star-forming complex. To obtain a quantitative estimate of the expansion velocity, we calculate the mean velocity within radial distance bins of 1~pc from the reference position. The variation of $v_{\text{exp}}$ with radial distance is shown in Fig. \ref{plt_vexp}. There is very little expansion up to a distance of $\sim$5~pc, which encompasses the central sub-cluster C-1. Previous work by \cite{2023ApJ...948....7D} shows very little expansion of $\rm \sim1~km~s^{-1}$ for sub-cluster located within the central region, based on their 3-D analysis of 68 cluster members with radial velocity information. However, Fig. \ref{plt_vexp} does not reveal expansion within the central cluster. This is because, in this analysis, we consider the relative velocities with respect to the mean values of C-1 itself. Thus, it is expected that stars within C-1 will exhibit both positive and negative $v_{\text{exp}}$, not reflecting the true expanding structure within C-1 itself. In this analysis, a clearer picture of expansion is obtained by moving away from cluster C-1. Beyond 5~pc, the expansion shows a gradual rise up to 12~pc, reaching an expansion velocity of $\rm 4.2 \pm 0.5~km~s^{-1}$. A dip is observed up to 13~pc, followed by another rise in $v_{\text{exp}}$. The expansion velocity increases to its maximum value of $\rm 5.7 \pm 0.5~km~s^{-1}$ at a distance of 21~pc, which is close to the periphery of the star-forming complex.

This variation of $v_{\text{exp}}$ with distance reveals the outward movement of distant stars from the central cluster. This provides evidence of a gravitationally unbound system, demonstrating a clear expanding feature. The gravitationally unbound and expanding nature of the complex has been reported in earlier works. \citet{2023AA...669A..22P} identified sub-clusters of the complex based on astrometry and observed the expanding nature by calculating pairwise projected velocity differences among the stars. From CO maps, \cite{1995ApJ...447..721P} suggested an expansion velocity of $\rm \sim5~km~s^{-1}$, which is also supported by \citet{2012MNRAS.426.2917G} and \citet{2019A&A...622A.118S}. Although we used a different approach to analyse the kinematics, our findings agree with the conclusions reported in previous studies of the star-forming complex.

\subsection{Hierarchical, sequential, and distributed mode of star formation} \label{seq_star_form}
The spatial distribution of the member sources (Fig. \ref{sd_clust}) and the association of sub-clusters with the BRCs and PDRs suggest ongoing feedback-driven star formation activity within the complex. Table \ref{stat_cluster} provides the mean stellar densities of all sub-clusters. Higher stellar densities are observed around the central massive star(s) HD 206267, reaching a maximum within sub-cluster C-1A and then peaking within sub-cluster C-1B, near the head of the BRC IC 1396 A. The stellar density gradually decreases and shows localized peaks in surrounding sub-clusters. This distribution indicates that the majority of star formation is occurring within the central part around the massive star(s), with density decreasing gradually towards the surrounding area. The localised peaks in stellar densities can be attributed to the local molecular conditions, where the star formation efficiency may vary. The presence of the massive star(s) HD 206267 drastically influencing the star formation within the central cluster, resulting in its higher stellar density. However, all the sub-clusters of IC 1396 have nearly similar average ages. This suggests a scenario where the molecular cloud undergoes a hierarchical fragmentation process, leading to star formation occurring within a similar or nearly simultaneous timeframe \citep{2003MNRAS.343..413B, 2018MNRAS.481..688G, 2022MNRAS.510.2097T}. Such possibilities of the star-formation process in IC 1396 have also been reported earlier \citep{2023ApJ...948....7D}. However, it is also possible that the fragmentation process may continue, forming much younger stars and leading to sequential hierarchical star formation within the molecular cloud.

The distribution of the identified stellar population in this work shows the presence of many sources distributed randomly in the whole star-forming complex of IC 1396 (see Fig. \ref{sd_clust}). These stars are lying in regions of stellar density less than 1.5~stars $\rm pc^{-2}$. This low density and the scattered distribution of stars are characteristic of a distributed mode of star formation, where stars form in a more isolated and spread-out manner rather than in dense, clustered regions \citep{2003ARA&A..41...57L}. This observation is similar to other regions known for distributed star formation, such as the Taurus Molecular Cloud \citep{2006ApJ...647.1180L, 2018AJ....156..271L}, and supports the notion that star formation can occur efficiently even in environments with relatively low stellar densities \citep{2009ApJS..181..321E}. Apart from distributed star formation, these low-mass objects might have been ejected from their parent sub-clusters through dynamic processes \citep{2002MNRAS.332L..65B,2003MNRAS.339..577B,2007A&A...466..943G}.

In IC 1396, stars are expected to move at a velocity of $\rm \sim 3~km~s^{-1}$ when they are ejected from their parent sub-clusters. This velocity is an average of values reported in previous studies of this complex. \citet{2023ApJ...948....7D} estimated the expansion velocity of the central cluster to be 1.11~$\rm km~s^{-1}$. From CO maps, \citet{1995ApJ...447..721P} obtained the expansion velocity of the entire complex to be around $\rm 5~km~s^{-1}$, a result supported by later studies \citep{2023AA...669A..22P,2012MNRAS.426.2917G,2019A&A...622A.118S}. Our current kinematic analysis also reveals an expanding feature of IC 1396.
An earlier study by \citet{1995A&A...304L...9S} estimated a velocity of $\rm \sim 3-4~km~s^{-1}$ for ejected stars from clusters. If we assume an ejected velocity of $\rm \sim 3~km~s^{-1}$ perpendicular to the line of sight, then over the average age of 4~Myr, a star might travel a distance of $\rm \sim 12~pc$. The radius of IC 1396 is $\rm 1.5^\circ$, corresponding to $\rm \sim 25~pc$. Thus, randomly distributed stars are expected to lie within IC 1396, even if they are ejected from the central cluster. Although we speculate about the ejection of stars, we are unable to determine the ejection rate, which requires a clear understanding of the total cluster mass and its distribution, necessitating more detailed analysis.

In this work, we provide a brief discussion of the possible modes of star formation processes in IC 1396. However, a much more dedicated analysis is essential to gain a deeper understanding of the ongoing star formation activities in the star-forming complex.

\section{Summary} \label{summ}
This work presents the deepest photometric observations of the star-forming complex IC 1396 with the Subaru-HSC. By combining the optical and NIR data from Pan-STARRS and UKIDSS catalogues with the HSC catalogue, we conduct a comprehensive membership analysis of this star-forming complex. In the following, we outline the significant findings of this study.

\begin{enumerate}
\item Our Subaru-HSC observations of IC 1396 were conducted with two pointings in the $\rm r_2$, $\rm i_2$, and Y-bands. We processed the optical data using \textsc{hscPipe} V6.7 to generate the photometric catalogue and images. After applying various quality flags, we obtained a clean HSC catalogue ensuring good photometric quality, where the magnitude error is less than 0.2 for all three bands. The complete catalogue comprises 859867 stars with photometric magnitudes in all three bands. The limiting magnitude of the catalogue is 27.6, 26.6, and 25.2~mag in the $\rm r_2$, $\rm i_2$, and Y-bands, respectively. The 90\% completeness of the catalogue is 24, 22, and 21~mag in the $\rm r_2$, $\rm i_2$, and Y-bands, respectively.

\item We expanded our dataset by including additional sources from the Pan-STARRS catalogue to cover the complete circular region of radius $1.5^{\circ}$ along with to compensate the brighter end of the HSC catalogue. The merged catalogue consists of 1322119 stars, of which 802100 have counterparts in the UKIDSS catalogue. The membership analysis was conducted using these stars.

\item We employed the RF classifier method of the ML algorithm and identify 2425 high-probability candidate members with $\rm P_{RF}\geqslant0.8$. The training set was largely limited to sources younger than approximately 10 Myr; consequently, the retrieved candidate members are also within this age range.

\item The CMD distribution of these 2425 candidate member stars indicates an average age of $\rm 2-4~Myr$, consistent with previous studies. We determined the limiting and 90\% mass completeness limit using the $\rm r_2$ band and isochrones from \citet{2015A&A...577A..42B}. The limiting mass of the members is approximately $\rm \sim 0.01-0.02~M_{\odot}$, with a 90\% mass completeness limit of $\rm \sim 0.04-0.05~M_{\odot}$. Notably, this study marks the detection of many low-mass member objects in IC 1396.

\item The spatial distribution of the member stars reveals the presence of several sub-clusters. Using the 1461 stars brighter than 20 mag in the $\rm r_2$ band, we identify sub-clusters in the complex. By applying the NN method, we detected {eight} sub-clusters. In addition to the central cluster, one sub-cluster (C-3) is observed on the diagonal connecting the BRC IC 1396 N and the central cluster. Moreover, the central cluster C-1 exhibits the presence of two sub-clusters, C-1A and C-1B, situated on two sides of the central massive star(s).

\item Our kinematic analysis, based on the astrometry of Gaia counterparts of candidate member sources, reveals the expanding nature of the star-forming complex. Additionally, we briefly discuss the possibility of hierarchical, sequential, and distributed modes of star formation activities in IC 1396.

\end{enumerate}

We wish to emphasise that the primary aim of this work is to identify the member populations of the star-forming complex IC 1396 using the deep, sensitive Subaru-HSC data. Therefore this work provides an opportunity to evaluate the strengths and limitations of machine learning techniques when applied to such astronomical datasets.

It is true that all membership detection methods have inherent limitations, including our own. In this work, we have derived several new parameters from various combinations of magnitudes and colours, which help mitigate some of these limitations, although not entirely. Due to these constraints, we are unable to detect candidate members of IC 1396 older than 10 Myr, primarily because of the absence of older candidate member populations in the training set. Additionally, the lack of sources in the training set towards the fainter end introduces a bias, and hence, statistically, the chance of non-detection of candidate member stars is higher at the fainter end compared to the brighter end. This limitation affects the detection of very low-mass objects such as brown dwarfs in the star-forming complex. Furthermore, the presence of nebular emission lines, along with foreground and background stars, can lead to potential contamination. Spectroscopic analyses will be crucial in further filtering out such contaminants identified through photometric studies.

We are keen to improve our detection method in future works. It would be interesting to test the results by combining physical parameters such as effective temperature, luminosity, age, and mass, with the existing parameters used in this study. Since machine learning methods are effective in handling large parameter sets, we expect that the use of more robust parameters could potentially help in better constraining the machine learning model. However, we must also focus on improving the training set, ensuring it includes both member and non-member populations without age restrictions, as failing to do so may introduce bias into the final outcome. We are eager to contribute to such membership analyses and make the method more robust by addressing known limitations, enabling it to work without age restrictions across various datasets and star-forming complexes in diverse environments.

\begin{acknowledgements}
We thank the anonymous referee for his/her detailed comments, which significantly improved the paper. This research is based on data collected at Subaru Telescope with Hyper Suprime-Cam, which is operated by the National Astronomical Observatory of Japan. We are honoured and grateful for the opportunity of observing the Universe from Mauna Kea, which has the cultural, historical and natural significance in Hawaii. We are grateful to The East Asian Observatory which is supported by The National Astronomical Observatory of Japan; Academia Sinica Institute of Astronomy and Astrophysics; the Korea Astronomy and Space Science Institute; the Operation, Maintenance and Upgrading Fund for Astronomical Telescopes and Facility Instruments, budgeted from the Ministry of Finance (MOF) of China and administrated by the Chinese Academy of Sciences (CAS), as well as the National Key R\&D Program of China (no. 2017YFA0402700). The authors thank the entire HSC staff and HSC helpdesk for their help. We would like to thank S. Mineo, H. Furusawa, Y. Yamada, and M. Kubo in HSC helpdesk team for useful discussions regarding the data reduction. We thank NAOJ for providing access to hanaco account which was used to perform some initial stages of data reduction. This work presents results from the European Space Agency (ESA) space mission Gaia. Gaia data are being processed by the Gaia Data Processing and Analysis Consortium (DPAC). Funding for the DPAC is provided by national institutions, in particular the institutions participating in the Gaia MultiLateral Agreement (MLA). The Gaia mission website is https://www.cosmos.esa.int/gaia. The Gaia archive website is https://archives.esac.esa.int/gaia. This publication makes use of data products from the Wide-field Infrared Survey Explorer, which is a joint project of the University of California, Los Angeles, and the Jet Propulsion Laboratory/California Institute of Technology, funded by the National Aeronautics and Space Administration. This research has made use of the SIMBAD database, operated at CDS, Strasbourg, France. This work made use of various packages of Python programming language. SRD acknowledges support from the Fondecyt Postdoctoral fellowship (project code 3220162) and ANID BASAL project FB210003. SRD acknowledges membership in the BRICS-SO. JJ acknowledges the financial support received through the DST-SERB grant SPG 2021/003850. We gratefully acknowledge the use of high performance computing facilities at IUCAA, Pune for the HSC data reduction. We acknowledge the use of the AI tool ChatGPT in this research work.
\end{acknowledgements}
\bibliographystyle{aa} 
\bibliography{IC1396} 

\begin{thebibliography}{140}
\expandafter\ifx\csname natexlab\endcsname\relax\def\natexlab#1{#1}\fi

\bibitem[{{Almendros-Abad} {et~al.}(2023){Almendros-Abad}, {Mu{\v{z}}i{\'c}},
  {Bouy}, {Bayo}, {Scholz}, {Pe{\~n}a Ram{\'\i}rez}, {Moitinho}, {Kubiak},
  {Sch{\"o}edel}, {Bara{\v{c}}}, {Br{\v{c}}i{\'c}}, {Ascenso}, \&
  {Jayawardhana}}]{2023A&A...677A..26A}
{Almendros-Abad}, V., {Mu{\v{z}}i{\'c}}, K., {Bouy}, H., {et~al.} 2023, \aap,
  677, A26

\bibitem[{{Baraffe} {et~al.}(2015){Baraffe}, {Homeier}, {Allard}, \&
  {Chabrier}}]{2015A&A...577A..42B}
{Baraffe}, I., {Homeier}, D., {Allard}, F., \& {Chabrier}, G. 2015, \aap, 577,
  A42

\bibitem[{{Barentsen} {et~al.}(2011){Barentsen}, {Vink}, {Drew}, {Greimel},
  {Wright}, {Drake}, {Martin}, {Valdivielso}, \&
  {Corradi}}]{2011MNRAS.415..103B}
{Barentsen}, G., {Vink}, J.~S., {Drew}, J.~E., {et~al.} 2011, \mnras, 415, 103

\bibitem[{{Bastian} {et~al.}(2010){Bastian}, {Covey}, \&
  {Meyer}}]{2010ARA&A..48..339B}
{Bastian}, N., {Covey}, K.~R., \& {Meyer}, M.~R. 2010, \araa, 48, 339

\bibitem[{{Bate} {et~al.}(2002){Bate}, {Bonnell}, \&
  {Bromm}}]{2002MNRAS.332L..65B}
{Bate}, M.~R., {Bonnell}, I.~A., \& {Bromm}, V. 2002, \mnras, 332, L65

\bibitem[{{Bate} {et~al.}(2003){Bate}, {Bonnell}, \&
  {Bromm}}]{2003MNRAS.339..577B}
{Bate}, M.~R., {Bonnell}, I.~A., \& {Bromm}, V. 2003, \mnras, 339, 577

\bibitem[{{Beltr{\'a}n} {et~al.}(2002){Beltr{\'a}n}, {Girart}, {Estalella},
  {Ho}, \& {Palau}}]{2002ApJ...573..246B}
{Beltr{\'a}n}, M.~T., {Girart}, J.~M., {Estalella}, R., {Ho}, P. T.~P., \&
  {Palau}, A. 2002, \apj, 573, 246

\bibitem[{{Beltr{\'a}n} {et~al.}(2009){Beltr{\'a}n}, {Massi}, {L{\'o}pez},
  {Girart}, \& {Estalella}}]{2009A&A...504...97B}
{Beltr{\'a}n}, M.~T., {Massi}, F., {L{\'o}pez}, R., {Girart}, J.~M., \&
  {Estalella}, R. 2009, \aap, 504, 97

\bibitem[{{Bonnell} {et~al.}(2003){Bonnell}, {Bate}, \&
  {Vine}}]{2003MNRAS.343..413B}
{Bonnell}, I.~A., {Bate}, M.~R., \& {Vine}, S.~G. 2003, \mnras, 343, 413

\bibitem[{{Bosch} {et~al.}(2018){Bosch}, {Armstrong}, {Bickerton}, {Furusawa},
  {Ikeda}, {Koike}, {Lupton}, {Mineo}, {Price}, {Takata}, {Tanaka}, {Yasuda},
  {AlSayyad}, {Becker}, {Coulton}, {Coupon}, {Garmilla}, {Huang}, {Krughoff},
  {Lang}, {Leauthaud}, {Lim}, {Lust}, {MacArthur}, {Mand elbaum}, {Miyatake},
  {Miyazaki}, {Murata}, {More}, {Okura}, {Owen}, {Swinbank}, {Strauss},
  {Yamada}, \& {Yamanoi}}]{2018PASJ...70S...5B}
{Bosch}, J., {Armstrong}, R., {Bickerton}, S., {et~al.} 2018, \pasj, 70, S5

\bibitem[{Breiman(2001)}]{breiman2001random}
Breiman, L. 2001, Machine Learning, 45, 5

\bibitem[{{Brink} {et~al.}(2013){Brink}, {Richards}, {Poznanski}, {Bloom},
  {Rice}, {Negahban}, \& {Wainwright}}]{2013MNRAS.435.1047B}
{Brink}, H., {Richards}, J.~W., {Poznanski}, D., {et~al.} 2013, \mnras, 435,
  1047

\bibitem[{{Cantat-Gaudin} {et~al.}(2018){Cantat-Gaudin}, {Jordi}, {Vallenari},
  {Bragaglia}, {Balaguer-N{\'u}{\~n}ez}, {Soubiran}, {Bossini}, {Moitinho},
  {Castro-Ginard}, {Krone-Martins}, {Casamiquela}, {Sordo}, \&
  {Carrera}}]{2018AA...618A..93C}
{Cantat-Gaudin}, T., {Jordi}, C., {Vallenari}, A., {et~al.} 2018, \aap, 618,
  A93

\bibitem[{{Caratti o Garatti} {et~al.}(2006){Caratti o Garatti}, {Giannini},
  {Nisini}, \& {Lorenzetti}}]{2006A&A...449.1077C}
{Caratti o Garatti}, A., {Giannini}, T., {Nisini}, B., \& {Lorenzetti}, D.
  2006, \aap, 449, 1077

\bibitem[{{Carpenter}(2001)}]{2001AJ....121.2851C}
{Carpenter}, J.~M. 2001, \aj, 121, 2851

\bibitem[{{Casertano} \& {Hut}(1985)}]{1985ApJ...298...80C}
{Casertano}, S. \& {Hut}, P. 1985, \apj, 298, 80

\bibitem[{{Chambers} \& {et al.}(2017)}]{2017yCat.2349....0C}
{Chambers}, K.~C. \& {et al.} 2017, VizieR Online Data Catalog, II/349

\bibitem[{{Chambers} {et~al.}(2016){Chambers}, {Magnier}, {Metcalfe},
  {Flewelling}, {Huber}, {Waters}, {Denneau}, {Draper}, {Farrow}, {Finkbeiner},
  {Holmberg}, {Koppenhoefer}, {Price}, {Rest}, {Saglia}, {Schlafly}, {Smartt},
  {Sweeney}, {Wainscoat}, {Burgett}, {Chastel}, {Grav}, {Heasley}, {Hodapp},
  {Jedicke}, {Kaiser}, {Kudritzki}, {Luppino}, {Lupton}, {Monet}, {Morgan},
  {Onaka}, {Shiao}, {Stubbs}, {Tonry}, {White}, {Ba{\~n}ados}, {Bell},
  {Bender}, {Bernard}, {Boegner}, {Boffi}, {Botticella}, {Calamida},
  {Casertano}, {Chen}, {Chen}, {Cole}, {Deacon}, {Frenk}, {Fitzsimmons},
  {Gezari}, {Gibbs}, {Goessl}, {Goggia}, {Gourgue}, {Goldman}, {Grant},
  {Grebel}, {Hambly}, {Hasinger}, {Heavens}, {Heckman}, {Henderson}, {Henning},
  {Holman}, {Hopp}, {Ip}, {Isani}, {Jackson}, {Keyes}, {Koekemoer}, {Kotak},
  {Le}, {Liska}, {Long}, {Lucey}, {Liu}, {Martin}, {Masci}, {McLean}, {Mindel},
  {Misra}, {Morganson}, {Murphy}, {Obaika}, {Narayan}, {Nieto-Santisteban},
  {Norberg}, {Peacock}, {Pier}, {Postman}, {Primak}, {Rae}, {Rai}, {Riess},
  {Riffeser}, {Rix}, {R{\"o}ser}, {Russel}, {Rutz}, {Schilbach}, {Schultz},
  {Scolnic}, {Strolger}, {Szalay}, {Seitz}, {Small}, {Smith}, {Soderblom},
  {Taylor}, {Thomson}, {Taylor}, {Thakar}, {Thiel}, {Thilker}, {Unger},
  {Urata}, {Valenti}, {Wagner}, {Walder}, {Walter}, {Watters}, {Werner},
  {Wood-Vasey}, \& {Wyse}}]{2016arXiv161205560C}
{Chambers}, K.~C., {Magnier}, E.~A., {Metcalfe}, N., {et~al.} 2016, arXiv
  e-prints, arXiv:1612.05560

\bibitem[{{Choudhury} {et~al.}(2010){Choudhury}, {Mookerjea}, \&
  {Bhatt}}]{2010ApJ...717.1067C}
{Choudhury}, R., {Mookerjea}, B., \& {Bhatt}, H.~C. 2010, \apj, 717, 1067

\bibitem[{{Concha-Ram{\'\i}rez} {et~al.}(2019){Concha-Ram{\'\i}rez}, {Vaher},
  \& {Portegies Zwart}}]{2019MNRAS.482..732C}
{Concha-Ram{\'\i}rez}, F., {Vaher}, E., \& {Portegies Zwart}, S. 2019, \mnras,
  482, 732

\bibitem[{{Contreras} {et~al.}(2002){Contreras}, {Sicilia-Aguilar},
  {Muzerolle}, {Calvet}, {Berlind}, \& {Hartmann}}]{2002AJ....124.1585C}
{Contreras}, M.~E., {Sicilia-Aguilar}, A., {Muzerolle}, J., {et~al.} 2002, \aj,
  124, 1585

\bibitem[{{Coupon} {et~al.}(2018){Coupon}, {Czakon}, {Bosch}, {Komiyama},
  {Medezinski}, {Miyazaki}, \& {Oguri}}]{2018PASJ...70S...7C}
{Coupon}, J., {Czakon}, N., {Bosch}, J., {et~al.} 2018, \pasj, 70, S7

\bibitem[{{Dale} {et~al.}(2012){Dale}, {Ercolano}, \&
  {Bonnell}}]{2012MNRAS.424..377D}
{Dale}, J.~E., {Ercolano}, B., \& {Bonnell}, I.~A. 2012, \mnras, 424, 377

\bibitem[{{Dale} {et~al.}(2013){Dale}, {Ercolano}, \&
  {Bonnell}}]{2013MNRAS.430..234D}
{Dale}, J.~E., {Ercolano}, B., \& {Bonnell}, I.~A. 2013, \mnras, 430, 234

\bibitem[{{Damian} {et~al.}(2023{\natexlab{a}}){Damian}, {Jose}, {Biller},
  {Herczeg}, {Albert}, {Allers}, {Zhang}, {Liu}, {Dubber}, {Paul}, {Chen},
  {Lalchand}, {Sharma}, \& {Oasa}}]{2023arXiv230317424D}
{Damian}, B., {Jose}, J., {Biller}, B., {et~al.} 2023{\natexlab{a}}, arXiv
  e-prints, arXiv:2303.17424

\bibitem[{{Damian} {et~al.}(2023{\natexlab{b}}){Damian}, {Jose}, {Biller}, \&
  {Paul}}]{2023arXiv230518147D}
{Damian}, B., {Jose}, J., {Biller}, B., \& {Paul}, K. 2023{\natexlab{b}}, arXiv
  e-prints, arXiv:2305.18147

\bibitem[{{Damian} {et~al.}(2021){Damian}, {Jose}, {Samal}, {Moraux}, {Das}, \&
  {Patra}}]{2021MNRAS.504.2557D}
{Damian}, B., {Jose}, J., {Samal}, M.~R., {et~al.} 2021, \mnras, 504, 2557

\bibitem[{{Das} {et~al.}(2023){Das}, {Gupta}, {Prakash}, {Samal}, \&
  {Jose}}]{2023ApJ...948....7D}
{Das}, S.~R., {Gupta}, S., {Prakash}, P., {Samal}, M., \& {Jose}, J. 2023,
  \apj, 948, 7

\bibitem[{{Das} {et~al.}(2021){Das}, {Jose}, {Samal}, {Zhang}, \&
  {Panwar}}]{2021MNRAS.500.3123D}
{Das}, S.~R., {Jose}, J., {Samal}, M.~R., {Zhang}, S., \& {Panwar}, N. 2021,
  \mnras, 500, 3123

\bibitem[{{Das} {et~al.}(2017){Das}, {Tej}, {Vig}, {Liu}, {Liu}, {Ishwara
  Chandra}, \& {Ghosh}}]{2017MNRAS.472.4750D}
{Das}, S.~R., {Tej}, A., {Vig}, S., {et~al.} 2017, \mnras, 472, 4750

\bibitem[{{de Grijs} {et~al.}(2002){de Grijs}, {Johnson}, {Gilmore}, \&
  {Frayn}}]{2002MNRAS.331..228D}
{de Grijs}, R., {Johnson}, R.~A., {Gilmore}, G.~F., \& {Frayn}, C.~M. 2002,
  \mnras, 331, 228

\bibitem[{{de Zeeuw} {et~al.}(1999){de Zeeuw}, {Hoogerwerf}, {de Bruijne},
  {Brown}, \& {Blaauw}}]{1999AJ....117..354D}
{de Zeeuw}, P.~T., {Hoogerwerf}, R., {de Bruijne}, J.~H.~J., {Brown}, A.~G.~A.,
  \& {Blaauw}, A. 1999, \aj, 117, 354

\bibitem[{{Dubath} {et~al.}(2011){Dubath}, {Rimoldini}, {S{\"u}veges},
  {Blomme}, {L{\'o}pez}, {Sarro}, {De Ridder}, {Cuypers}, {Guy}, {Lecoeur},
  {Nienartowicz}, {Jan}, {Beck}, {Mowlavi}, {De Cat}, {Lebzelter}, \&
  {Eyer}}]{2011MNRAS.414.2602D}
{Dubath}, P., {Rimoldini}, L., {S{\"u}veges}, M., {et~al.} 2011, \mnras, 414,
  2602

\bibitem[{{Elmegreen}(2006)}]{2006astro.ph.10687E}
{Elmegreen}, B.~G. 2006, arXiv e-prints, astro

\bibitem[{{Elmegreen} \& {Scalo}(2004)}]{2004ARA&A..42..211E}
{Elmegreen}, B.~G. \& {Scalo}, J. 2004, \araa, 42, 211

\bibitem[{{Evans} {et~al.}(2009){Evans}, {Dunham}, {J{\o}rgensen}, {Enoch},
  {Mer{\'\i}n}, {van Dishoeck}, {Alcal{\'a}}, {Myers}, {Stapelfeldt}, {Huard},
  {Allen}, {Harvey}, {van Kempen}, {Blake}, {Koerner}, {Mundy}, {Padgett}, \&
  {Sargent}}]{2009ApJS..181..321E}
{Evans}, Neal~J., I., {Dunham}, M.~M., {J{\o}rgensen}, J.~K., {et~al.} 2009,
  \apjs, 181, 321

\bibitem[{{Ferraro} {et~al.}(2016){Ferraro}, {Massari}, {Dalessandro},
  {Lanzoni}, {Origlia}, {Rich}, \& {Mucciarelli}}]{2016ApJ...828...75F}
{Ferraro}, F.~R., {Massari}, D., {Dalessandro}, E., {et~al.} 2016, \apj, 828,
  75

\bibitem[{{Flewelling} {et~al.}(2020){Flewelling}, {Magnier}, {Chambers},
  {Heasley}, {Holmberg}, {Huber}, {Sweeney}, {Waters}, {Calamida}, {Casertano},
  {Chen}, {Farrow}, {Hasinger}, {Henderson}, {Long}, {Metcalfe}, {Narayan},
  {Nieto-Santisteban}, {Norberg}, {Rest}, {Saglia}, {Szalay}, {Thakar},
  {Tonry}, {Valenti}, {Werner}, {White}, {Denneau}, {Draper}, {Hodapp},
  {Jedicke}, {Kaiser}, {Kudritzki}, {Price}, {Wainscoat}, {Chastel}, {McLean},
  {Postman}, \& {Shiao}}]{2020ApJS..251....7F}
{Flewelling}, H.~A., {Magnier}, E.~A., {Chambers}, K.~C., {et~al.} 2020, \apjs,
  251, 7

\bibitem[{{Froebrich} {et~al.}(2005){Froebrich}, {Scholz}, {Eisl{\"o}ffel}, \&
  {Murphy}}]{2005A&A...432..575F}
{Froebrich}, D., {Scholz}, A., {Eisl{\"o}ffel}, J., \& {Murphy}, G.~C. 2005,
  \aap, 432, 575

\bibitem[{{Furusawa} {et~al.}(2018){Furusawa}, {Koike}, {Takata}, {Okura},
  {Miyatake}, {Lupton}, {Bickerton}, {Price}, {Bosch}, {Yasuda}, {Mineo},
  {Yamada}, {Miyazaki}, {Nakata}, {Koshida}, {Komiyama}, {Utsumi},
  {Kawanomoto}, {Jeschke}, {Noumaru}, {Schubert}, {Iwata}, {Finet},
  {Fujiyoshi}, {Tajitsu}, {Terai}, \& {Lee}}]{2018PASJ...70S...3F}
{Furusawa}, H., {Koike}, M., {Takata}, T., {et~al.} 2018, \pasj, 70, S3

\bibitem[{{Gaia Collaboration} {et~al.}(2018){Gaia Collaboration}, {Helmi},
  {van Leeuwen}, {McMillan}, {Massari}, {Antoja}, {Robin}, {Lindegren},
  {Bastian}, {Arenou}, {Babusiaux}, {Biermann}, {Breddels}, {Hobbs}, {Jordi},
  {Pancino}, {Reyl{\'e}}, {Veljanoski}, {Brown}, {Vallenari}, {Prusti}, {de
  Bruijne}, {Bailer-Jones}, {Evans}, {Eyer}, {Jansen}, {Klioner}, {Lammers},
  {Luri}, {Mignard}, {Panem}, {Pourbaix}, {Randich}, {Sartoretti}, {Siddiqui},
  {Soubiran}, {Walton}, {Cropper}, {Drimmel}, {Katz}, {Lattanzi}, {Bakker},
  {Cacciari}, {Casta{\~n}eda}, {Chaoul}, {Cheek}, {De Angeli}, {Fabricius},
  {Guerra}, {Holl}, {Masana}, {Messineo}, {Mowlavi}, {Nienartowicz}, {Panuzzo},
  {Portell}, {Riello}, {Seabroke}, {Tanga}, {Th{\'e}venin}, {Gracia-Abril},
  {Comoretto}, {Garcia-Reinaldos}, {Teyssier}, {Altmann}, {Andrae}, {Audard},
  {Bellas-Velidis}, {Benson}, {Berthier}, {Blomme}, {Burgess}, {Busso},
  {Carry}, {Cellino}, {Clementini}, {Clotet}, {Creevey}, {Davidson}, {De
  Ridder}, {Delchambre}, {Dell'Oro}, {Ducourant},
  {Fern{\'a}ndez-Hern{\'a}ndez}, {Fouesneau}, {Fr{\'e}mat}, {Galluccio},
  {Garc{\'\i}a-Torres}, {Gonz{\'a}lez-N{\'u}{\~n}ez}, {Gonz{\'a}lez-Vidal},
  {Gosset}, {Guy}, {Halbwachs}, {Hambly}, {Harrison}, {Hern{\'a}ndez},
  {Hestroffer}, {Hodgkin}, {Hutton}, {Jasniewicz}, {Jean-Antoine-Piccolo},
  {Jordan}, {Korn}, {Krone-Martins}, {Lanzafame}, {Lebzelter}, {L{\"o}ffler},
  {Manteiga}, {Marrese}, {Mart{\'\i}n-Fleitas}, {Moitinho}, {Mora}, {Muinonen},
  {Osinde}, {Pauwels}, {Petit}, {Recio-Blanco}, {Richards}, {Rimoldini},
  {Sarro}, {Siopis}, {Smith}, {Sozzetti}, {S{\"u}veges}, {Torra}, {van Reeven},
  {Abbas}, {Abreu Aramburu}, {Accart}, {Aerts}, {Altavilla}, {{\'A}lvarez},
  {Alvarez}, {Alves}, {Anderson}, {Andrei}, {Anglada Varela}, {Antiche},
  {Arcay}, {Astraatmadja}, {Bach}, {Baker}, {Balaguer-N{\'u}{\~n}ez}, {Balm},
  {Barache}, {Barata}, {Barbato}, {Barblan}, {Barklem}, {Barrado}, {Barros},
  {Barstow}, {Bartholom{\'e} Mu{\~n}oz}, {Bassilana}, {Becciani}, {Bellazzini},
  {Berihuete}, {Bertone}, {Bianchi}, {Bienaym{\'e}}, {Blanco-Cuaresma}, {Boch},
  {Boeche}, {Bombrun}, {Borrachero}, {Bossini}, {Bouquillon}, {Bourda},
  {Bragaglia}, {Bramante}, {Bressan}, {Brouillet}, {Br{\"u}semeister},
  {Brugaletta}, {Bucciarelli}, {Burlacu}, {Busonero}, {Butkevich}, {Buzzi},
  {Caffau}, {Cancelliere}, {Cannizzaro}, {Cantat-Gaudin}, {Carballo},
  {Carlucci}, {Carrasco}, {Casamiquela}, {Castellani}, {Castro-Ginard},
  {Charlot}, {Chemin}, {Chiavassa}, {Cocozza}, {Costigan}, {Cowell}, {Crifo},
  {Crosta}, {Crowley}, {Cuypers}, {Dafonte}, {Damerdji}, {Dapergolas}, {David},
  {David}, {de Laverny}, {De Luise}, {De March}, {de Martino}, {de Souza}, {de
  Torres}, {Debosscher}, {del Pozo}, {Delbo}, {Delgado}, {Delgado}, {Di
  Matteo}, {Diakite}, {Diener}, {Distefano}, {Dolding}, {Drazinos},
  {Dur{\'a}n}, {Edvardsson}, {Enke}, {Eriksson}, {Esquej}, {Eynard Bontemps},
  {Fabre}, {Fabrizio}, {Faigler}, {Falc{\~a}o}, {Farr{\`a}s Casas}, {Federici},
  {Fedorets}, {Fernique}, {Figueras}, {Filippi}, {Findeisen}, {Fonti},
  {Fraile}, {Fraser}, {Fr{\'e}zouls}, {Gai}, {Galleti}, {Garabato},
  {Garc{\'\i}a-Sedano}, {Garofalo}, {Garralda}, {Gavel}, {Gavras}, {Gerssen},
  {Geyer}, {Giacobbe}, {Gilmore}, {Girona}, {Giuffrida}, {Glass}, {Gomes},
  {Granvik}, {Gueguen}, {Guerrier}, {Guiraud}, {Guti{\'e}rrez-S{\'a}nchez},
  {Hofmann}, {Holland}, {Huckle}, {Hypki}, {Icardi}, {Jan{\ss}en}, {Jevardat de
  Fombelle}, {Jonker}, {Juh{\'a}sz}, {Julbe}, {Karampelas}, {Kewley}, {Klar},
  {Kochoska}, {Kohley}, {Kolenberg}, {Kontizas}, {Kontizas}, {Koposov},
  {Kordopatis}, {Kostrzewa-Rutkowska}, {Koubsky}, {Lambert}, {Lanza}, {Lasne},
  {Lavigne}, {Le Fustec}, {Le Poncin-Lafitte}, {Lebreton}, {Leccia}, {Leclerc},
  {Lecoeur-Taibi}, {Lenhardt}, {Leroux}, {Liao}, {Licata}, {Lindstr{\o}m},
  {Lister}, {Livanou}, {Lobel}, {L{\'o}pez}, {Managau}, {Mann}, {Mantelet},
  {Marchal}, {Marchant}, {Marconi}, {Marinoni}, {Marschalk{\'o}}, {Marshall},
  {Martino}, {Marton}, {Mary}, {Matijevi{\v{c}}}, {Mazeh}, {Messina},
  {Michalik}, {Millar}, {Molina}, {Molinaro}, {Moln{\'a}r}, {Montegriffo},
  {Mor}, {Morbidelli}, {Morel}, {Morris}, {Mulone}, {Muraveva}, {Musella},
  {Nelemans}, {Nicastro}, {Noval}, {O'Mullane}, {Ord{\'e}novic},
  {Ord{\'o}{\~n}ez-Blanco}, {Osborne}, {Pagani}, {Pagano}, {Pailler},
  {Palacin}, {Palaversa}, {Panahi}, {Pawlak}, {Piersimoni}, {Pineau}, {Plachy},
  {Plum}, {Poggio}, {Poujoulet}, {Pr{\v{s}}a}, {Pulone}, {Racero}, {Ragaini},
  {Rambaux}, {Ramos-Lerate}, {Regibo}, {Riclet}, {Ripepi}, {Riva}, {Rivard},
  {Rixon}, {Roegiers}, {Roelens}, {Romero-G{\'o}mez}, {Rowell}, {Royer},
  {Ruiz-Dern}, {Sadowski}, {Sagrist{\`a} Sell{\'e}s}, {Sahlmann}, {Salgado},
  {Salguero}, {Sanna}, {Santana-Ros}, {Sarasso}, {Savietto}, {Schultheis},
  {Sciacca}, {Segol}, {Segovia}, {S{\'e}gransan}, {Shih}, {Siltala}, {Silva},
  {Smart}, {Smith}, {Solano}, {Solitro}, {Sordo}, {Soria Nieto}, {Souchay},
  {Spagna}, {Spoto}, {Stampa}, {Steele}, {Steidelm{\"u}ller}, {Stephenson},
  {Stoev}, {Suess}, {Surdej}, {Szabados}, {Szegedi-Elek}, {Tapiador}, {Taris},
  {Tauran}, {Taylor}, {Teixeira}, {Terrett}, {Teyssandier}, {Thuillot},
  {Titarenko}, {Torra Clotet}, {Turon}, {Ulla}, {Utrilla}, {Uzzi}, {Vaillant},
  {Valentini}, {Valette}, {van Elteren}, {Van Hemelryck}, {van Leeuwen},
  {Vaschetto}, {Vecchiato}, {Viala}, {Vicente}, {Vogt}, {von Essen}, {Voss},
  {Votruba}, {Voutsinas}, {Walmsley}, {Weiler}, {Wertz}, {Wevems},
  {Wyrzykowski}, {Yoldas}, {{\v{Z}}erjal}, {Ziaeepour}, {Zorec}, {Zschocke},
  {Zucker}, {Zurbach}, \& {Zwitter}}]{2018A&A...616A..12G}
{Gaia Collaboration}, {Helmi}, A., {van Leeuwen}, F., {et~al.} 2018, \aap, 616,
  A12

\bibitem[{{Gaia Collaboration} {et~al.}(2016){Gaia Collaboration}, {Prusti},
  {de Bruijne}, {Brown}, {Vallenari}, {Babusiaux}, {Bailer-Jones}, {Bastian},
  {Biermann}, {Evans}, {Eyer}, {Jansen}, {Jordi}, {Klioner}, {Lammers},
  {Lindegren}, {Luri}, {Mignard}, {Milligan}, {Panem}, {Poinsignon},
  {Pourbaix}, {Randich}, {Sarri}, {Sartoretti}, {Siddiqui}, {Soubiran},
  {Valette}, {van Leeuwen}, {Walton}, {Aerts}, {Arenou}, {Cropper}, {Drimmel},
  {H{\o}g}, {Katz}, {Lattanzi}, {O'Mullane}, {Grebel}, {Holland}, {Huc},
  {Passot}, {Bramante}, {Cacciari}, {Casta{\~n}eda}, {Chaoul}, {Cheek}, {De
  Angeli}, {Fabricius}, {Guerra}, {Hern{\'a}ndez}, {Jean-Antoine-Piccolo},
  {Masana}, {Messineo}, {Mowlavi}, {Nienartowicz}, {Ord{\'o}{\~n}ez-Blanco},
  {Panuzzo}, {Portell}, {Richards}, {Riello}, {Seabroke}, {Tanga},
  {Th{\'e}venin}, {Torra}, {Els}, {Gracia-Abril}, {Comoretto},
  {Garcia-Reinaldos}, {Lock}, {Mercier}, {Altmann}, {Andrae}, {Astraatmadja},
  {Bellas-Velidis}, {Benson}, {Berthier}, {Blomme}, {Busso}, {Carry},
  {Cellino}, {Clementini}, {Cowell}, {Creevey}, {Cuypers}, {Davidson}, {De
  Ridder}, {de Torres}, {Delchambre}, {Dell'Oro}, {Ducourant}, {Fr{\'e}mat},
  {Garc{\'\i}a-Torres}, {Gosset}, {Halbwachs}, {Hambly}, {Harrison}, {Hauser},
  {Hestroffer}, {Hodgkin}, {Huckle}, {Hutton}, {Jasniewicz}, {Jordan},
  {Kontizas}, {Korn}, {Lanzafame}, {Manteiga}, {Moitinho}, {Muinonen},
  {Osinde}, {Pancino}, {Pauwels}, {Petit}, {Recio-Blanco}, {Robin}, {Sarro},
  {Siopis}, {Smith}, {Smith}, {Sozzetti}, {Thuillot}, {van Reeven}, {Viala},
  {Abbas}, {Abreu Aramburu}, {Accart}, {Aguado}, {Allan}, {Allasia},
  {Altavilla}, {{\'A}lvarez}, {Alves}, {Anderson}, {Andrei}, {Anglada Varela},
  {Antiche}, {Antoja}, {Ant{\'o}n}, {Arcay}, {Atzei}, {Ayache}, {Bach},
  {Baker}, {Balaguer-N{\'u}{\~n}ez}, {Barache}, {Barata}, {Barbier}, {Barblan},
  {Baroni}, {Barrado y Navascu{\'e}s}, {Barros}, {Barstow}, {Becciani},
  {Bellazzini}, {Bellei}, {Bello Garc{\'\i}a}, {Belokurov}, {Bendjoya},
  {Berihuete}, {Bianchi}, {Bienaym{\'e}}, {Billebaud}, {Blagorodnova},
  {Blanco-Cuaresma}, {Boch}, {Bombrun}, {Borrachero}, {Bouquillon}, {Bourda},
  {Bouy}, {Bragaglia}, {Breddels}, {Brouillet}, {Br{\"u}semeister},
  {Bucciarelli}, {Budnik}, {Burgess}, {Burgon}, {Burlacu}, {Busonero}, {Buzzi},
  {Caffau}, {Cambras}, {Campbell}, {Cancelliere}, {Cantat-Gaudin}, {Carlucci},
  {Carrasco}, {Castellani}, {Charlot}, {Charnas}, {Charvet}, {Chassat},
  {Chiavassa}, {Clotet}, {Cocozza}, {Collins}, {Collins}, {Costigan}, {Crifo},
  {Cross}, {Crosta}, {Crowley}, {Dafonte}, {Damerdji}, {Dapergolas}, {David},
  {David}, {De Cat}, {de Felice}, {de Laverny}, {De Luise}, {De March}, {de
  Martino}, {de Souza}, {Debosscher}, {del Pozo}, {Delbo}, {Delgado},
  {Delgado}, {di Marco}, {Di Matteo}, {Diakite}, {Distefano}, {Dolding}, {Dos
  Anjos}, {Drazinos}, {Dur{\'a}n}, {Dzigan}, {Ecale}, {Edvardsson}, {Enke},
  {Erdmann}, {Escolar}, {Espina}, {Evans}, {Eynard Bontemps}, {Fabre},
  {Fabrizio}, {Faigler}, {Falc{\~a}o}, {Farr{\`a}s Casas}, {Faye}, {Federici},
  {Fedorets}, {Fern{\'a}ndez-Hern{\'a}ndez}, {Fernique}, {Fienga}, {Figueras},
  {Filippi}, {Findeisen}, {Fonti}, {Fouesneau}, {Fraile}, {Fraser}, {Fuchs},
  {Furnell}, {Gai}, {Galleti}, {Galluccio}, {Garabato}, {Garc{\'\i}a-Sedano},
  {Gar{\'e}}, {Garofalo}, {Garralda}, {Gavras}, {Gerssen}, {Geyer}, {Gilmore},
  {Girona}, {Giuffrida}, {Gomes}, {Gonz{\'a}lez-Marcos},
  {Gonz{\'a}lez-N{\'u}{\~n}ez}, {Gonz{\'a}lez-Vidal}, {Granvik}, {Guerrier},
  {Guillout}, {Guiraud}, {G{\'u}rpide}, {Guti{\'e}rrez-S{\'a}nchez}, {Guy},
  {Haigron}, {Hatzidimitriou}, {Haywood}, {Heiter}, {Helmi}, {Hobbs},
  {Hofmann}, {Holl}, {Holland}, {Hunt}, {Hypki}, {Icardi}, {Irwin}, {Jevardat
  de Fombelle}, {Jofr{\'e}}, {Jonker}, {Jorissen}, {Julbe}, {Karampelas},
  {Kochoska}, {Kohley}, {Kolenberg}, {Kontizas}, {Koposov}, {Kordopatis},
  {Koubsky}, {Kowalczyk}, {Krone-Martins}, {Kudryashova}, {Kull}, {Bachchan},
  {Lacoste-Seris}, {Lanza}, {Lavigne}, {Le Poncin-Lafitte}, {Lebreton},
  {Lebzelter}, {Leccia}, {Leclerc}, {Lecoeur-Taibi}, {Lemaitre}, {Lenhardt},
  {Leroux}, {Liao}, {Licata}, {Lindstr{\o}m}, {Lister}, {Livanou}, {Lobel},
  {L{\"o}ffler}, {L{\'o}pez}, {Lopez-Lozano}, {Lorenz}, {Loureiro},
  {MacDonald}, {Magalh{\~a}es Fernandes}, {Managau}, {Mann}, {Mantelet},
  {Marchal}, {Marchant}, {Marconi}, {Marie}, {Marinoni}, {Marrese},
  {Marschalk{\'o}}, {Marshall}, {Mart{\'\i}n-Fleitas}, {Martino}, {Mary},
  {Matijevi{\v{c}}}, {Mazeh}, {McMillan}, {Messina}, {Mestre}, {Michalik},
  {Millar}, {Miranda}, {Molina}, {Molinaro}, {Molinaro}, {Moln{\'a}r},
  {Moniez}, {Montegriffo}, {Monteiro}, {Mor}, {Mora}, {Morbidelli}, {Morel},
  {Morgenthaler}, {Morley}, {Morris}, {Mulone}, {Muraveva}, {Musella},
  {Narbonne}, {Nelemans}, {Nicastro}, {Noval}, {Ord{\'e}novic},
  {Ordieres-Mer{\'e}}, {Osborne}, {Pagani}, {Pagano}, {Pailler}, {Palacin},
  {Palaversa}, {Parsons}, {Paulsen}, {Pecoraro}, {Pedrosa}, {Pentik{\"a}inen},
  {Pereira}, {Pichon}, {Piersimoni}, {Pineau}, {Plachy}, {Plum}, {Poujoulet},
  {Pr{\v{s}}a}, {Pulone}, {Ragaini}, {Rago}, {Rambaux}, {Ramos-Lerate},
  {Ranalli}, {Rauw}, {Read}, {Regibo}, {Renk}, {Reyl{\'e}}, {Ribeiro},
  {Rimoldini}, {Ripepi}, {Riva}, {Rixon}, {Roelens}, {Romero-G{\'o}mez},
  {Rowell}, {Royer}, {Rudolph}, {Ruiz-Dern}, {Sadowski}, {Sagrist{\`a}
  Sell{\'e}s}, {Sahlmann}, {Salgado}, {Salguero}, {Sarasso}, {Savietto},
  {Schnorhk}, {Schultheis}, {Sciacca}, {Segol}, {Segovia}, {Segransan},
  {Serpell}, {Shih}, {Smareglia}, {Smart}, {Smith}, {Solano}, {Solitro},
  {Sordo}, {Soria Nieto}, {Souchay}, {Spagna}, {Spoto}, {Stampa}, {Steele},
  {Steidelm{\"u}ller}, {Stephenson}, {Stoev}, {Suess}, {S{\"u}veges}, {Surdej},
  {Szabados}, {Szegedi-Elek}, {Tapiador}, {Taris}, {Tauran}, {Taylor},
  {Teixeira}, {Terrett}, {Tingley}, {Trager}, {Turon}, {Ulla}, {Utrilla},
  {Valentini}, {van Elteren}, {Van Hemelryck}, {van Leeuwen}, {Varadi},
  {Vecchiato}, {Veljanoski}, {Via}, {Vicente}, {Vogt}, {Voss}, {Votruba},
  {Voutsinas}, {Walmsley}, {Weiler}, {Weingrill}, {Werner}, {Wevers},
  {Whitehead}, {Wyrzykowski}, {Yoldas}, {{\v{Z}}erjal}, {Zucker}, {Zurbach},
  {Zwitter}, {Alecu}, {Allen}, {Allende Prieto}, {Amorim},
  {Anglada-Escud{\'e}}, {Arsenijevic}, {Azaz}, {Balm}, {Beck}, {Bernstein},
  {Bigot}, {Bijaoui}, {Blasco}, {Bonfigli}, {Bono}, {Boudreault}, {Bressan},
  {Brown}, {Brunet}, {Bunclark}, {Buonanno}, {Butkevich}, {Carret}, {Carrion},
  {Chemin}, {Ch{\'e}reau}, {Corcione}, {Darmigny}, {de Boer}, {de Teodoro}, {de
  Zeeuw}, {Delle Luche}, {Domingues}, {Dubath}, {Fodor}, {Fr{\'e}zouls},
  {Fries}, {Fustes}, {Fyfe}, {Gallardo}, {Gallegos}, {Gardiol}, {Gebran},
  {Gomboc}, {G{\'o}mez}, {Grux}, {Gueguen}, {Heyrovsky}, {Hoar}, {Iannicola},
  {Isasi Parache}, {Janotto}, {Joliet}, {Jonckheere}, {Keil}, {Kim},
  {Klagyivik}, {Klar}, {Knude}, {Kochukhov}, {Kolka}, {Kos}, {Kutka}, {Lainey},
  {LeBouquin}, {Liu}, {Loreggia}, {Makarov}, {Marseille}, {Martayan},
  {Martinez-Rubi}, {Massart}, {Meynadier}, {Mignot}, {Munari}, {Nguyen},
  {Nordlander}, {Ocvirk}, {O'Flaherty}, {Olias Sanz}, {Ortiz}, {Osorio},
  {Oszkiewicz}, {Ouzounis}, {Palmer}, {Park}, {Pasquato}, {Peltzer}, {Peralta},
  {P{\'e}turaud}, {Pieniluoma}, {Pigozzi}, {Poels}, {Prat}, {Prod'homme},
  {Raison}, {Rebordao}, {Risquez}, {Rocca-Volmerange}, {Rosen}, {Ruiz-Fuertes},
  {Russo}, {Sembay}, {Serraller Vizcaino}, {Short}, {Siebert}, {Silva},
  {Sinachopoulos}, {Slezak}, {Soffel}, {Sosnowska}, {Strai{\v{z}}ys}, {ter
  Linden}, {Terrell}, {Theil}, {Tiede}, {Troisi}, {Tsalmantza}, {Tur},
  {Vaccari}, {Vachier}, {Valles}, {Van Hamme}, {Veltz}, {Virtanen}, {Wallut},
  {Wichmann}, {Wilkinson}, {Ziaeepour}, \& {Zschocke}}]{2016A&A...595A...1G}
{Gaia Collaboration}, {Prusti}, T., {de Bruijne}, J.~H.~J., {et~al.} 2016,
  \aap, 595, A1

\bibitem[{{Gao}(2018{\natexlab{a}})}]{2018ApJ...869....9G}
{Gao}, X. 2018{\natexlab{a}}, \apj, 869, 9

\bibitem[{{Gao}(2018{\natexlab{b}})}]{2018AJ....156..121G}
{Gao}, X. 2018{\natexlab{b}}, \aj, 156, 121

\bibitem[{{Getman} {et~al.}(2007){Getman}, {Feigelson}, {Garmire}, {Broos}, \&
  {Wang}}]{2007ApJ...654..316G}
{Getman}, K.~V., {Feigelson}, E.~D., {Garmire}, G., {Broos}, P., \& {Wang}, J.
  2007, \apj, 654, 316

\bibitem[{{Getman} {et~al.}(2012){Getman}, {Feigelson}, {Sicilia-Aguilar},
  {Broos}, {Kuhn}, \& {Garmire}}]{2012MNRAS.426.2917G}
{Getman}, K.~V., {Feigelson}, E.~D., {Sicilia-Aguilar}, A., {et~al.} 2012,
  \mnras, 426, 2917

\bibitem[{{Goodwin} \& {Whitworth}(2007)}]{2007A&A...466..943G}
{Goodwin}, S.~P. \& {Whitworth}, A. 2007, \aap, 466, 943

\bibitem[{{Green} {et~al.}(2019){Green}, {Schlafly}, {Zucker}, {Speagle}, \&
  {Finkbeiner}}]{2019ApJ...887...93G}
{Green}, G.~M., {Schlafly}, E., {Zucker}, C., {Speagle}, J.~S., \&
  {Finkbeiner}, D. 2019, \apj, 887, 93

\bibitem[{{Green} {et~al.}(2018){Green}, {Schlafly}, {Finkbeiner}, {Rix},
  {Martin}, {Burgett}, {Draper}, {Flewelling}, {Hodapp}, {Kaiser}, {Kudritzki},
  {Magnier}, {Metcalfe}, {Tonry}, {Wainscoat}, \&
  {Waters}}]{2018MNRAS.478..651G}
{Green}, G.~M., {Schlafly}, E.~F., {Finkbeiner}, D., {et~al.} 2018, \mnras,
  478, 651

\bibitem[{{Green} {et~al.}(2015){Green}, {Schlafly}, {Finkbeiner}, {Rix},
  {Martin}, {Burgett}, {Draper}, {Flewelling}, {Hodapp}, {Kaiser}, {Kudritzki},
  {Magnier}, {Metcalfe}, {Price}, {Tonry}, \&
  {Wainscoat}}]{2015ApJ...810...25G}
{Green}, G.~M., {Schlafly}, E.~F., {Finkbeiner}, D.~P., {et~al.} 2015, \apj,
  810, 25

\bibitem[{{Grudi{\'c}} {et~al.}(2018){Grudi{\'c}}, {Guszejnov}, {Hopkins},
  {Lamberts}, {Boylan-Kolchin}, {Murray}, \& {Schmitz}}]{2018MNRAS.481..688G}
{Grudi{\'c}}, M.~Y., {Guszejnov}, D., {Hopkins}, P.~F., {et~al.} 2018, \mnras,
  481, 688

\bibitem[{{Gupta} {et~al.}(2024){Gupta}, {Jose}, {Das}, {Guo}, {Damian},
  {Prakash}, \& {Samal}}]{2024MNRAS.tmp..456G}
{Gupta}, S., {Jose}, J., {Das}, S.~R., {et~al.} 2024, \mnras

\bibitem[{{Gupta} {et~al.}(2021){Gupta}, {Jose}, {More}, {Das}, {Herczeg},
  {Samal}, {Guo}, {Prakash}, {Damian}, {Takami}, {Takahashi}, {Ogura}, {Terai},
  \& {Pyo}}]{2021MNRAS.508.3388G}
{Gupta}, S., {Jose}, J., {More}, S., {et~al.} 2021, \mnras, 508, 3388

\bibitem[{{Hewett} {et~al.}(2006){Hewett}, {Warren}, {Leggett}, \&
  {Hodgkin}}]{2006MNRAS.367..454H}
{Hewett}, P.~C., {Warren}, S.~J., {Leggett}, S.~K., \& {Hodgkin}, S.~T. 2006,
  \mnras, 367, 454

\bibitem[{{Huang} {et~al.}(2018){Huang}, {Leauthaud}, {Murata}, {Bosch},
  {Price}, {Lupton}, {Mand elbaum}, {Lackner}, {Bickerton}, {Miyazaki},
  {Coupon}, \& {Tanaka}}]{2018PASJ...70S...6H}
{Huang}, S., {Leauthaud}, A., {Murata}, R., {et~al.} 2018, \pasj, 70, S6

\bibitem[{{Ikeda} {et~al.}(2008){Ikeda}, {Sugitani}, {Watanabe}, {Fukuda},
  {Tamura}, {Nakajima}, {Pickles}, {Nagashima}, {Nagayama}, {Nakaya}, {Nakano},
  \& {Nagata}}]{2008AJ....135.2323I}
{Ikeda}, H., {Sugitani}, K., {Watanabe}, M., {et~al.} 2008, \aj, 135, 2323

\bibitem[{{Iye} {et~al.}(2004){Iye}, {Karoji}, {Ando}, {Kaifu}, {Kodaira},
  {Aoki}, {Aoki}, {Chikada}, {Doi}, {Ebizuka}, {Elms}, {Fujihara}, {Furusawa},
  {Fuse}, {Gaessler}, {Harasawa}, {Hayano}, {Hayashi}, {Hayashi}, {Ichikawa},
  {Imanishi}, {Ishida}, {Kamata}, {Kanzawa}, {Kashikawa}, {Kawabata},
  {Kobayashi}, {Komiyama}, {Kosugi}, {Kurakami}, {Letawsky}, {Mikami},
  {Miyashita}, {Miyazaki}, {Mizumoto}, {Morino}, {Motohara}, {Murakawa},
  {Nakagiri}, {Nakamura}, {Nakaya}, {Nariai}, {Nishimura}, {Noguchi},
  {Noguchi}, {Noumaru}, {Ogasawara}, {Ohshima}, {Ohyama}, {Okita}, {Omata},
  {Otsubo}, {Oya}, {Potter}, {Saito}, {Sasaki}, {Sato}, {Scarla}, {Schubert},
  {Sekiguchi}, {Sekiguchi}, {Shelton}, {Simpson}, {Suto}, {Tajitsu}, {Takami},
  {Takata}, {Takato}, {Tamae}, {Tamura}, {Tanaka}, {Terada}, {Torii},
  {Uraguchi}, {Usuda}, {Weber}, {Winegar}, {Yagi}, {Yamada}, {Yamashita},
  {Yamashita}, {Yasuda}, {Yoshida}, \& {Yutani}}]{2004PASJ...56..381I}
{Iye}, M., {Karoji}, H., {Ando}, H., {et~al.} 2004, \pasj, 56, 381

\bibitem[{{Jose} {et~al.}(2020){Jose}, {Biller}, {Albert}, {Dubber}, {Allers},
  {Herczeg}, {Liu}, {Pearson}, {Lalchand}, {Chen}, {Bonnefoy}, {Artigau},
  {Delorme}, {Chiang}, {Zhang}, \& {Oasa}}]{2020ApJ...892..122J}
{Jose}, J., {Biller}, B.~A., {Albert}, L., {et~al.} 2020, \apj, 892, 122

\bibitem[{{Jose} {et~al.}(2017){Jose}, {Herczeg}, {Samal}, {Fang}, \&
  {Panwar}}]{2017ApJ...836...98J}
{Jose}, J., {Herczeg}, G.~J., {Samal}, M.~R., {Fang}, Q., \& {Panwar}, N. 2017,
  \apj, 836, 98

\bibitem[{{Jose} {et~al.}(2016){Jose}, {Kim}, {Herczeg}, {Samal}, {Bieging},
  {Meyer}, \& {Sherry}}]{2016ApJ...822...49J}
{Jose}, J., {Kim}, J.~S., {Herczeg}, G.~J., {et~al.} 2016, \apj, 822, 49

\bibitem[{{Karnath} {et~al.}(2019){Karnath}, {Prchlik}, {Gutermuth}, {Allen},
  {Megeath}, {Pipher}, {Wolk}, \& {Jeffries}}]{2019ApJ...871...46K}
{Karnath}, N., {Prchlik}, J.~J., {Gutermuth}, R.~A., {et~al.} 2019, \apj, 871,
  46

\bibitem[{{Kawanomoto} {et~al.}(2018){Kawanomoto}, {Uraguchi}, {Komiyama},
  {Miyazaki}, {Furusawa}, {Finet}, {Hattori}, {Wang}, {Yasuda}, \&
  {Suzuki}}]{2018PASJ...70...66K}
{Kawanomoto}, S., {Uraguchi}, F., {Komiyama}, Y., {et~al.} 2018, \pasj, 70, 66

\bibitem[{{Khalaj} \& {Baumgardt}(2013)}]{2013MNRAS.434.3236K}
{Khalaj}, P. \& {Baumgardt}, H. 2013, \mnras, 434, 3236

\bibitem[{{Komiyama} {et~al.}(2018){Komiyama}, {Obuchi}, {Nakaya}, {Kamata},
  {Kawanomoto}, {Utsumi}, {Miyazaki}, {Uraguchi}, {Furusawa}, {Morokuma},
  {Uchida}, {Miyatake}, {Mineo}, {Fujimori}, {Aihara}, {Karoji}, {Gunn}, \&
  {Wang}}]{2018PASJ...70S...2K}
{Komiyama}, Y., {Obuchi}, Y., {Nakaya}, H., {et~al.} 2018, \pasj, 70, S2

\bibitem[{{Kroupa}(2008)}]{2008IAUS..246...13K}
{Kroupa}, P. 2008, in Dynamical Evolution of Dense Stellar Systems, ed.
  E.~{Vesperini}, M.~{Giersz}, \& A.~{Sills}, Vol. 246, 13--22

\bibitem[{{Kruijssen}(2012)}]{2012MNRAS.426.3008K}
{Kruijssen}, J.~M.~D. 2012, \mnras, 426, 3008

\bibitem[{{Kuhn} {et~al.}(2019){Kuhn}, {Hillenbrand}, {Sills}, {Feigelson}, \&
  {Getman}}]{2019ApJ...870...32K}
{Kuhn}, M.~A., {Hillenbrand}, L.~A., {Sills}, A., {Feigelson}, E.~D., \&
  {Getman}, K.~V. 2019, \apj, 870, 32

\bibitem[{{Lada} \& {Lada}(2003)}]{2003ARA&A..41...57L}
{Lada}, C.~J. \& {Lada}, E.~A. 2003, \araa, 41, 57

\bibitem[{{Larson}(1981)}]{1981MNRAS.194..809L}
{Larson}, R.~B. 1981, \mnras, 194, 809

\bibitem[{{Lawrence} {et~al.}(2007){Lawrence}, {Warren}, {Almaini}, {Edge},
  {Hambly}, {Jameson}, {Lucas}, {Casali}, {Adamson}, {Dye}, {Emerson},
  {Foucaud}, {Hewett}, {Hirst}, {Hodgkin}, {Irwin}, {Lodieu}, {McMahon},
  {Simpson}, {Smail}, {Mortlock}, \& {Folger}}]{2007MNRAS.379.1599L}
{Lawrence}, A., {Warren}, S.~J., {Almaini}, O., {et~al.} 2007, \mnras, 379,
  1599

\bibitem[{{Li} {et~al.}(2020){Li}, {Shao}, {Li}, {Yu}, {Zhong}, \&
  {Chen}}]{2020ApJ...901...49L}
{Li}, L., {Shao}, Z., {Li}, Z.-Z., {et~al.} 2020, \apj, 901, 49

\bibitem[{{Lin} {et~al.}(2018){Lin}, {Chen}, {Wang}, {Wang}, {Yoshida}, {Ip},
  {Miyazaki}, \& {Terai}}]{2018PASJ...70S..39L}
{Lin}, H.-W., {Chen}, Y.-T., {Wang}, J.-H., {et~al.} 2018, \pasj, 70, S39

\bibitem[{{Liu} {et~al.}(2017){Liu}, {Deng}, {Wang}, \&
  {Wang}}]{2017ApJ...843..104L}
{Liu}, C., {Deng}, N., {Wang}, J. T.~L., \& {Wang}, H. 2017, \apj, 843, 104

\bibitem[{{Lucas} {et~al.}(2008){Lucas}, {Hoare}, {Longmore}, {Schr{\"o}der},
  {Davis}, {Adamson}, {Bandyopadhyay}, {de Grijs}, {Smith}, {Gosling},
  {Mitchison}, {G{\'a}sp{\'a}r}, {Coe}, {Tamura}, {Parker}, {Irwin}, {Hambly},
  {Bryant}, {Collins}, {Cross}, {Evans}, {Gonzalez-Solares}, {Hodgkin},
  {Lewis}, {Read}, {Riello}, {Sutorius}, {Lawrence}, {Drew}, {Dye}, \&
  {Thompson}}]{2008MNRAS.391..136L}
{Lucas}, P.~W., {Hoare}, M.~G., {Longmore}, A., {et~al.} 2008, \mnras, 391, 136

\bibitem[{{Luhman}(2018)}]{2018AJ....156..271L}
{Luhman}, K.~L. 2018, \aj, 156, 271

\bibitem[{{Luhman} {et~al.}(2006){Luhman}, {Whitney}, {Meade}, {Babler},
  {Indebetouw}, {Bracker}, \& {Churchwell}}]{2006ApJ...647.1180L}
{Luhman}, K.~L., {Whitney}, B.~A., {Meade}, M.~R., {et~al.} 2006, \apj, 647,
  1180

\bibitem[{{Mac Low} \& {Klessen}(2004)}]{2004RvMP...76..125M}
{Mac Low}, M.-M. \& {Klessen}, R.~S. 2004, Reviews of Modern Physics, 76, 125

\bibitem[{{Mahmudunnobe} {et~al.}(2021){Mahmudunnobe}, {Hasan}, {Raja}, \&
  {Hasan}}]{2021arXiv210305826M}
{Mahmudunnobe}, M., {Hasan}, P., {Raja}, M., \& {Hasan}, S.~N. 2021, arXiv
  e-prints, arXiv:2103.05826

\bibitem[{{Ma{\'\i}z Apell{\'a}niz} \& {Barb{\'a}}(2020)}]{2020A&A...636A..28M}
{Ma{\'\i}z Apell{\'a}niz}, J. \& {Barb{\'a}}, R.~H. 2020, \aap, 636, A28

\bibitem[{{McKee} \& {Ostriker}(2007)}]{2007ARA&A..45..565M}
{McKee}, C.~F. \& {Ostriker}, E.~C. 2007, \araa, 45, 565

\bibitem[{{Meng} {et~al.}(2019){Meng}, {Rieke}, {Kim}, {Sicilia-Aguilar},
  {Cross}, {Esplin}, {Rebull}, \& {Hodapp}}]{2019ApJ...878....7M}
{Meng}, H. Y.~A., {Rieke}, G.~H., {Kim}, J.~S., {et~al.} 2019, \apj, 878, 7

\bibitem[{{Mercer} {et~al.}(2009){Mercer}, {Miller}, {Calvet}, {Hartmann},
  {Hernandez}, {Sicilia-Aguilar}, \& {Gutermuth}}]{2009AJ....138....7M}
{Mercer}, E.~P., {Miller}, J.~M., {Calvet}, N., {et~al.} 2009, \aj, 138, 7

\bibitem[{{Meylan}(2000)}]{2000ASPC..211..215M}
{Meylan}, G. 2000, in Astronomical Society of the Pacific Conference Series,
  Vol. 211, Massive Stellar Clusters, ed. A.~{Lan{\c{c}}on} \& C.~M. {Boily},
  215

\bibitem[{{Miyazaki} {et~al.}(2018){Miyazaki}, {Komiyama}, {Kawanomoto}, {Doi},
  {Furusawa}, {Hamana}, {Hayashi}, {Ikeda}, {Kamata}, {Karoji}, {Koike},
  {Kurakami}, {Miyama}, {Morokuma}, {Nakata}, {Namikawa}, {Nakaya}, {Nariai},
  {Obuchi}, {Oishi}, {Okada}, {Okura}, {Tait}, {Takata}, {Tanaka}, {Tanaka},
  {Terai}, {Tomono}, {Uraguchi}, {Usuda}, {Utsumi}, {Yamada}, {Yamanoi},
  {Aihara}, {Fujimori}, {Mineo}, {Miyatake}, {Oguri}, {Uchida}, {Tanaka},
  {Yasuda}, {Takada}, {Murayama}, {Nishizawa}, {Sugiyama}, {Chiba}, {Futamase},
  {Wang}, {Chen}, {Ho}, {Liaw}, {Chiu}, {Ho}, {Lai}, {Lee}, {Jeng}, {Iwamura},
  {Armstrong}, {Bickerton}, {Bosch}, {Gunn}, {Lupton}, {Loomis}, {Price},
  {Smith}, {Strauss}, {Turner}, {Suzuki}, {Miyazaki}, {Muramatsu}, {Yamamoto},
  {Endo}, {Ezaki}, {Ito}, {Kawaguchi}, {Sofuku}, {Taniike}, {Akutsu}, {Dojo},
  {Kasumi}, {Matsuda}, {Imoto}, {Miwa}, {Suzuki}, {Takeshi}, \&
  {Yokota}}]{2018PASJ...70S...1M}
{Miyazaki}, S., {Komiyama}, Y., {Kawanomoto}, S., {et~al.} 2018, \pasj, 70, S1

\bibitem[{{Morales-Calder{\'o}n} {et~al.}(2009){Morales-Calder{\'o}n},
  {Stauffer}, {Rebull}, {Whitney}, {Barrado y Navascu{\'e}s}, {Ardila}, {Song},
  {Brooke}, {Hartmann}, \& {Calvet}}]{2009ApJ...702.1507M}
{Morales-Calder{\'o}n}, M., {Stauffer}, J.~R., {Rebull}, L., {et~al.} 2009,
  \apj, 702, 1507

\bibitem[{{Mu{\v{z}}i{\'c}} {et~al.}(2013){Mu{\v{z}}i{\'c}}, {Scholz}, {Geers},
  {Jayawardhana}, {Tamura}, {Dawson}, \& {Ray}}]{2013MmSAI..84..931M}
{Mu{\v{z}}i{\'c}}, K., {Scholz}, A., {Geers}, V.~C., {et~al.} 2013, \memsai,
  84, 931

\bibitem[{{Nakano} {et~al.}(2012){Nakano}, {Sugitani}, {Watanabe}, {Fukuda},
  {Ishihara}, \& {Ueno}}]{2012AJ....143...61N}
{Nakano}, M., {Sugitani}, K., {Watanabe}, M., {et~al.} 2012, \aj, 143, 61

\bibitem[{{Nisini} {et~al.}(2001){Nisini}, {Massi}, {Vitali}, {Giannini},
  {Lorenzetti}, {Di Paola}, {Codella}, {D'Alessio}, \&
  {Speziali}}]{2001A&A...376..553N}
{Nisini}, B., {Massi}, F., {Vitali}, F., {et~al.} 2001, \aap, 376, 553

\bibitem[{{Nony} {et~al.}(2021){Nony}, {Robitaille}, {Motte}, {Gonzalez},
  {Joncour}, {Moraux}, {Men'shchikov}, {Didelon}, {Louvet}, {Buckner},
  {Schneider}, {Lumsden}, {Bontemps}, {Pouteau}, {Cunningham}, {Fiorellino},
  {Oudmaijer}, {Andr{\'e}}, \& {Thomasson}}]{2021A&A...645A..94N}
{Nony}, T., {Robitaille}, J.~F., {Motte}, F., {et~al.} 2021, \aap, 645, A94

\bibitem[{{Olczak} {et~al.}(2011){Olczak}, {Spurzem}, \&
  {Henning}}]{2011A&A...532A.119O}
{Olczak}, C., {Spurzem}, R., \& {Henning}, T. 2011, \aap, 532, A119

\bibitem[{{Pandey} {et~al.}(2022){Pandey}, {Sharma}, {Dewangan}, {Ojha},
  {Panwar}, {Das}, {Bisen}, {Ghosh}, \& {Sinha}}]{2022ApJ...926...25P}
{Pandey}, R., {Sharma}, S., {Dewangan}, L.~K., {et~al.} 2022, \apj, 926, 25

\bibitem[{{Pang} {et~al.}(2020){Pang}, {Li}, {Tang}, {Pasquato}, \&
  {Kouwenhoven}}]{2020ApJ...900L...4P}
{Pang}, X., {Li}, Y., {Tang}, S.-Y., {Pasquato}, M., \& {Kouwenhoven}, M.~B.~N.
  2020, \apjl, 900, L4

\bibitem[{{Panwar} {et~al.}(2014){Panwar}, {Chen}, {Pandey}, {Samal}, {Ogura},
  {Ojha}, {Jose}, \& {Bhatt}}]{2014MNRAS.443.1614P}
{Panwar}, N., {Chen}, W.~P., {Pandey}, A.~K., {et~al.} 2014, \mnras, 443, 1614

\bibitem[{{Panwar} {et~al.}(2017){Panwar}, {Samal}, {Pandey}, {Jose}, {Chen},
  {Ojha}, {Ogura}, {Singh}, \& {Yadav}}]{2017MNRAS.468.2684P}
{Panwar}, N., {Samal}, M.~R., {Pandey}, A.~K., {et~al.} 2017, \mnras, 468, 2684

\bibitem[{{Patel} {et~al.}(1998){Patel}, {Goldsmith}, {Heyer}, {Snell}, \&
  {Pratap}}]{1998ApJ...507..241P}
{Patel}, N.~A., {Goldsmith}, P.~F., {Heyer}, M.~H., {Snell}, R.~L., \&
  {Pratap}, P. 1998, \apj, 507, 241

\bibitem[{{Patel} {et~al.}(1995){Patel}, {Goldsmith}, {Snell}, {Hezel}, \&
  {Xie}}]{1995ApJ...447..721P}
{Patel}, N.~A., {Goldsmith}, P.~F., {Snell}, R.~L., {Hezel}, T., \& {Xie}, T.
  1995, \apj, 447, 721

\bibitem[{{Patra} {et~al.}(2022){Patra}, {Evans}, {Kim}, {Heyer}, {Kauffmann},
  {Jose}, {Samal}, \& {Das}}]{2022AJ....164..129P}
{Patra}, S., {Evans}, Neal~J., I., {Kim}, K.-T., {et~al.} 2022, \aj, 164, 129

\bibitem[{Pedregosa {et~al.}(2012)Pedregosa, Varoquaux, Gramfort, Michel,
  Thirion, Grisel, Blondel, Prettenhofer, Weiss, Dubourg, VanderPlas, Passos,
  Cournapeau, Brucher, Perrot, \& Duchesnay}]{DBLP:journals/corr/abs-1201-0490}
Pedregosa, F., Varoquaux, G., Gramfort, A., {et~al.} 2012, CoRR, abs/1201.0490
  [\eprint[arXiv]{1201.0490}]

\bibitem[{{Pelayo-Bald{\'a}rrago} {et~al.}(2023){Pelayo-Bald{\'a}rrago},
  {Sicilia-Aguilar}, {Fang}, {Roccatagliata}, {Kim}, \&
  {Garc{\'\i}a-{\'A}lvarez}}]{2023AA...669A..22P}
{Pelayo-Bald{\'a}rrago}, M.~E., {Sicilia-Aguilar}, A., {Fang}, M., {et~al.}
  2023, \aap, 669, A22

\bibitem[{{Peter} {et~al.}(2012){Peter}, {Feldt}, {Henning}, \&
  {Hormuth}}]{2012A&A...538A..74P}
{Peter}, D., {Feldt}, M., {Henning}, T., \& {Hormuth}, F. 2012, \aap, 538, A74

\bibitem[{{Piskunov} {et~al.}(2018){Piskunov}, {Just}, {Kharchenko}, {Berczik},
  {Scholz}, {Reffert}, \& {Yen}}]{2018A&A...614A..22P}
{Piskunov}, A.~E., {Just}, A., {Kharchenko}, N.~V., {et~al.} 2018, \aap, 614,
  A22

\bibitem[{{Platais} {et~al.}(1998){Platais}, {Kozhurina-Platais}, \& {van
  Leeuwen}}]{1998AJ....116.2423P}
{Platais}, I., {Kozhurina-Platais}, V., \& {van Leeuwen}, F. 1998, \aj, 116,
  2423

\bibitem[{{Plewa}(2018)}]{2018MNRAS.476.3974P}
{Plewa}, P.~M. 2018, \mnras, 476, 3974

\bibitem[{{Pottasch}(1956)}]{1956BAN....13...77P}
{Pottasch}, S.~R. 1956, \bain, 13, 77

\bibitem[{{Ramachandran} {et~al.}(2017){Ramachandran}, {Das}, {Tej}, {Vig},
  {Ghosh}, \& {Ojha}}]{2017MNRAS.465.4753R}
{Ramachandran}, V., {Das}, S.~R., {Tej}, A., {et~al.} 2017, \mnras, 465, 4753

\bibitem[{{Reach} {et~al.}(2004){Reach}, {Rho}, {Young}, {Muzerolle},
  {Fajardo-Acosta}, {Hartmann}, {Sicilia-Aguilar}, {Allen}, {Carey},
  {Cuillandre}, {Jarrett}, {Lowrance}, {Marston}, {Noriega-Crespo}, \&
  {Hurt}}]{2004ApJS..154..385R}
{Reach}, W.~T., {Rho}, J., {Young}, E., {et~al.} 2004, \apjs, 154, 385

\bibitem[{{Rebull} {et~al.}(2023){Rebull}, {Anderson}, {Hall}, {Kirkpatrick},
  {Koenig}, {Odden}, {Rodriguez}, {Sanchez}, {Senson}, {Urbanowski}, {Austin},
  {Blood}, {Kerman}, {Long}, \& {Roosa}}]{2023AJ....166...87R}
{Rebull}, L.~M., {Anderson}, R.~L., {Hall}, G., {et~al.} 2023, \aj, 166, 87

\bibitem[{{Rebull} {et~al.}(2013){Rebull}, {Johnson}, {Gibbs}, {Linahan},
  {Sartore}, {Laher}, {Legassie}, {Armstrong}, {Allen}, {McGehee}, {Padgett},
  {Aryal}, {Badura}, {Canakapalli}, {Carlson}, {Clark}, {Ezyk}, {Fagan},
  {Killingstad}, {Koop}, {McCanna}, {Nishida}, {Nuthmann}, {O'Bryan},
  {Pullinger}, {Rameswaram}, {Ravelomanantsoa}, {Sprow}, \&
  {Tilley}}]{2013AJ....145...15R}
{Rebull}, L.~M., {Johnson}, C.~H., {Gibbs}, J.~C., {et~al.} 2013, \aj, 145, 15

\bibitem[{{Reiter} \& {Parker}(2022)}]{2022EPJP..137.1071R}
{Reiter}, M. \& {Parker}, R.~J. 2022, European Physical Journal Plus, 137, 1071

\bibitem[{{Ribas} {et~al.}(2017){Ribas}, {Espaillat}, {Mac{\'\i}as}, {Bouy},
  {Andrews}, {Calvet}, {Naylor}, {Riviere-Marichalar}, {van der Wiel}, \&
  {Wilner}}]{2017ApJ...849...63R}
{Ribas}, {\'A}., {Espaillat}, C.~C., {Mac{\'\i}as}, E., {et~al.} 2017, \apj,
  849, 63

\bibitem[{{Robitaille} {et~al.}(2019){Robitaille}, {Rice}, {Beaumont},
  {Ginsburg}, {MacDonald}, \& {Rosolowsky}}]{2019ascl.soft07016R}
{Robitaille}, T., {Rice}, T., {Beaumont}, C., {et~al.} 2019, {astrodendro:
  Astronomical data dendrogram creator}, Astrophysics Source Code Library,
  record ascl:1907.016

\bibitem[{{Samal} {et~al.}(2012){Samal}, {Pandey}, {Ojha}, {Chauhan}, {Jose},
  \& {Pandey}}]{2012ApJ...755...20S}
{Samal}, M.~R., {Pandey}, A.~K., {Ojha}, D.~K., {et~al.} 2012, \apj, 755, 20

\bibitem[{{Samal} {et~al.}(2014){Samal}, {Zavagno}, {Deharveng}, {Molinari},
  {Ojha}, {Paradis}, {Tig{\'e}}, {Pandey}, \& {Russeil}}]{2014A&A...566A.122S}
{Samal}, M.~R., {Zavagno}, A., {Deharveng}, L., {et~al.} 2014, \aap, 566, A122

\bibitem[{{Saurin} {et~al.}(2012){Saurin}, {Bica}, \&
  {Bonatto}}]{2012MNRAS.421.3206S}
{Saurin}, T.~A., {Bica}, E., \& {Bonatto}, C. 2012, \mnras, 421, 3206

\bibitem[{{Schmeja}(2011)}]{2011AN....332..172S}
{Schmeja}, S. 2011, Astronomische Nachrichten, 332, 172

\bibitem[{{Schmeja} {et~al.}(2008){Schmeja}, {Kumar}, \&
  {Ferreira}}]{2008MNRAS.389.1209S}
{Schmeja}, S., {Kumar}, M.~S.~N., \& {Ferreira}, B. 2008, \mnras, 389, 1209

\bibitem[{{Schwartz} {et~al.}(1991){Schwartz}, {Wilking}, \&
  {Giulbudagian}}]{1991ApJ...370..263S}
{Schwartz}, R.~D., {Wilking}, B.~A., \& {Giulbudagian}, A.~L. 1991, \apj, 370,
  263

\bibitem[{{Sharpless}(1959)}]{1959ApJS....4..257S}
{Sharpless}, S. 1959, \apjs, 4, 257

\bibitem[{{Sicilia-Aguilar} {et~al.}(2006{\natexlab{a}}){Sicilia-Aguilar},
  {Hartmann}, {Calvet}, {Megeath}, {Muzerolle}, {Allen}, {D'Alessio},
  {Mer{\'\i}n}, {Stauffer}, {Young}, \& {Lada}}]{2006ApJ...638..897S}
{Sicilia-Aguilar}, A., {Hartmann}, L., {Calvet}, N., {et~al.}
  2006{\natexlab{a}}, \apj, 638, 897

\bibitem[{{Sicilia-Aguilar} {et~al.}(2004){Sicilia-Aguilar}, {Hartmann},
  {Brice{\~n}o}, {Muzerolle}, \& {Calvet}}]{2004AJ....128..805S}
{Sicilia-Aguilar}, A., {Hartmann}, L.~W., {Brice{\~n}o}, C., {Muzerolle}, J.,
  \& {Calvet}, N. 2004, \aj, 128, 805

\bibitem[{{Sicilia-Aguilar} {et~al.}(2006{\natexlab{b}}){Sicilia-Aguilar},
  {Hartmann}, {F{\"u}r{\'e}sz}, {Henning}, {Dullemond}, \&
  {Brandner}}]{2006AJ....132.2135S}
{Sicilia-Aguilar}, A., {Hartmann}, L.~W., {F{\"u}r{\'e}sz}, G., {et~al.}
  2006{\natexlab{b}}, \aj, 132, 2135

\bibitem[{{Sicilia-Aguilar} {et~al.}(2005){Sicilia-Aguilar}, {Hartmann},
  {Hern{\'a}ndez}, {Brice{\~n}o}, \& {Calvet}}]{2005AJ....130..188S}
{Sicilia-Aguilar}, A., {Hartmann}, L.~W., {Hern{\'a}ndez}, J., {Brice{\~n}o},
  C., \& {Calvet}, N. 2005, \aj, 130, 188

\bibitem[{{Sicilia-Aguilar} {et~al.}(2010){Sicilia-Aguilar}, {Henning}, \&
  {Hartmann}}]{2010ApJ...710..597S}
{Sicilia-Aguilar}, A., {Henning}, T., \& {Hartmann}, L.~W. 2010, \apj, 710, 597

\bibitem[{{Sicilia-Aguilar} {et~al.}(2013{\natexlab{a}}){Sicilia-Aguilar},
  {Kim}, {Sobolev}, {Getman}, {Henning}, \& {Fang}}]{2013A&A...559A...3S}
{Sicilia-Aguilar}, A., {Kim}, J.~S., {Sobolev}, A., {et~al.}
  2013{\natexlab{a}}, \aap, 559, A3

\bibitem[{{Sicilia-Aguilar} {et~al.}(2013{\natexlab{b}}){Sicilia-Aguilar},
  {Kim}, {Sobolev}, {Getman}, {Henning}, \& {Fang}}]{2013AA...559A...3S}
{Sicilia-Aguilar}, A., {Kim}, J.~S., {Sobolev}, A., {et~al.}
  2013{\natexlab{b}}, \aap, 559, A3

\bibitem[{{Sicilia-Aguilar} {et~al.}(2019){Sicilia-Aguilar}, {Patel}, {Fang},
  {Roccatagliata}, {Getman}, \& {Goldsmith}}]{2019A&A...622A.118S}
{Sicilia-Aguilar}, A., {Patel}, N., {Fang}, M., {et~al.} 2019, \aap, 622, A118

\bibitem[{{Sicilia-Aguilar} {et~al.}(2014){Sicilia-Aguilar}, {Roccatagliata},
  {Getman}, {Henning}, {Mer{\'\i}n}, {Eiroa}, {Rivi{\`e}re-Marichalar}, \&
  {Currie}}]{2014A&A...562A.131S}
{Sicilia-Aguilar}, A., {Roccatagliata}, V., {Getman}, K., {et~al.} 2014, \aap,
  562, A131

\bibitem[{{Silverberg} {et~al.}(2021){Silverberg}, {G{\"u}nther}, {Kim},
  {Principe}, \& {Wolk}}]{2021AJ....162..279S}
{Silverberg}, S.~M., {G{\"u}nther}, H.~M., {Kim}, J.~S., {Principe}, D.~A., \&
  {Wolk}, S.~J. 2021, \aj, 162, 279

\bibitem[{{Skrutskie} {et~al.}(2006){Skrutskie}, {Cutri}, {Stiening},
  {Weinberg}, {Schneider}, {Carpenter}, {Beichman}, {Capps}, {Chester},
  {Elias}, {Huchra}, {Liebert}, {Lonsdale}, {Monet}, {Price}, {Seitzer},
  {Jarrett}, {Kirkpatrick}, {Gizis}, {Howard}, {Evans}, {Fowler}, {Fullmer},
  {Hurt}, {Light}, {Kopan}, {Marsh}, {McCallon}, {Tam}, {Van Dyk}, \&
  {Wheelock}}]{2006AJ....131.1163S}
{Skrutskie}, M.~F., {Cutri}, R.~M., {Stiening}, R., {et~al.} 2006, \aj, 131,
  1163

\bibitem[{{Sterzik} \& {Durisen}(1995)}]{1995A&A...304L...9S}
{Sterzik}, M.~F. \& {Durisen}, R.~H. 1995, \aap, 304, L9

\bibitem[{{Stickland}(1995)}]{1995Obs...115..180S}
{Stickland}, D.~J. 1995, The Observatory, 115, 180

\bibitem[{{Sugitani} {et~al.}(1991){Sugitani}, {Fukui}, \&
  {Ogura}}]{1991ApJS...77...59S}
{Sugitani}, K., {Fukui}, Y., \& {Ogura}, K. 1991, \apjs, 77, 59

\bibitem[{{Tonry} {et~al.}(2012){Tonry}, {Stubbs}, {Lykke}, {Doherty},
  {Shivvers}, {Burgett}, {Chambers}, {Hodapp}, {Kaiser}, {Kudritzki},
  {Magnier}, {Morgan}, {Price}, \& {Wainscoat}}]{2012ApJ...750...99T}
{Tonry}, J.~L., {Stubbs}, C.~W., {Lykke}, K.~R., {et~al.} 2012, \apj, 750, 99

\bibitem[{{Torniamenti} {et~al.}(2022){Torniamenti}, {Pasquato}, {Di Cintio},
  {Ballone}, {Iorio}, {Artale}, \& {Mapelli}}]{2022MNRAS.510.2097T}
{Torniamenti}, S., {Pasquato}, M., {Di Cintio}, P., {et~al.} 2022, \mnras, 510,
  2097

\bibitem[{{Trumpler}(1930)}]{1930LicOB..14..154T}
{Trumpler}, R.~J. 1930, Lick Observatory Bulletin, 420, 154

\bibitem[{{Walch} {et~al.}(2015){Walch}, {Girichidis}, {Naab}, {Gatto},
  {Glover}, {W{\"u}nsch}, {Klessen}, {Clark}, {Peters}, {Derigs}, \&
  {Baczynski}}]{2015MNRAS.454..238W}
{Walch}, S., {Girichidis}, P., {Naab}, T., {et~al.} 2015, \mnras, 454, 238

\bibitem[{{Wang} \& {Chen}(2019)}]{2019ApJ...877..116W}
{Wang}, S. \& {Chen}, X. 2019, \apj, 877, 116

\bibitem[{{Wright} {et~al.}(2019){Wright}, {Jeffries}, {Jackson}, {Bayo},
  {Bonito}, {Damiani}, {Kalari}, {Lanzafame}, {Pancino}, {Parker},
  {Prisinzano}, {Randich}, {Vink}, {Alfaro}, {Bergemann}, {Franciosini},
  {Gilmore}, {Gonneau}, {Hourihane}, {Jofr{\'e}}, {Koposov}, {Lewis},
  {Magrini}, {Micela}, {Morbidelli}, {Sacco}, {Worley}, \&
  {Zaggia}}]{2019MNRAS.486.2477W}
{Wright}, N.~J., {Jeffries}, R.~D., {Jackson}, R.~J., {et~al.} 2019, \mnras,
  486, 2477

\bibitem[{{Zhang} {et~al.}(2018){Zhang}, {Liu}, {Best}, {Magnier}, {Aller},
  {Chambers}, {Draper}, {Flewelling}, {Hodapp}, {Kaiser}, {Kudritzki},
  {Metcalfe}, {Wainscoat}, \& {Waters}}]{2018ApJ...858...41Z}
{Zhang}, Z., {Liu}, M.~C., {Best}, W. M.~J., {et~al.} 2018, \apj, 858, 41

\bibitem[{{Zucker} {et~al.}(2023){Zucker}, {Alves}, {Goodman}, {Meingast}, \&
  {Galli}}]{2023ASPC..534...43Z}
{Zucker}, C., {Alves}, J., {Goodman}, A., {Meingast}, S., \& {Galli}, P. 2023,
  in Astronomical Society of the Pacific Conference Series, Vol. 534,
  Protostars and Planets VII, ed. S.~{Inutsuka}, Y.~{Aikawa}, T.~{Muto},
  K.~{Tomida}, \& M.~{Tamura}, 43

\end{thebibliography}
%

\begin{appendix} 

\section{HSC images of IC 1396} \label{hsc_img_full}

\begin{figure*}
\centering
\includegraphics[scale=0.38]{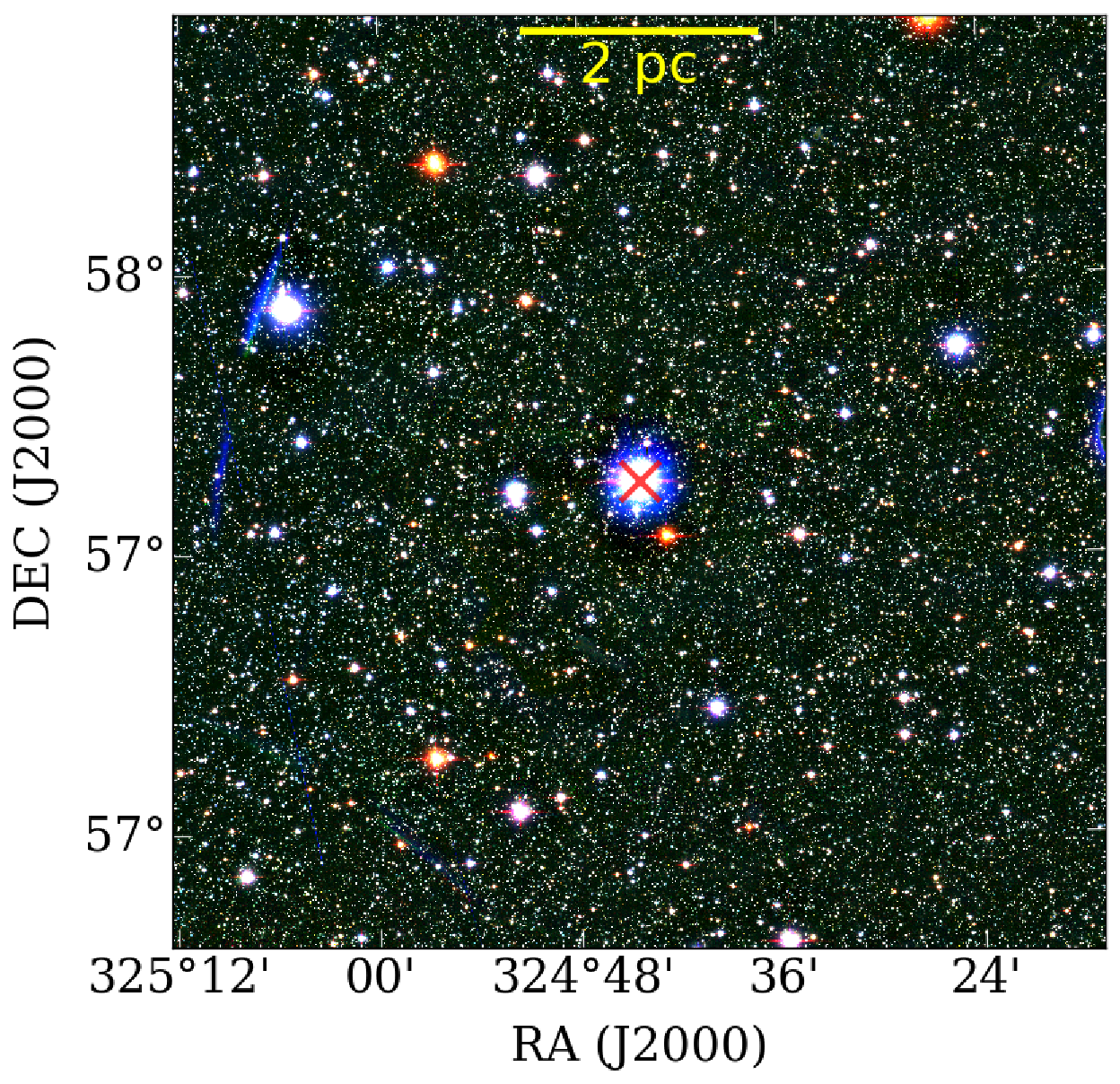}
\includegraphics[scale=0.38]{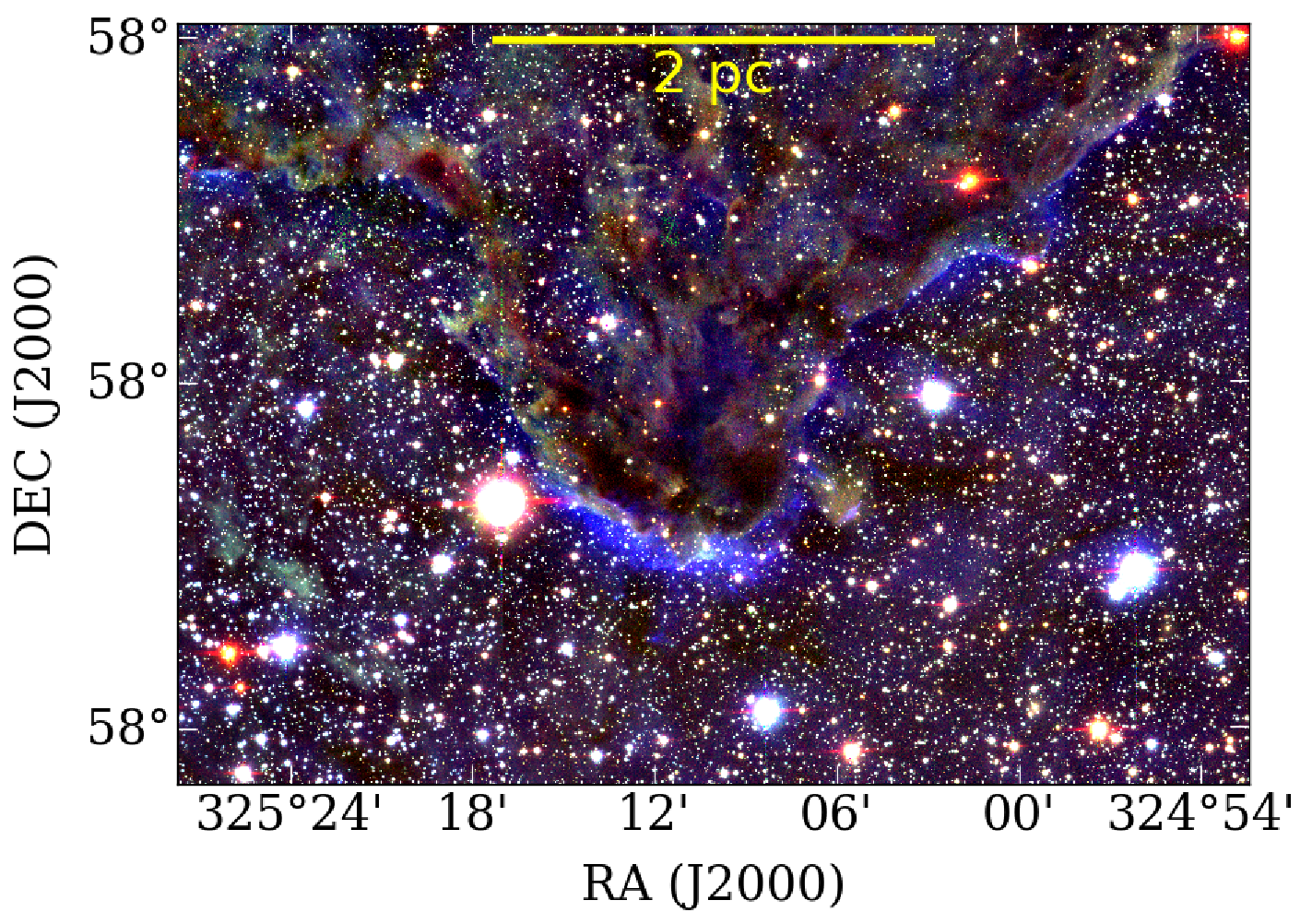}
\includegraphics[scale=0.4]{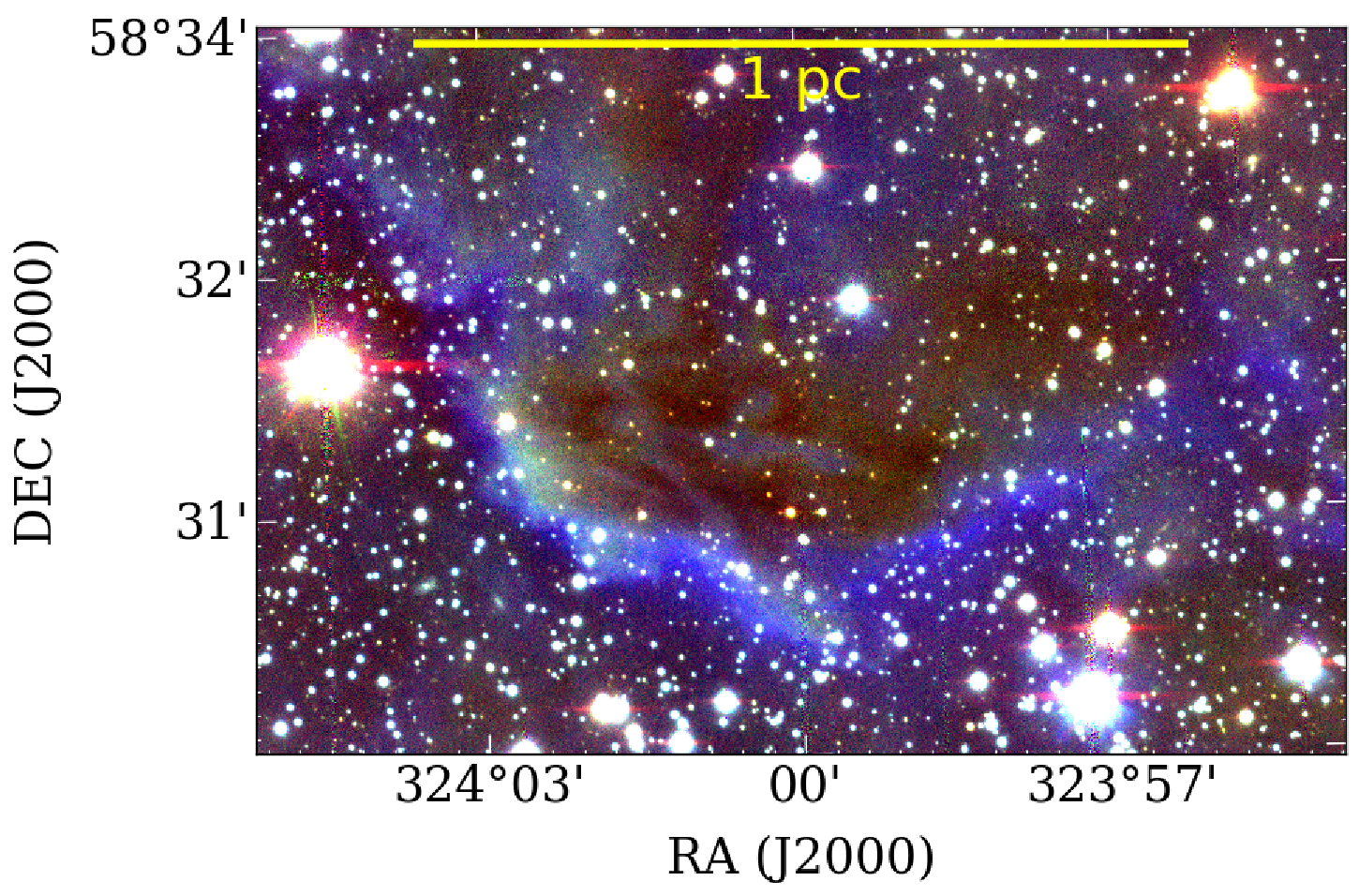}
\includegraphics[scale=0.4]{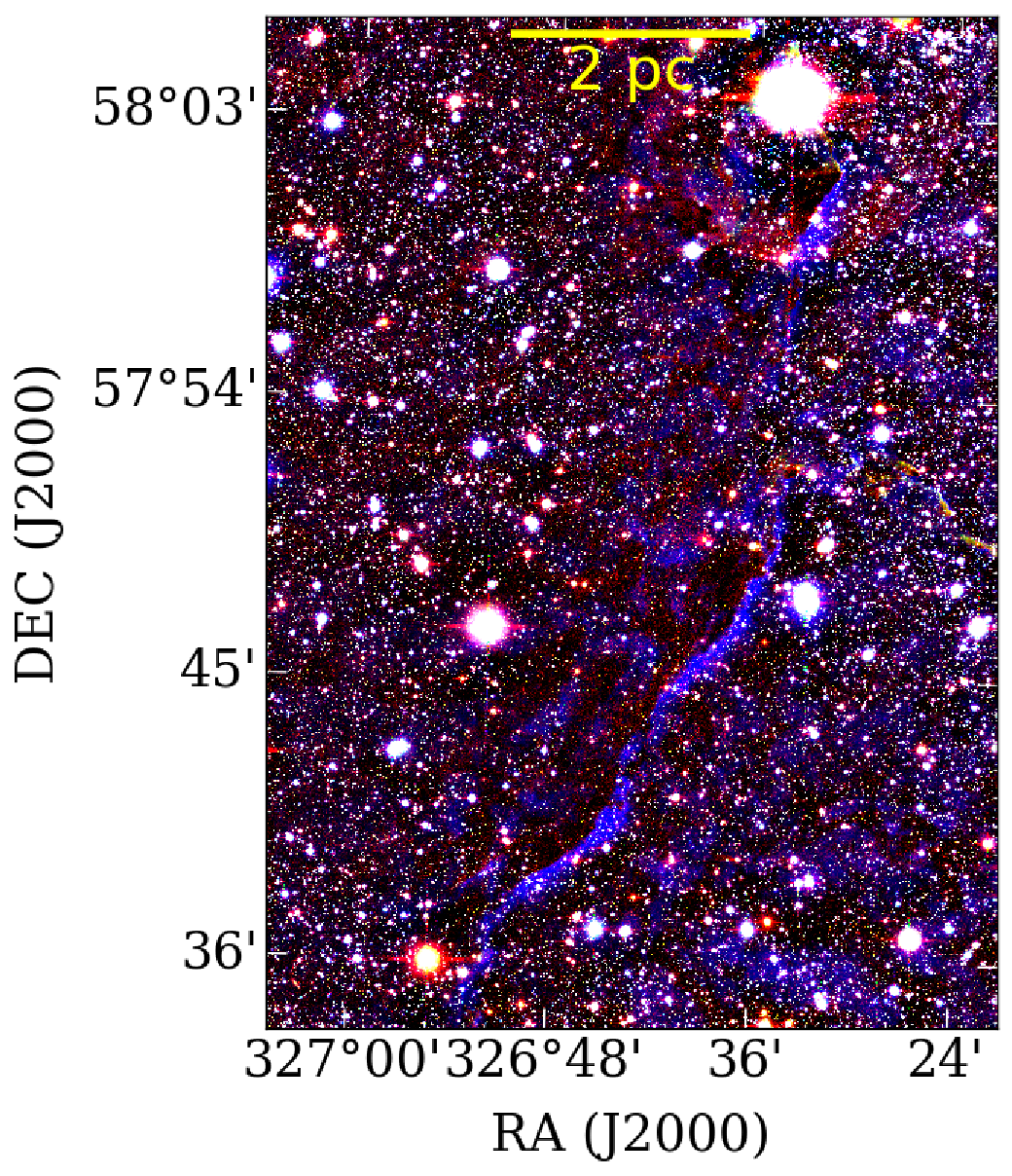}
\caption{Same as Fig. \ref{rgb_ic1396a}, but for the regions of the central part of the complex ({\it top left}), region towards BRCs IC 1396 N ({\it top right}), SFO 35 ({\it bottom left}), and IC 1396 G ({\it bottom right}), respectively. Scale bars of 2~pc and 1~pc are shown on the images. }
\label{rgb_imags}
\end{figure*}

\section{Efficiency test of RF classifier} \label{rf_eff}

\begin{figure}
\centering
\includegraphics[scale=0.4]{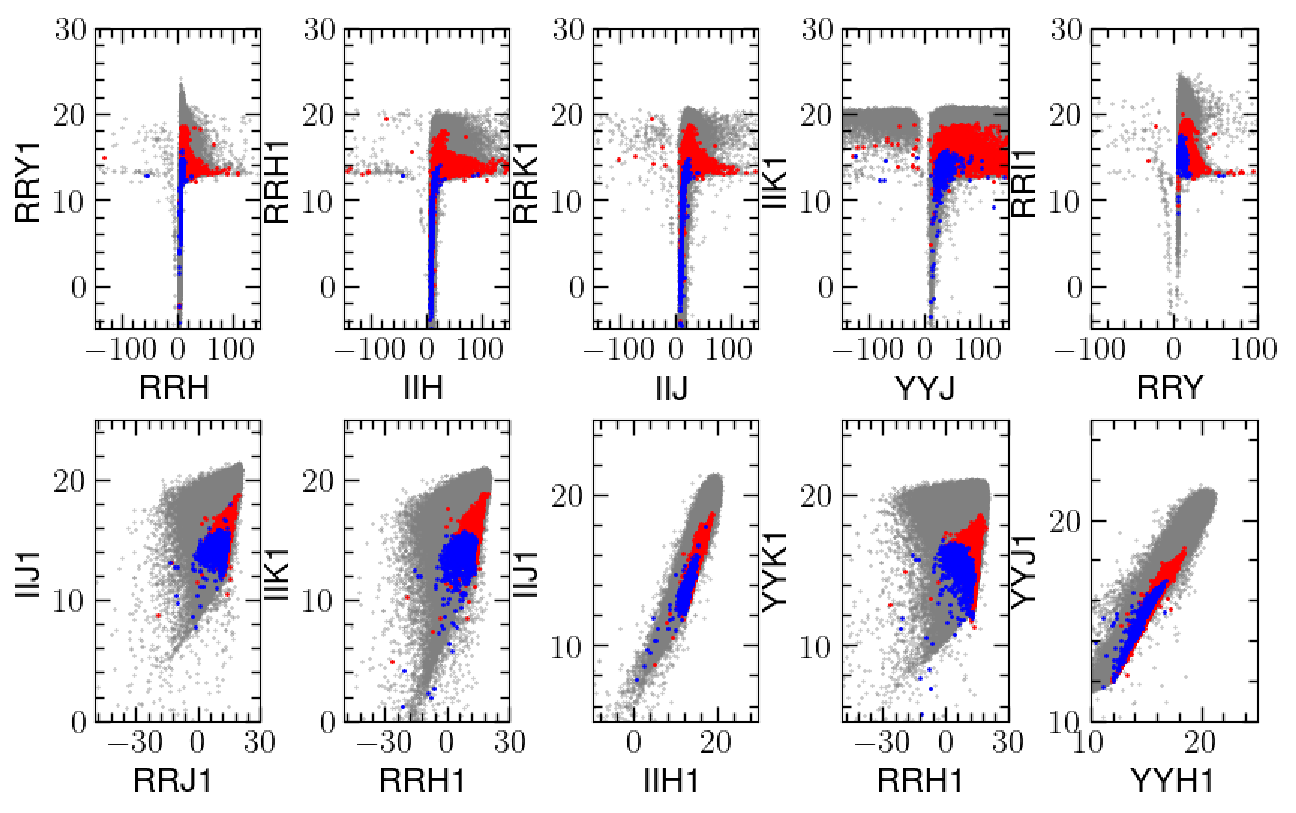}
\caption{Same as Fig. \ref{pair_const}, but with a few more combinations of the additional new parameters derived in this work.}
\label{pair_const_ten}
\end{figure}

\subsection{For $30\arcmin$ region} \label{eff_small}
\begin{figure}
\centering
\includegraphics[scale=0.3]{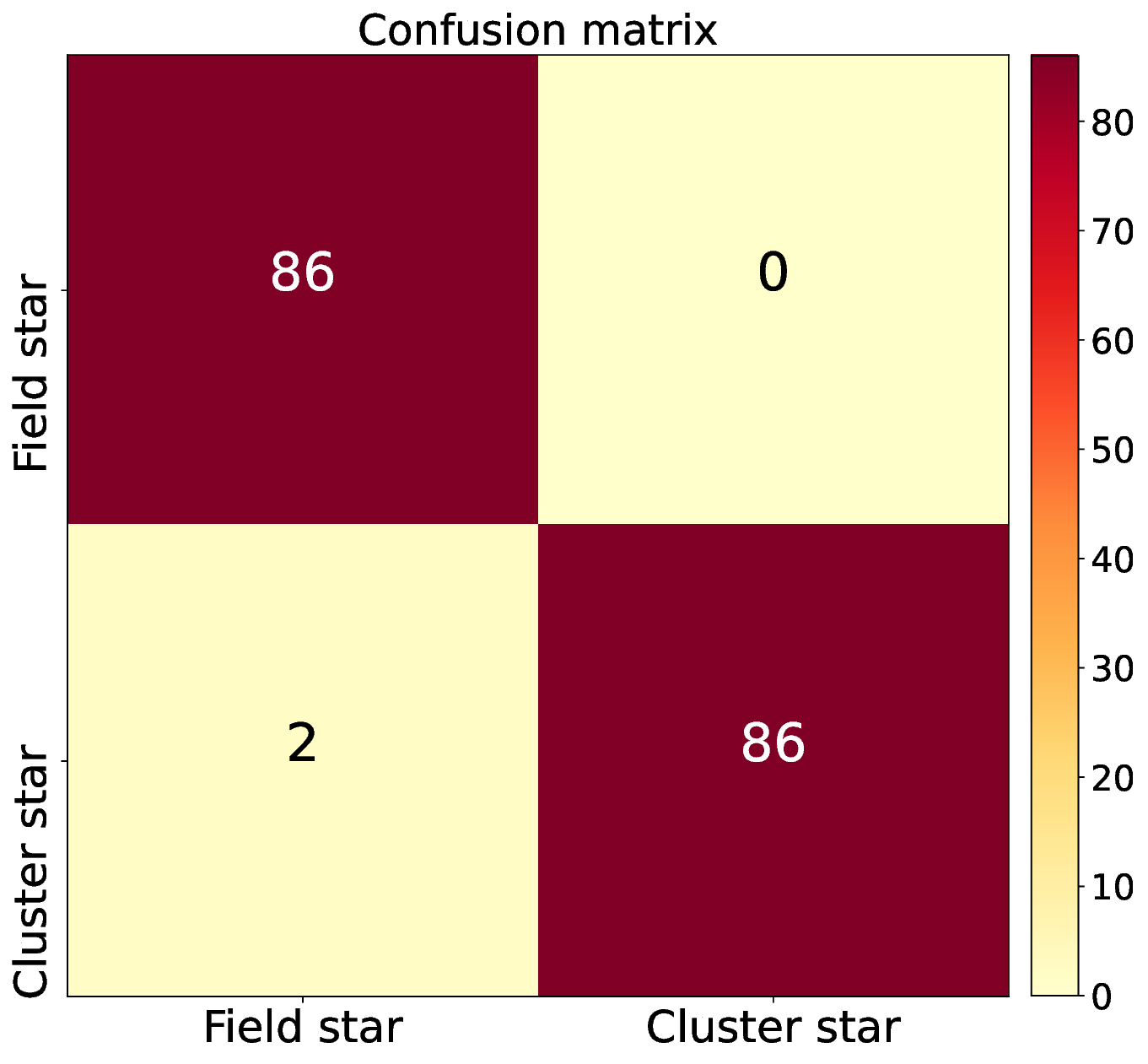}
\caption{The figure shows the confusion matrix generated by the RF method for the $30\arcmin$ circular region. The cluster stars and field stars are displayed in the plot. The colour bar represents the variation in the number of objects within each cell of the confusion matrix. }
\label{rf_confusion_30}
\end{figure}

For the $30\arcmin$ region, the training set consists of 435 stars, of which 236 are member stars, and 199 are non-member stars. These are the HSC+UKIDSS counterparts of the Gaia-DR3-based sources obtained by \citet{2023ApJ...948....7D}. We use only the twenty-four newly derived parameters to train the machine in this step. The reason for using only these new parameters is discussed in Section \ref{memb_analy}. Of the total 435 stars, 60\% are used for training, and 40\% are used to test its accuracy. With this setup, we train the machine and see a high accuracy of 0.99. In Fig. \ref{rf_confusion_30}, we show the confusion matrix, which demonstrates the high accuracy level of the machine during training. Out of the 174 stars used for testing, the machine successfully identified most of the member and non-member stars. The machine is confused to identify the true nature of only two sources. This indicates the higher accuracy of the machine in training. In Table \ref{rf_impt_30}, we provide the relative importance of the parameters used in this analysis. Here, we only provide the top ten parameters. As discussed in Section \ref{memb_analy}, we can see the effectiveness of these new parameters in segregating the member and non-member sources.

\begin{table}
\centering
\caption{The relative importance of input parameters used in the RF classifier during the training process for $30\arcmin$ circle. The top ten parameters are listed here.}
\label{rf_impt_30}
\begin{tabular}{cc}
\\ \hline \hline
Parameter & Releative importance \\
\hline
RRK1  &  0.220289 \\
RRH1  &  0.123324 \\
RRJ1  &  0.098757 \\
IIH   & 0.089294 \\
IIY   & 0.071514 \\
RRH   & 0.059053 \\
IIJ   & 0.050878 \\
RRY   & 0.038637 \\
IIK1  &  0.038527 \\
RRY1  &  0.035775 \\
\hline
\end{tabular}
\end{table}

\subsection{For the whole complex}

\begin{figure}
\centering
\includegraphics[scale=0.3]{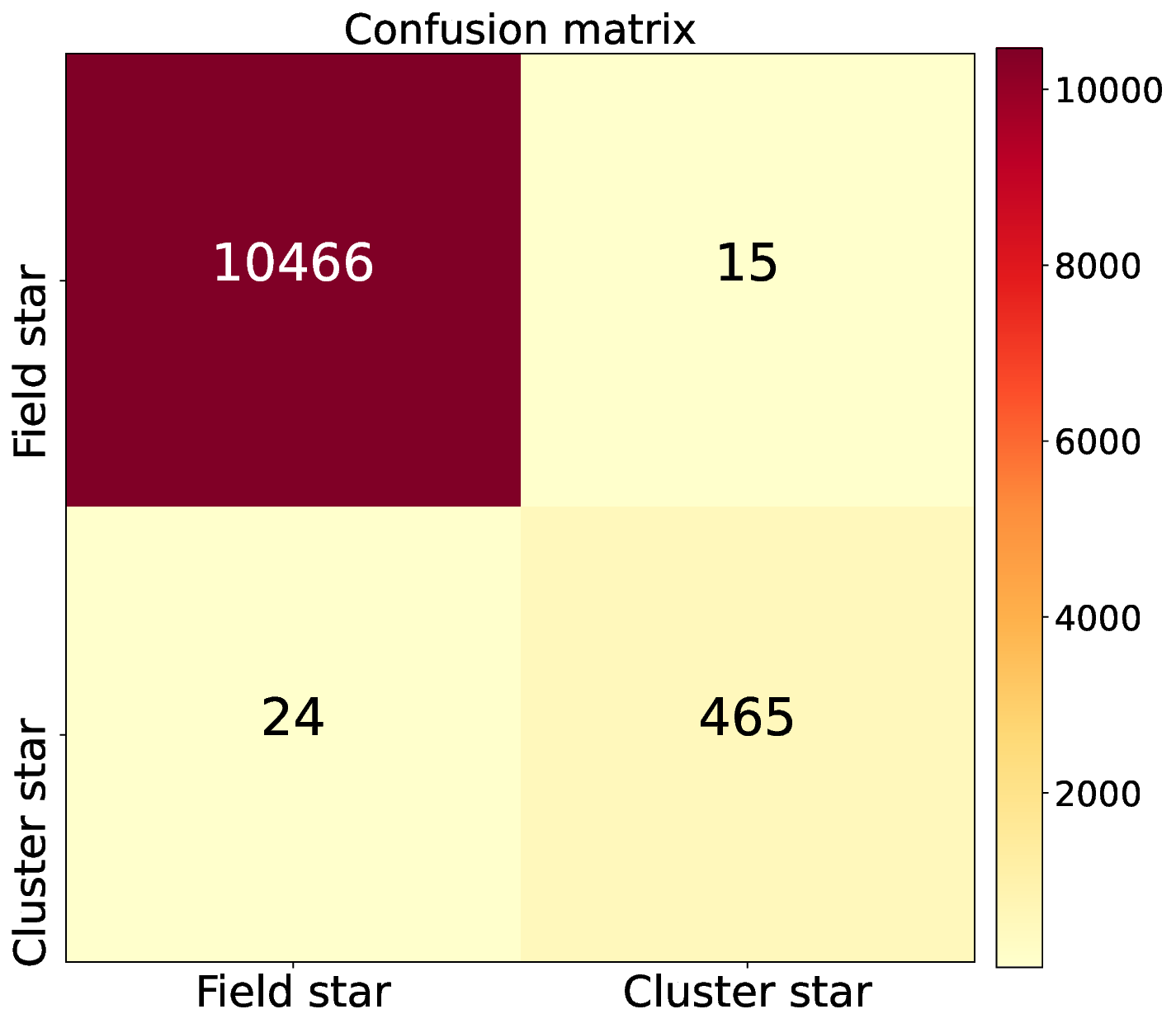}
\caption{The figure shows the confusion matrix generated by the RF method for the entire star-forming complex. The cluster stars and field stars are displayed in the plot. The colour bar represents the variation in the number of objects within each cell of the confusion matrix. }
\label{rf_confusion}
\end{figure}

As explained in Section \ref{memb_analy}, for the entire complex, our training set comprises 27423 stars, with 1250 being member stars and 26173 being non-member or field stars. The RF classifier algorithm efficiently handles large dimensions; therefore, in this analysis, we utilize the complete set of forty-two parameters, as detailed in Section \ref{memb_analy}. Consequently, we construct the RF classifier using the forty-two-dimensional reliable training set and assess its accuracy. For this purpose, out of the 27423 stars in the input training set, 60\% are utilized to train the RF classifier, with the remaining 40\% used for testing accuracy. Thus, in this analysis, out of the 27423 stars, 16453 stars are employed for training, and the remaining 10970 stars are utilized to test the  efficiency of the machine in distinguishing member stars from field stars. The machine autonomously selects the training and test sets randomly. Our training procedure yields a high accuracy of 0.996 with the RF method. We present the confusion matrix in Fig. \ref{rf_confusion}, demonstrating the high accuracy of the RF method. This confusion matrix illustrates how the machine identifies sources based on training. From the confusion matrix, we observe that out of the 10970 stars used for testing the accuracy of the machine, and it successfully identifies 10466 field stars and 465 cluster member stars. During classification, the machine encounters confusion with only a few field and cluster member stars. This exercise highlights the effectiveness of the RF method. Table \ref{rf_impt} provides the relative importance of the input parameters as determined by RF while providing the membership probability. Here, we present the top ten parameters. Similarly, the new parameters exhibit high relative importance, indicating their effectiveness in significant segregation between member stars and non-member stars.

\begin{table}
\centering
\caption{The relative importance of input parameters used in the RF classifier during the training process for the entire star-forming complex IC 1396. The top ten parameters are listed here.}
\label{rf_impt}
\begin{tabular}{cc}
\\ \hline \hline
Parameter & Releative importance \\
\hline
IIH1  &  0.087797  \\
RRY1  &  0.083987  \\
IIK1  &  0.064884  \\
IIY   & 0.049824  \\
RRH   & 0.046862  \\
IIH   & 0.039322  \\
RRK1  &  0.037958  \\
RRH1  &  0.036737  \\
IIJ1  &  0.036561  \\
IIY1  &  0.036218  \\
\hline
\end{tabular}
\end{table}

\section{CMD plots} \label{cmd_ext_307}
\begin{figure}
\includegraphics[scale=0.45]{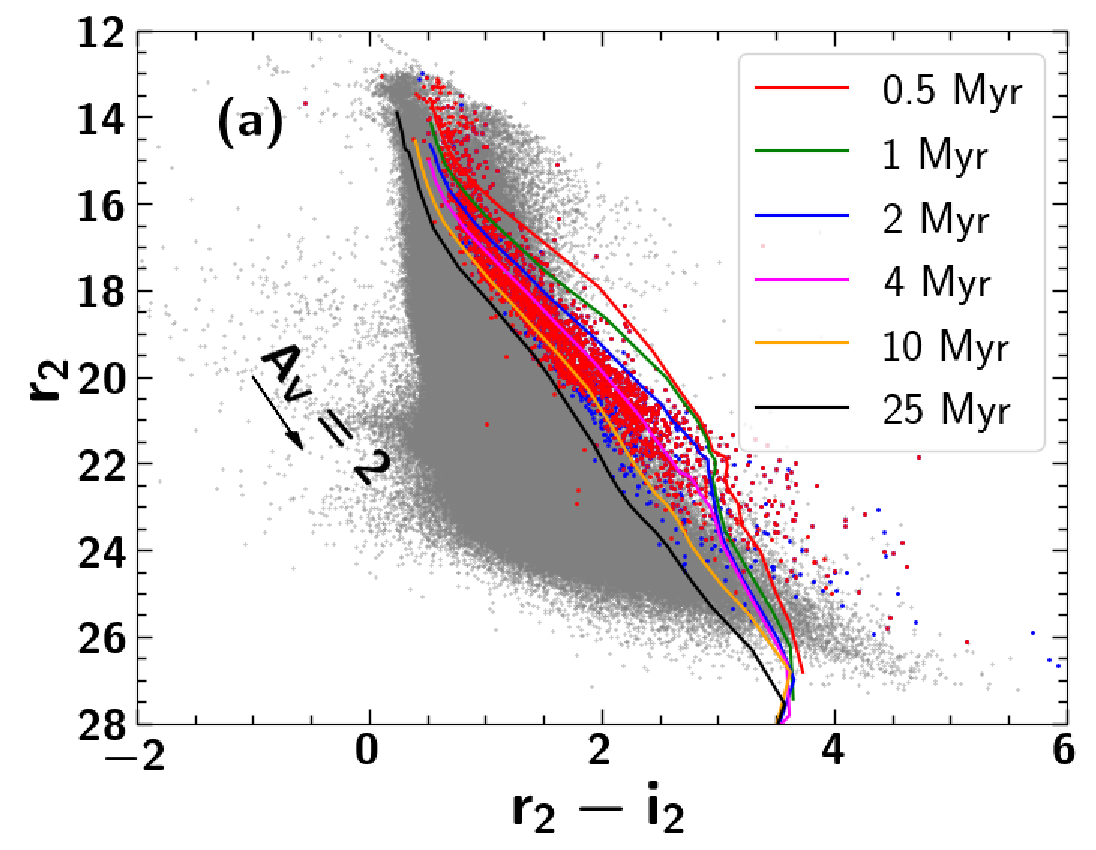}
\includegraphics[scale=0.45]{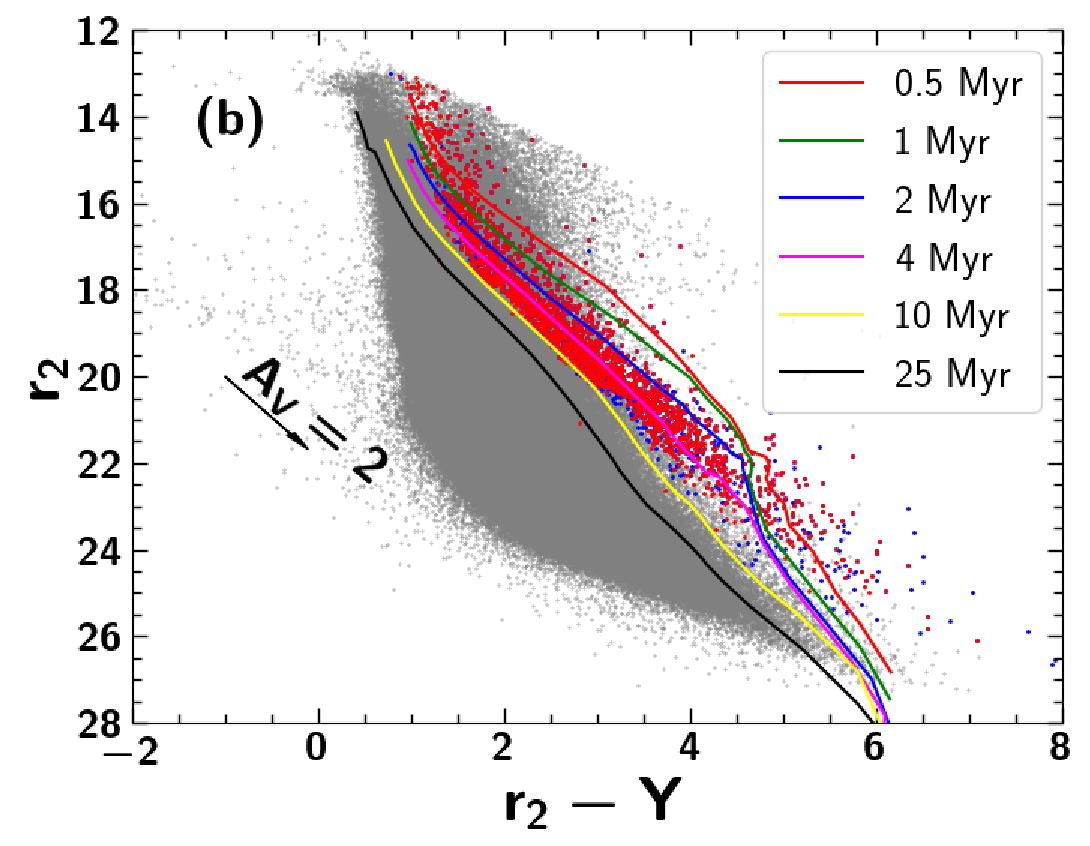}
\caption{Same as Fig. \ref{cm_hsc}. Grey dots are the full HSC+UKIDSS 802100 stars, and the blue dots are the 2425 stars detected in this work. The red dots are the 2310 stars retrieved by including $\rm A_V$ into the analysis. }
\label{cm_hsc_AV}
\end{figure}

\begin{figure}
\centering
\includegraphics[scale=0.5]{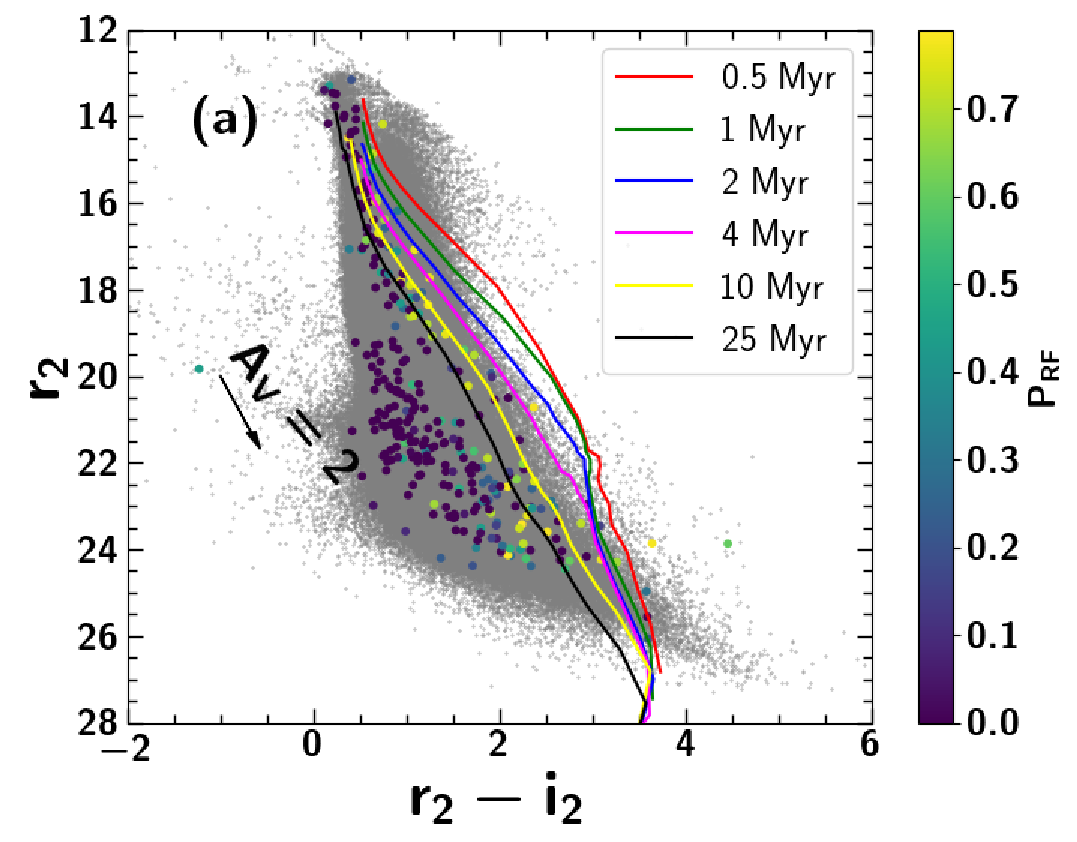}
\includegraphics[scale=0.5]{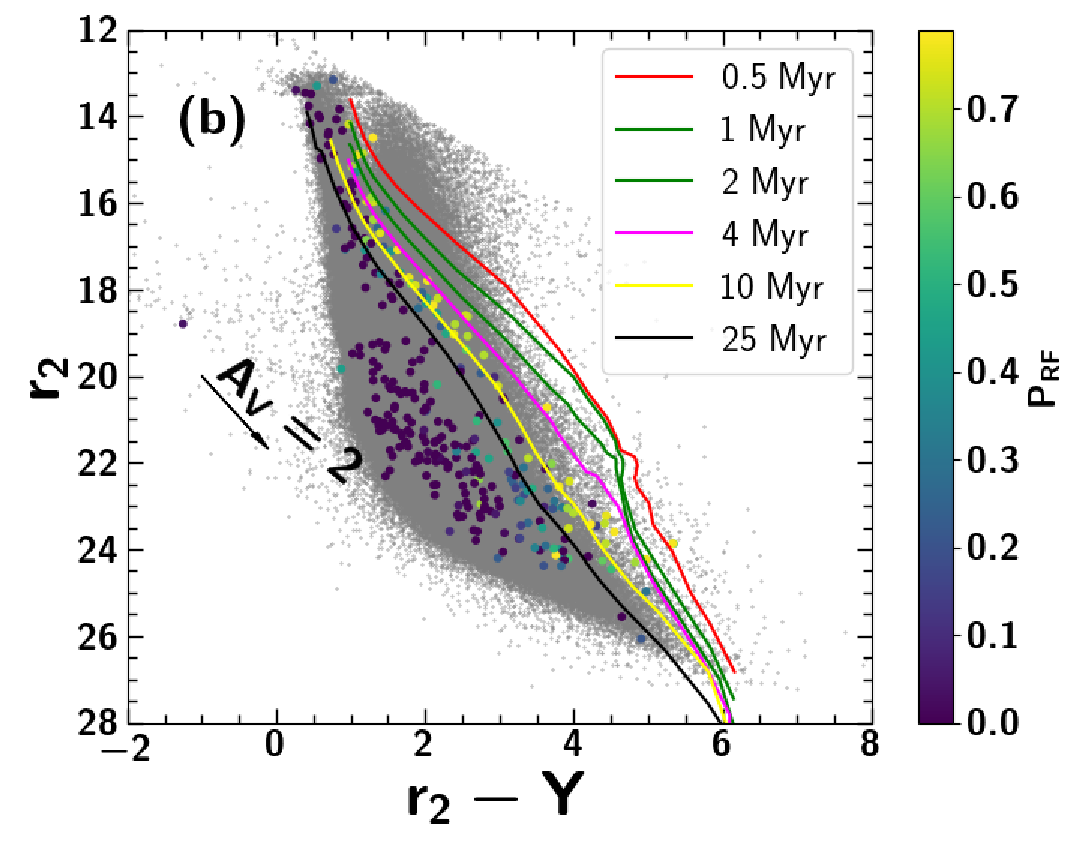}
\caption{Same as Fig. \ref{cm_hsc}. Grey dots are the full HSC+UKIDSS 802100 stars, and the dots are the 307 non-Gaia based literature stars, which have $\rm P_{RF}<0.8$ in our membership analysis. Colour of the sources demonstrate their $\rm P_{RF}$ values.}
\label{cmd_307}
\end{figure}

\end{appendix}

\end{document}